\documentclass[twocolumn, times]{aastex63}

\usepackage[version=4]{mhchem}
\usepackage{multirow}
\usepackage{graphicx}
\usepackage{color}
\usepackage{hyperref}

\begin{document}

\title{\large{Unveiling the Chemical Complexity and C/O Ratio of the HD 163296 Protoplanetary Disk: Constraints from Multi-line ALMA Observations of Organics, Nitriles, Sulfur-bearing, and Deuterated Molecules}}

\correspondingauthor{Liton Majumdar}
\email{liton@niser.ac.in, dr.liton.majumdar@gmail.com}

\author[0009-0008-2350-5876]{Parashmoni Kashyap}
\author [0000-0001-7031-8039] {Liton Majumdar}
\affiliation{Exoplanets and Planetary Formation Group, School of Earth and Planetary Sciences, National Institute of Science Education and Research, Jatni 752050, Odisha, India}
\affiliation{Homi Bhabha National Institute, Training School Complex, Anushaktinagar, Mumbai 400094, India}

\author[0000-0003-4179-6394]{Edwin A. Bergin}
\affiliation{University of Michigan, 323 West Hall, 1085 South University Ave., Ann Arbor, MI 48109, USA}

\author[0000-0003-0787-1610]{Geoffrey A. Blake}
\affiliation{Division of Geological \& Planetary Sciences, California Institute of Technology, Pasadena, CA 91125, USA}

\author[0000-0001-6124-5974]{Karen Willacy} 
\affiliation{Jet Propulsion Laboratory, California Institute of Technology, 4800 Oak Grove Dr. Pasadena, CA, 91109, USA}

\author[0000-0003-3773-1870]{St{\'e}phane Guilloteau}
\affiliation{Laboratoire d'Astrophysique de Bordeaux, Universit\'e de Bordeaux, CNRS, B18N, All\'ee Geoffroy Saint-Hilaire, F-33615 Pessac, France}

\author[0000-0003-2341-5922]{Anne Dutrey}
\affiliation{Laboratoire d'Astrophysique de Bordeaux, Universit\'e de Bordeaux, CNRS, B18N, All\'ee Geoffroy Saint-Hilaire, F-33615 Pessac, France}

 \author[0000-0003-4603-7119]{Sheng-Yuan Liu}
 \affiliation{Academia Sinica Institute of Astronomy and Astrophysics, PO Box 23-141, Taipei 106, Taiwan}
 
\author[0000-0002-1493-300X]{Thomas Henning}
\affiliation{Max Planck Institute for Astronomy, Königstuhl 17, D-69117 Heidelberg, Germany}  

\author[0000-0002-6622-8396]{Paul F. Goldsmith}
\author[0000-0002-0500-4700]{Dariusz C. Lis }
\affiliation{Jet Propulsion Laboratory, California Institute of Technology, 4800 Oak Grove Dr. Pasadena, CA, 91109, USA}

\author[0000-0003-3412-9454]{S. Maitrey}
\affiliation{Exoplanets and Planetary Formation Group, School of Earth and Planetary Sciences, National Institute of Science Education and Research, Jatni 752050, Odisha, India}
\affiliation{Homi Bhabha National Institute, Training School Complex, Anushaktinagar, Mumbai 400094, India}

\author[0000-0001-8292-1943]{Neal Turner}
\author[0000-0002-6858-5063]{Raghvendra Sahai}
\affiliation{Jet Propulsion Laboratory, California Institute of Technology, 4800 Oak Grove Dr. Pasadena, CA, 91109, USA} 

\author[0000-0002-3024-5864]{Chin-Fei Lee}
\affiliation{Academia Sinica Institute of Astronomy and Astrophysics, PO Box 23-141, Taipei 106, Taiwan}

\author[0000-0003-0769-8627]{Masao Saito}
\affiliation{National Astronomical Observatory of Japan, National Institutes of Natural Sciences, 2-21-1 Osawa, Mitaka, Tokyo 181-8588, Japan}
\affiliation{Graduate Institute for Advanced Studies, SOKENDAI, 2-21-1 Osawa, Mitaka, Tokyo 181-8588, Japan}

\begin{abstract}
 The physical and chemical conditions within a protoplanetary disk play a crucial role in determining its chemical composition, which is subsequently inherited by any forming planets. To probe these conditions, high-resolution molecular line observations, coupled with modelling, are essential. In this study, we investigate the chemistry of the nearby, massive, and relatively line-rich protoplanetary disk around HD 163296 using high-resolution observations from ALMA across Bands 3, 4, 6, and 7. We constrain the disk-averaged and radial distributions of column density and excitation temperature for the detected molecules using the new retrieval code \textsc{DRive}. The disk chemistry is modelled using the astrochemical code \textsc{PEGASIS}, with variations in the initial elemental C/O ratio. Our modelling, informed by molecular observations of \ce{HCO+}, \ce{DCO+}, \ce{HCN}, \ce{DCN}, \ce{CS}, \ce{HC3N}, \ce{H2CO}, \ce{CH3OH}, \ce{HNCO}, and \ce{NH2CHO}, allows us to place strong constraints on the C/O ratio, with a best-fit value of 1.1 that is broadly consistent with previous estimates. We present the highest-resolution DCO$^+$ emission map of this disk to date, revealing triple-ringed chemical substructures that closely align with the dust continuum rings. Additionally, our results provide the first and most stringent upper limits on the column densities of \ce{NH2CHO} and \ce{HNCO} in this protoplanetary disk, measured at $< 7 \times 10^{11}$~cm$^{-2}$ and $< 1 \times 10^{11}$~cm$^{-2}$, respectively. Our chemical models suggest that \ce{NH2CHO} and \ce{HNCO} predominantly form on grain surfaces within the disk. However, physico-chemical desorption mechanisms are inefficient at releasing these species into detectable gas-phase abundances, yet they remain promising targets for future ALMA observations.

\end{abstract}

\keywords{Protoplanetary disks (1300); Planet formation (1241); Astrochemistry (75)}

\section{Introduction}
A protoplanetary disk sets up the conditions for nascent planetary systems. These conditions are dictated by the interplay of inherited material from previous evolutionary stages and in situ chemical and physical evolution. Even though it is not very straightforward \citep{Molliere2022}, several studies have linked protoplanetary disk composition to the exoplanetary atmospheric composition, aiming to understand its implications for planet formation within the disk \citep[e.g.,][]{Bitsch2022, Schneider2021, Mordasini2016, Thiabaud2015, Madhusudhan2014, Oberg2011_2}. Hence, accurately modelling the chemistry of protoplanetary disks, informed by high-resolution observations, is crucial for advancing our understanding of planet formation.

To date, over 30 gas-phase molecules (excluding isotopologues) have been detected in protoplanetary disks at millimetre wavelengths \citep{oberg23, McGuire2022}. Most of these are relatively simple diatomic or triatomic species, although a few larger molecules have also been observed \citep{Booth2024a, Booth2024b, Loomis2020, Favre2018, Walsh2016, Oberg2015Natur, Qi2013}. The most complex organic molecule detected so far is dimethyl ether (\ce{CH3OCH3}), a nine-atom species found in the disk around MWC 480 \citep{yamato24} and Oph IRS 48 \citep{Brunken2022}. These detections have largely been enabled by the Atacama Large Millimeter/submillimeter Array (ALMA). Angular resolution is crucial because planet formation occurs on AU scales that would otherwise remain unresolved. The Molecules with ALMA at Planet-forming Scales (MAPS) ALMA Large Program \citep{Oberg2021} provided the highest-resolution molecular line observations to date for five disks: IM Lup, GM Aur, AS 209, MWC 480, and HD 163296, all of which exhibit diverse substructures \citep{Law2021_MAPSIII}.

The chemistry of these molecules within a disk is governed by a complex interplay of several factors, including the disk’s density structure, dust properties, stellar characteristics, UV and X-ray radiation, cosmic ray penetration, and ionisation processes \citep{Henning13}. Additionally, the initial chemical composition inherited from the parent molecular cloud plays a crucial role in shaping the disk’s molecular budget. At the onset of cloud evolution, a significant fraction of key elements, such as carbon, oxygen, and nitrogen, exists in refractory forms, shaping the chemical landscape of the emerging protoplanetary system. \citet{Oberg2021review} highlighted that the total budgets of oxygen and nitrogen in interstellar sources remain unaccounted for, underscoring the need for better constraints on elemental abundance ratios. Addressing these uncertainties is essential for developing a more complete picture of the molecular inventory that nascent planets inherit, ultimately influencing their potential for habitability.

The common assumption is that the elemental abundance ratios in a protoplanetary disk mirror those of its central star, given their shared origin in the same parent molecular cloud. However, this assumption may falter for elements that are mostly contained in volatile species, since their gas phase abundances are susceptible to their respective snowlines \citet{Oberg2021review}. Carbon and Oxygen are two such elements that are known to occur mostly in volatile species in the solar nebula \citep{Lodders2003}. Accurately constraining the gas-phase carbon-to-oxygen (C/O) ratio can thus provide valuable insights into the physical and chemical processes in protoplanetary disks.

In this study, we investigate the chemistry in the relatively nearby and massive protoplanetary disk around HD 163296, located approximately 101 pc away \citep{Gaia2018}, with a disk mass estimated to range from 0.1 to 0.5~$M_{\odot}$ \citep{Isella2007, Tilling2012, Muro2018, Powell2019, Kama2020}. This disk is known for its rich molecular inventory beyond the water snowline \citep[e.g.][]{Oberg2021, Bergner2019, Salinas2017, Qi2011, Thi2004}, making it a prime candidate for studying complex chemistry in protoplanetary environments. HD 163296 also features three prominent continuum gaps at 49, 87, and 145 au \citep[see the continuum image in the bottom right corner of Figure~\ref{fig:m0};][]{Huang2018a}, with multiple studies suggesting the presence of forming planets within these gaps \citep{Isella2016, Isella2018, Pinte2018, Teague2018, Teague2019, Pinte2020, Rodenkirch2021, Calahan2021, Alarcón2022, Calcino2022, Izquierdo2022, Pezzotta2025}. The large and massive nature of the HD 163296 disk, along with its striking rings and gaps, makes it a compelling analogue to systems that may form wide-orbit giant planets. These features are reminiscent of directly imaged planetary systems such as HR 8799, where multiple massive planets are observed at large orbital separations \citep{Marois08, Marois10}. Recent atmospheric studies of directly imaged exoplanets have placed strong constraints on key elemental ratios, such as C/O and metallicity, providing important insights into their formation pathways \citep{Molliere20}.
These comparisons underscore the potential of systems like HD 163296 to link observed disk substructures with exoplanetary compositions and architectures.

A number of studies have used ALMA data to investigate specific aspects of disk chemistry in HD 163296. For example, \citet{Zhang2021} used CO observations to model disk structures, while \citet{Guzman2021} examined the elemental budgets of carbon, nitrogen, and oxygen using \ce{HCN}, \ce{C2H}, and \ce{H2CO}. \citet{Bosman2021} used \ce{C2H} observations to constrain the C/O ratio, and \citet{Bergner2021} investigated photochemistry through CN and HCN. \citet{LeGal2021} and \citet{Ma2024} explored sulfur chemistry and C/O ratios using CS, \ce{C2S}, and \ce{SO}, establishing CS/SO as a promising C/O tracer. Similarly, \citet{Aikawa2021} modelled \ce{HCO+} to probe ionization structures, and \citet{Calahan2023} recently analyzed the chemistry of HD 163296 using \ce{CH3CN}, HCN, and \ce{HC3N}. \citet{Salinas2017, Salinas2018} investigated warm and cold formation pathways in the disk using \ce{DCO+}, \ce{DCN}, and \ce{N2D+}, while \citet{Cataldi2021} examined disk deuteration using these molecules along with their non-deuterated counterparts. Importantly, each of these studies focuses on a small set of molecules, typically two or three, chosen to address specific scientific questions. However, we observe dozens of molecular species in protoplanetary disks simultaneously. There is thus a pressing need to develop a more comprehensive understanding of disk chemistry that considers a larger pool of molecular species, rather than limiting analyses to just a few at a time.

We investigate multiple molecular transitions from 10 molecular species observed toward the protoplanetary disk around HD 163296 (see Table \ref{tab:disk_averaged}), including isocyanic acid (\ce{HNCO}) and the complex organic molecule formamide (\ce{NH2CHO}). Formamide is particularly significant due to its inclusion of the amide bond [\ce{-NH-C(=O)-}], making it a key precursor in amino acid synthesis and thus a molecule of major relevance for prebiotic chemistry. Although \ce{NH2CHO} has been widely detected in protostellar environments and cometary materials \citep{Colzi2021, Lis97, Morvan2000}, its presence in protoplanetary disks remains largely unconfirmed. To date, the only tentative detection of \ce{NH2CHO} in such an environment has been reported by \citet{Fadul2025} in the disk around the outbursting star V883 Ori.

Isocyanic acid (\ce{HNCO}), the most stable isomer among the four simplest molecules containing hydrogen, nitrogen, carbon, and oxygen, also plays a key role in astrochemical processes. It has been observed in a wide range of astrophysical environments, including interstellar molecular clouds \citep[e.g.,][]{Snyder1972, Turner1991, Martín2008}, hot cores \citep[e.g.,][]{MacDonald1996, Helmich1997, Bisschop2008, Nazari2022, Taniguchi2023}, molecular outflows \citep[e.g.,][]{Rodriguez2010}, hot corinos \citep[e.g.,][]{Lopez2015, Bergner2017}, and comets \citep{Lis97, Biver2006, Biver24}. However, despite its ubiquity and significance, \ce{HNCO} has not yet been detected in a protoplanetary disk, leaving an important gap in our understanding of its role in disk chemistry.

To address these gaps, we analyze high-resolution ALMA observations using two in-house tools: the disk retrieval code \textsc{DRive} and the chemical modelling code \textsc{PEGASIS}. In particular, we apply an extensive chemical network that is based on the \texttt{KIDA 2024} database, incorporating deuterium fractionation for molecules with up to six atoms. The remainder of this paper is organized as follows. In Section \ref{sec:obs}, we describe the observations and the reduction procedures employed. We derive constraints on the disk-averaged and radial distributions of column densities and excitation temperatures for the detected molecules, and we report column density upper limits for the undetected molecules, as detailed in Section \ref{sec:obs_constrainsts}. We model the disk chemistry with varying carbon-to-oxygen (C/O) ratios to identify the best-fitting model that explains the observed molecular lines of ten species, including organics, nitriles, sulfates, and deuterated molecules. The physical and chemical details of the explored models are thoroughly discussed in Section \ref{sec:thermo_chem_model}. Our results and their implications are presented in Section \ref{sec:discussion}, with key findings summarized in Section \ref{sec:conclusions}.

\begin{deluxetable*}{lcccccccccc}
\tablewidth{0pt}
\tabletypesize{\scriptsize}
\tablecaption{Spectroscopic Properties of the Targeted Lines \label{tab:line-props}}
\tablehead{
\colhead{Species} & \colhead{Transition} & \colhead{Frequency} & \colhead{log$_{10}$(A$_{ul}$)} & \colhead{g$_u$} & \colhead{E$_{u}$} & \colhead{r$_{\text{range}}$} & \colhead{beam$_{\text{maj}}\times$ beam$_{\text{min}}$, PA} & \colhead{$\delta v$} & \colhead{rms} & \colhead{Catalog}\\[-1ex] 
\colhead{} & \colhead{} & \colhead{[GHz]} & \colhead{[s$^{-1}$]} & \colhead{} & \colhead{[K]} & \colhead{[$''$]} & \colhead{[$'' \times '', ^{\circ}$]} & \colhead{[km s$^{-1}$]}& \colhead{[mJy beam$^{-1}$]} & \colhead{} }
\startdata
\multicolumn{11}{c}{2018.1.00181.S (Band 4)} \\ \hline
\ce{NH2CHO} & J(K$_a$, K$_c$) = 7(0,7) - 6(0,6) & 146.871614 & -3.6590 & 13 & 28.33 & 0.9, 2.1 & $1.23 \times 0.94, 106.53$ & 0.13 & 1.5 & JPL\\
 & J(K$_a$, K$_c$) = 6(1,5) - 5(1,4) & 131.618004 & -3.8223 & 11 & 25.11 & 0.9, 2.1 & $1.37 \times 1.06, 105.29$ & 0.13 & 1.3 & JPL\\
HNCO & J(K$_a$, K$_c$) = 6(0,6) - 5(0,5) & 131.885734 & -4.5119 & 13 & 22.15 & 0.9, 2.1 & $1.36 \times 1.05, 106.41$ & 0.14 & 1.3 & CDMS\\
 & J(K$_a$, K$_c$) = 6(1,5) - 5(1,4) & 132.356701 & -4.5245 & 13 & 65.51 & 0.9, 2.1 & $1.36 \times 1.05, 106.59$ & 0.14 & 1.3 & CDMS\\
\ce{CH3OH} & J(K$_a$, K$_c$,F) = 3(2,2,0) - 2(2,2,0) & 145.124332 & -5.1617 & 28 & 51.64 & 0.9, 2.1 & $1.24 \times 0.96, 106.47$ & 0.23 & 0.9 & CDMS \\
 & J(K$_a$, K$_c$,F) = 3(2,1,1) - 2(2,1,1) & 145.126191 & -5.1694 & 28 & 36.17 & 0.9, 2.1 & $1.24 \times 0.96, 106.47$ & 0.25 & 1.0 & CDMS \\
 & J(K$_a$, K$_c$,F) = 3(1,1,1) - 2(1,1,1) & 145.131864 & -4.9491 & 28 & 34.98 & 0.9, 2.1 & $1.23 \times 0.96, 106.47$ & 0.25 & 1.0 & CDMS \\
 & J(K$_a$, K$_c$,F) = 3(2,0,0) - 2(2,0,0) & 145.133415 & -5.1615 & 28 & 51.64 & 0.9, 2.1 & $1.23 \times 0.96, 106.47$ & 0.25 & 1.0 & CDMS \\
\ce{p-H2CO} & J(K$_a$,K$_c$) = 2(0,2)-1(0,1) & 145.602949 & -4.1072 & 5 & 10.48 & 0.3, 4.0 & $0.73 \times 0.55, 117.46$ & 0.13 & 1.9 & CDMS \\
\ce{HC3N} & J = 16-15 & 145.560960 & -3.6155 & 33 & 59.38 & 0.0, 1.5 & $0.73 \times 0.55, 117.46$ & 0.13 & 1.9 & CDMS \\
\ce{DCO+} & J = 2-1 & 144.077289 & -3.6740 & 5 & 10.37 & 0.0, 3.0 & $0.74 \times 0.56, 117.39$ & 0.25 & 1.5 & CDMS \\
DCN & J = 2-1 & 144.828001 & -3.8978 & 15 & 10.43 & 0.0, 2.5 & $1.13 \times 0.87, 108.21$ & 0.25 & 1.1 & CDMS \\
CS & J = 3-2 & 146.969029 & -4.2167 & 7 & 14.11 & 0.0, 2.0 & $0.72 \times 0.54, 117.95$ & 0.13 & 2.3 & CDMS \\ \hline
\multicolumn{11}{c}{2018.1.01055.L (Band 6)} \\ \hline
\ce{p-H2CO} & J(K$_a$,K$_c$) = 3(0,3)-2(0,2) & 218.222192 & -3.5501 & 7 & 20.96 & 0.3, 4.0 & $0.15 \times 0.10, 105.17$ & 0.2 & 0.9 & CDMS \\
\ce{HC3N} & J = 29-28 & 263.792308 & -2.8349 & 59 & 189.92 & 0.0, 1.5 & $0.14 \times 0.10, 104.06$ & 0.2 & 0.8 & CDMS \\
HCN & J(F) = 3(3)-2(2) & 265.886434 & -3.1292 & 7 & 25.50 & 0.0, 2.0 & $0.14 \times 0.09, 103.17$ & 0.2 & 1.0 & CDMS \\
 & J(F) = 3(3)-2(3) & 265.884891 & -4.0322 & 7 & 25.50 & 0.0, 2.0 & $0.14 \times 0.09, 103.17$ & 0.2 & 1.0 & CDMS \\
 & J(F) = 3(2)-2(2) & 265.888522 & -3.8861 & 5 & 25.50 & 0.0, 2.0 & $0.14 \times 0.09, 103.17$ & 0.2 & 1.0 & CDMS \\
DCN & J = 3-2 & 217.238538 & -3.3396 & 21 & 20.85 & 0.0, 2.5 & $0.15 \times 0.10, 105.14$ & 0.2 & 1.0 & CDMS \\ \hline
\multicolumn{11}{c}{2018.1.01055.L (Band 3)} \\ \hline
\ce{HC3N} & J = 11-10 & 100.076392 & -4.1096 & 23 & 28.82 & 0.0, 1.5 & $0.31 \times 0.23, 93.39$ & 0.5 & 0.6 & CDMS \\
\ce{HCO+} & J = 1-0 & 89.188525 & -4.3782 & 3 & 4.28 & 0.1, 3.0 & $0.34 \times 0.26, 93.99$ & 0.5 & 0.6 & CDMS \\
HCN & J(F) = 1(1)-0(1) & 88.630416 & -4.6184 & 3 & 4.30 & 0.0, 2.0 & $0.34 \times 0.26, 94.38$ & 0.5 & 0.6 & CDMS \\
 & J(F) = 1(2)-0(1) & 88.631848 & -4.6185 & 5 & 4.30 & 0.0, 2.0 & $0.34 \times 0.26, 94.38$ & 0.5 & 0.6 & CDMS \\
 & J(F) = 1(0)-0(1) & 88.633936 & -4.6184 & 1 & 4.30 & 0.0, 2.0 & $0.34 \times 0.26, 94.39$ & 0.5 & 0.6 & CDMS \\
CS & J = 2-1 & 97.980953 & -4.7749 & 5 & 7.05 & 0.0, 1.3 & $0.31 \times 0.24, 91.79$ & 0.5 & 0.5 & CDMS \\ \hline
\multicolumn{11}{c}{2011.0.000010.SV (Band 7)} \\ \hline
\ce{HCO+} & J = 4-3 & 356.734223 & -2.4471 & 9 & 42.80 & 0.1, 3.0 & $0.48 \times 0.36, 84.82$ & 0.1 & 17.0 & CDMS \\ \hline
\multicolumn{11}{c}{2021.1.00138.S (Band 7)} \\ \hline
\ce{DCO+} & J = 5-4 & 360.169778 & -2.4248 & 11 & 51.86 & 0.0, 3.0 & $0.31 \times 0.27, 108.94$ & 0.2 & 7.6 & CDMS \\
\enddata
\tablecomments{
Here, E$_{u}$ is the upper energy level of the transition, A$_{ul}$ is the Einstein coefficient for spontaneous transition, and g$_u$ is the upper state degeneracy. r$_{\text{range}}$ are the radial ranges for which integrated spectra (Figure \ref{fig:avgspec}) are generated. The spectroscopic data are obtained from the Cologne Database for Molecular Spectroscopy, CDMS \citep{Muller2001, Muller2005, Endres2016} and Jet Propulsion Laboratory, JPL Molecular Database \citep{pickett1998}.}
\end{deluxetable*}

\section{Observations} \label{sec:obs}
We investigated the presence of ALMA Band 4 transitions of \ce{NH2CHO}, \ce{HNCO}, \ce{CH3OH}, \ce{HC3N}, \ce{H2CO}, \ce{DCN}, \ce{DCO+} and \ce{CS} in the protoplanetary disk around HD 163296 (International Celestial Reference System 17:56:21.2880, -21:57:21.870) as a part of the project \#2018.1.00181.S (PI: Liton Majumdar). The observations were carried out over four executions on April 1st, 2019, and January 11th, March 3rd and 4th, 2020, with a total on-source integration time of 50 minutes. The mean precipitable water vapour (PWV) during the observations ranged from 2.4 to 5.5 mm, while the maximum recoverable scale (MRS) spanned 10.3$''$ to 12.0$''$. The interferometer arrangement incorporates 44 antennas with baselines ranging from 15 to 783 m. We obtain a velocity resolution of around 0.27 km s$^{-1}$ in this observation. To ensure that we have at least two transitions for each molecule included in our project, we complemented our observations with high-resolution data available from MAPS and other datasets of comparable resolution. Additionally, we also considered the non-deuterated counterparts of \ce{DCO+} and \ce{DCN}, namely \ce{HCO+} and \ce{HCN}, to better understand the role of deuteration. We utilised ALMA Band 3 and 6 transitions of \ce{HC3N}, \ce{H2CO}, \ce{HCO+}, \ce{HCN}, \ce{DCN} and \ce{CS} taken from the publicly available data provided by MAPS ALMA Large Program \citep{Oberg2021}. We also extracted the ALMA Band 7 transitions of \ce{DCO+} and \ce{HCO+} from the archival datasets \#2021.1.00138.S (PI: Ilse Cleeves) and \#2011.0.000010.SV, respectively. Dataset \#2021.1.00138.S was observed over two executions on May 15, 2022, using 44 antennas with baselines ranging from 15 to 740 m, for a total integration time of 1 hour and 44 minutes. The observations were carried out under excellent conditions, with a mean PWV of 0.4–0.5 mm, yielding a MRS of 4.2$''$ and a velocity resolution of around $ 0.20$ km s$^{-1}$. Observations for \#2011.0.000010.SV were conducted over four executions on July 11, 22, and 6, 2012, using 17 to 20 antennas with baselines ranging from 13 to 402 m, resulting in a total on-source time of 59 minutes. Here, the mean PWV levels span from 0.3 to 1.8 mm, and the velocity resolution obtained is around $0.10$ km s$^{-1}$. The spectroscopic properties of the targeted transitions from our program, along with the observational details of the four additional observing programs used in this study, are shown in Table~\ref{tab:line-props}. 

Our molecular selection, as discussed above, was guided by the overarching science goal of this work: to probe the chemical complexity of the HD 163296 disk, including simple and complex nitrogen- and oxygen-bearing organics, the C/O ratio, ionisation, deuteration fraction, and the underlying physical and chemical structure. Specifically, the molecules included in program \# 2018.1.00181.S (PI: Liton Majumdar) were chosen to provide a balanced set of tracers: (i) nitrogen-bearing simple and complex organics (\ce{HNCO}, \ce{NH2CHO},), (ii) grain-surface versus gas-phase processes and their connection to oxygen-bearing organics (\ce{H2CO}, \ce{CH3OH}), (iii) C/O ratio diagnostics (\ce{HC3N}, \ce{HCN}, \ce{DCN}, \ce{CS}), (iv) ionization and deuterium chemistry (\ce{HCO+}, \ce{DCO+}), and (v) sulfur chemistry (\ce{CS}). Complementary MAPS and archival datasets were incorporated to secure multiple transitions for each species, thereby strengthening the robustness of excitation analyses and abundance determinations. Collectively, this molecular suite provides a comprehensive framework to constrain the C/O ratio, assess deuteration, and connect the disk’s chemical inventory to its physical environment.

\subsection{Data Reduction} \label{sec:data_redcn}
All datasets were calibrated using the Common Astronomy Software Applications \citep[CASA,][]{CASATeam2022} pipeline. For project \#2018.1.00181.S, the quasar J1924–2914 was used as both the flux and bandpass calibrator, while J1733–1304 served as the phase calibrator. In project \#2021.1.00138.S, J1924–2914 again acted as the flux and bandpass calibrator, with J1742–1517 serving as the phase calibrator. For the science verification dataset \#2011.0.000010.SV, calibration was carried out using J1733–130 as the gain calibrator, J1924–292 as the bandpass calibrator, and Neptune as the flux calibrator. Continuum visibility datasets from each execution in every dataset were independently fitted with a power-law disk model projected onto the sky plane, using eight free parameters in our retrieval code \texttt{DRive}, which is described in Appendix~\ref{sec:drive}. The power law disk is modelled via four free parameters: the intensity at $0.1''$ (I$_{0}$), an outer radius (r$_{\text{out}}$), radial flux exponent (q) and background intensity (I$_{\texttt{b}}$) defined in the \texttt{diskrender.toy\_models.PowerLawDisk} function in \texttt{DRive}. The projection onto the sky plane was made using the \texttt{diskrender.utils.project\_coordinates} function, which accounts for the rest of the free parameters: inclination (inc), position angle (PA) and central offsets (x$_{c}$, y$_{c}$). We explore the parameter space using \texttt{ultranest}\footnote{\url{https://johannesbuchner.github.io/UltraNest/}} \citep{Buchner2021} with the nested sampling Monte Carlo algorithm \citep{Bunchner2016,Buchner2019}. Each execution was corrected for constrained central offsets using the \texttt{phaseshift} task in CASA. The coordinates were then realigned to a common pointing direction across all executions using the \texttt{fixplanets} task in CASA. We then performed several rounds of continuum self-phase calibration for each execution separately using the python code \texttt{auto\_selfcal}\footnote{\url{https://github.com/jjtobin/auto_selfcal.git}} and applied the solutions to each execution. For project 2018.1.00181.S, four rounds of phase self-calibration were carried out with solution intervals of ‘inf\_EB’, ‘inf’, 200 s, and 67 s. For project 2021.1.00138.S, we performed five rounds with solution intervals of ‘inf\_EB’, ‘inf’, 139 s, 42 s, and 12 s. For project 2011.0.000010.SV, we downloaded the self-calibrated data directly from the ALMA archive and carried out an additional two rounds of phase self-calibration with solution intervals of ‘inf\_EB’ and ‘inf’. These procedures improved the peak continuum SNRs by approximately 595 \%, 340 \%, and 585 \% for the three datasets, respectively. Self-calibrated executions with identical spectral setups were combined for each dataset using the \texttt{concat} task in CASA. We performed continuum subtraction on the combined datasets by fitting a first-order polynomial to the carefully selected line-free channels using the \texttt{uvcontsub} task in CASA. We constructed the images of the targeted lines from the continuum-subtracted visibility data using the \texttt{tclean} task in CASA with \texttt{hogbom} deconvolving algorithm and \texttt{briggs} weighting scheme. For weak lines, we set the robust parameter to 2 to enhance the signal-to-noise ratio, whereas for lines with a stronger signal-to-noise ratio, we used \texttt{robust} = 0 to achieve high angular resolution at sufficient SNR. Masks used during the cleaning process were generated using the Python script \texttt{keplerian\_mask.py}\footnote{\url{https://github.com/richteague/keplerian_mask.git}}, adopting a stellar mass of $2.0\, M_{\odot}$ and a systemic velocity of $5.8 \, \mathrm{km\,s^{-1}}$ \citep{Teague2021}, with a disk inclination of $46.7^{\circ}$ and position angle of $313.3^{\circ}$ \citep{Huang2018b}. The radial ranges considered for generating the masks are based on visual inspection and can be found in $r_{\text{range}}$ column in Table \ref{tab:line-props}. The generated channel maps of the observed transitions are shown in appendix \ref{app:ch_maps}. The channel maps of the MAPS transitions can be found in \citet{Oberg2021}.

\subsection{Observational Results} \label{sec:obs_results}
We first investigated our observations using the matched filtering technique \citep{loomis2018detecting} with the help of the Python package \texttt{VISIBLE} \citep{loomis2018visible}. Matched filter responses were further normalised with respect to $\sigma$, where $\sigma$ is the standard deviation of 10,000 samples randomly collected from a visually identified signal-free region of filter response spectra. For this purpose, we used the Keplerian masks generated during the cleaning process as filters. We tried different combinations of inner and outer radii for the Keplerian masks and settled on the values (r$_\text{range}$ column in Table \ref{tab:line-props}) for which the matched filter response is maximum. While most of the targeted molecules are clearly detected, the filter responses for \ce{HNCO} and \ce{CH3OH} indicate non-detections (Figure \ref{fig:matched}). For \ce{NH2CHO}, the 7(0,7)–6(0,6) transition shows a signal close to $3\sigma$, whereas the 6(1,5)–5(1,4) transition reaches only about $2\sigma$. We therefore report \ce{NH2CHO} as a non-detection, since a signal-to-noise ratio (SNR) greater than $5\sigma$ is required to claim a secure detection of COMs, preferably in more than one transition \citep{Walsh2016, Favre2018, Booth21}.

\begin{figure*}
    \centering
    \includegraphics[width=\linewidth]{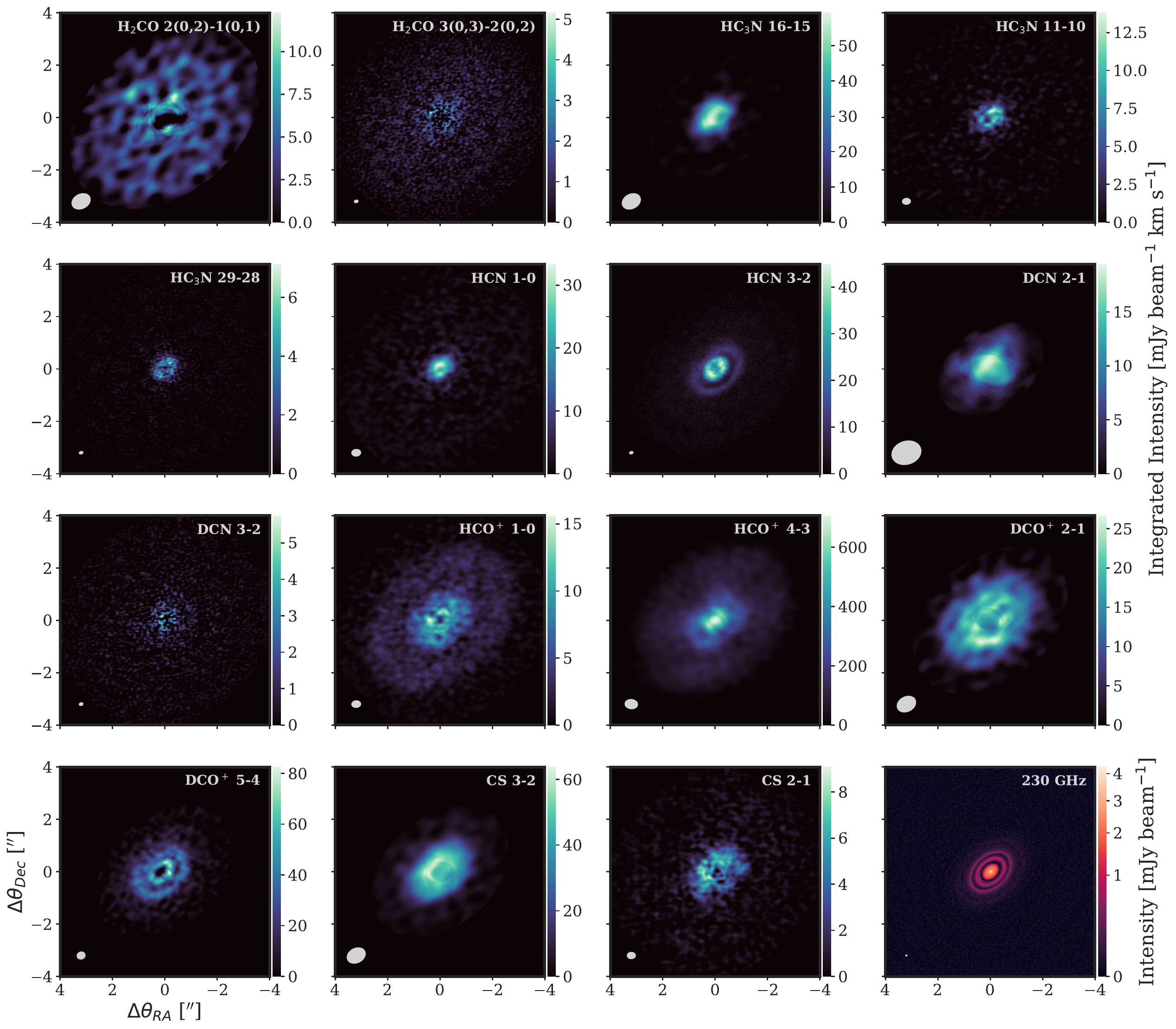}
    \caption{Integrated intensity maps of the detected transitions. A continuum image highlighting the rings and gaps is displayed at the bottom right. The synthesised beam is indicated at the bottom left of each panel.}
    \label{fig:m0}
\end{figure*}

We generated integrated intensity (zeroth moment) maps of the detected molecules using the Python package \texttt{bettermoments} \citep{bettermoments2018} while including only those pixels that are inside the Keplerian masks. Moment 0 maps of the detected molecules are shown in Figure \ref{fig:m0}. We have also included a continuum image at 230 GHz, taken from the Disk Substructures at High Angular Resolution Project \citep[DSHARP][]{Andrews2018} ALMA Large Program at the bottom right corner of Figure \ref{fig:m0}.
To better understand the radial trends in the emission distribution of the detected molecules, we generated azimuthally averaged radial distributions of velocity-integrated intensity from the zeroth moment maps. The disk's radial range was divided into annular rings with a width of one-fourth of the beam's major axis before calculating radial profiles using the \texttt{radial\_profile} function in the \texttt{GoFish} Python package \citep{GoFish}. The radial profiles for the detected molecules are shown in the left-most column of Figure \ref{fig:radial}. We clearly see a triple-ringed morphology in the case of \ce{DCO+}, \ce{H2CO} and a double-ringed one in the case of \ce{HCN}. 

\begin{figure*}
    \centering
    \includegraphics[width=0.85\linewidth]{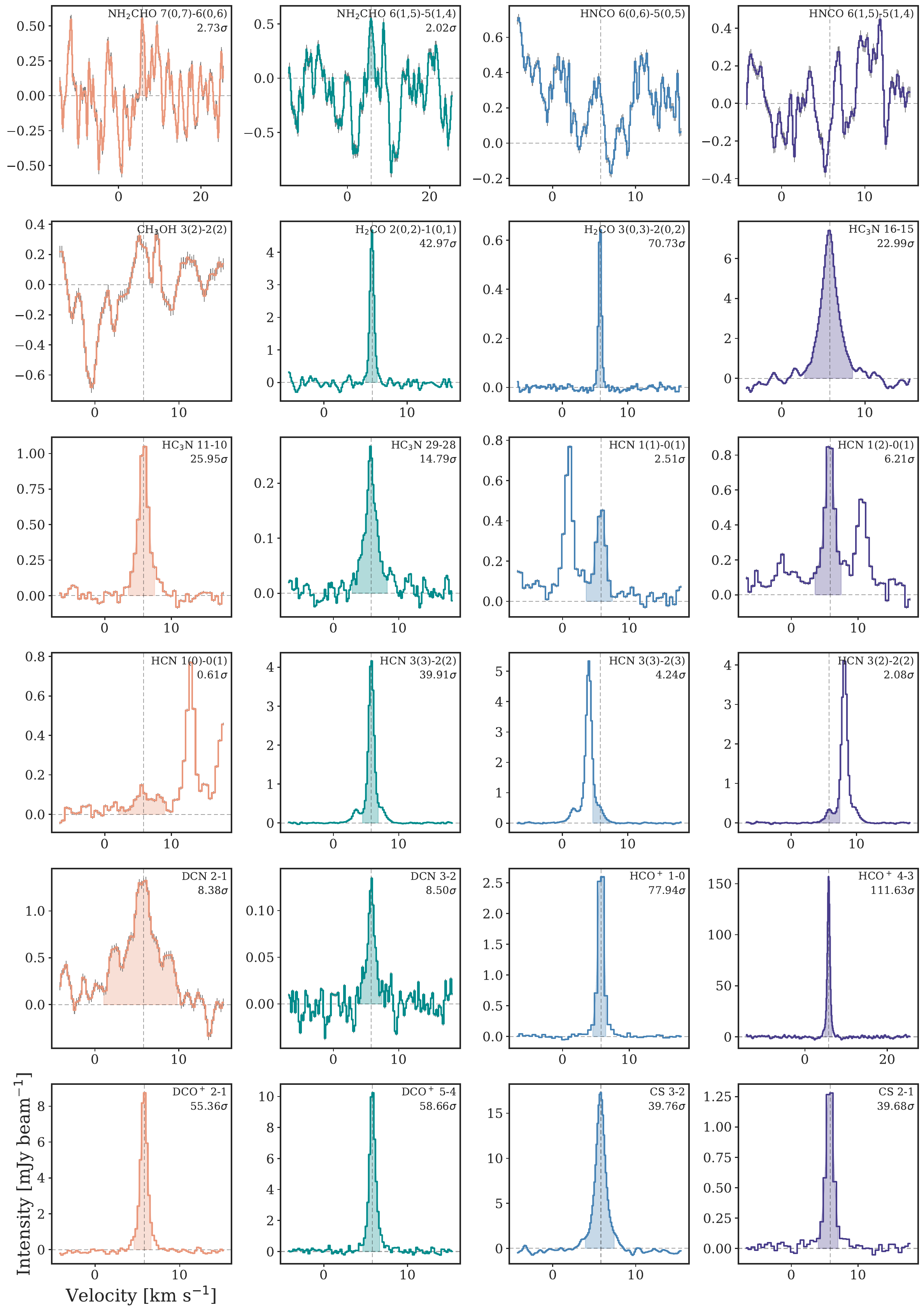}
    \caption{Keplerian-deprojected, disk-averaged spectra of all targeted transitions, plotted as a function of velocity in the LSRK frame. The vertical dashed line indicates the systemic velocity at 5.8 km s$^{-1}$, while the horizontal dashed line marks the zero level of the averaged intensity. The signal strength of each transition is noted in the top right corner of each panel. The shaded region represents the velocity range considered as signal for the signal-to-noise ratio calculation.}
    \label{fig:avgspec}
\end{figure*}

\begin{figure*}
    \centering
    \includegraphics[width=0.9\linewidth]{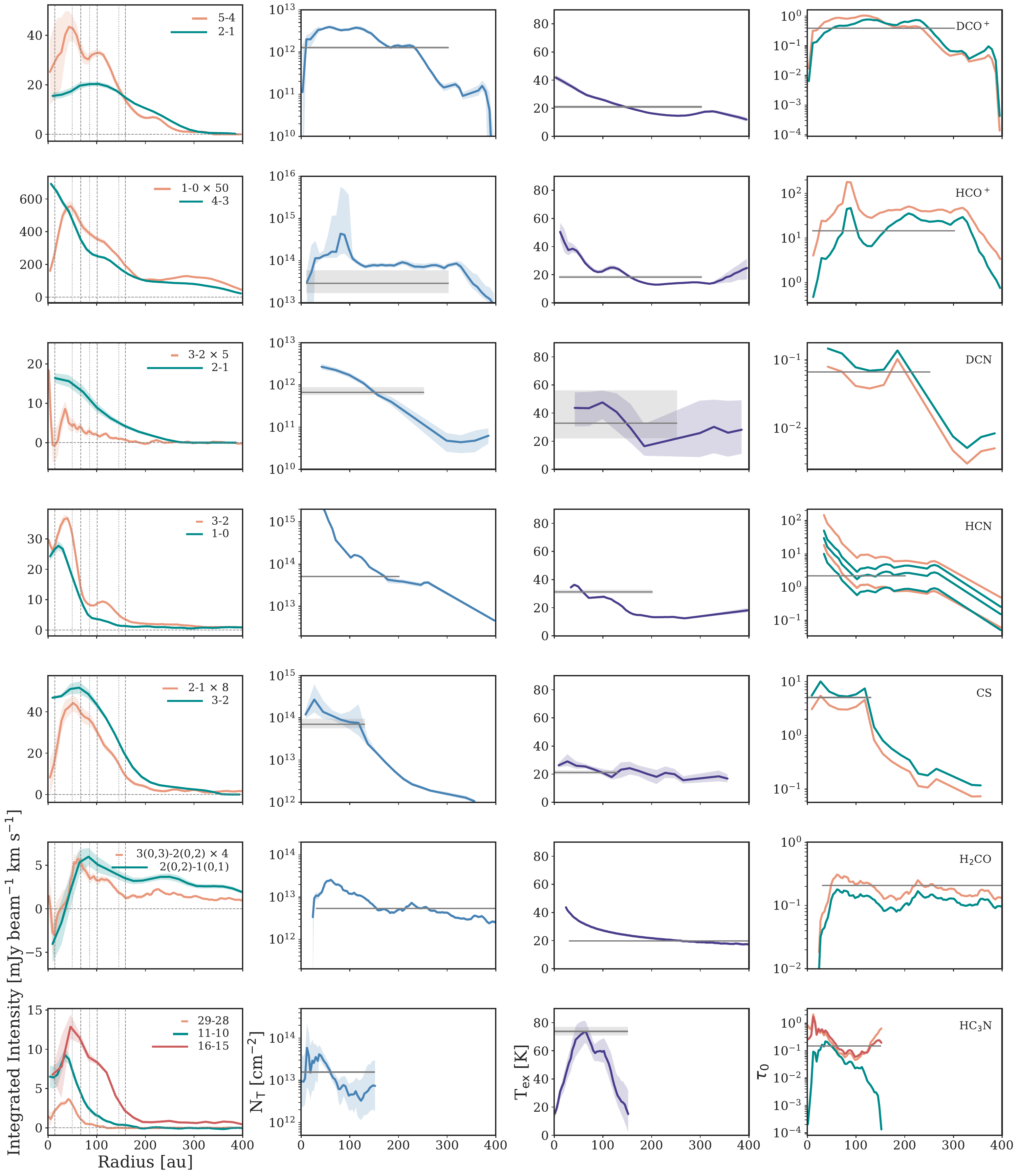}
    \caption{Azimuthally averaged radial profiles (first column) and radial distributions of column density (second column), excitation temperature (third column) and the corresponding optical depth, $\tau_0$ (fourth column) derived from our LTE-Slab model. In the first column, the horizontal line at the upper right of each subplot indicates the beam’s major axis. Dashed vertical lines mark continuum emission peaks at 14 au, 67 au, 101 au, and 159 au, while dotted vertical lines denote continuum depressions at 49 au, 85 au, and 145 au \citep{Huang2018a}. In the second, third, and fourth columns, grey horizontal lines show the disk-averaged column density, excitation temperature, and the $\tau_0$ for the transition that has the largest central-optical depth, respectively; their widths indicate the radial range used to generate the integrated spectrum. In the first three columns, shaded regions of the same colour represent the associated uncertainties. Note that the 10 \% calibration error and spectral-autocorrelation errors are not shown in the radial integrated-intensity profile but are included in the error bars of the column density and excitation temperature (second and third columns). Optical depths in the fourth column are calculated using equation \eqref{eq:tau_0_final} at median values of column density and excitation temperatures.}
    \label{fig:radial}
\end{figure*}

We extracted Keplerian deprojected disk averaged intensity spectra for all the molecules by employing the shifting and stacking technique using the \texttt{average\_spectrum} function in the \texttt{GoFish} Python package. We assume a plane disk for the velocity shift, i.e. $z/r=0$. Radial ranges considered for averaging are noted in the r$_{\text{range}}$ column in Table \ref{tab:line-props}. Even in the averaged spectra, we observe close to $3 \sigma$ signal for \ce{NH2CHO} (shaded region in the top left corner of Figure \ref{fig:avgspec}).

\section{Constraining the Excitation Conditions Using \texttt{DRive}} \label{sec:obs_constrainsts}
A LTE spectral line model within the \texttt{DRive} retrieval framework is used to constrain both disk-averaged and radially resolved excitation conditions for all molecules considered. A detailed description of the \texttt{DRive} retrieval framework and its full capabilities is provided in Appendix~\ref{sec:drive}.

\subsection{Disk Averaged Constraints} \label{sec:drive-diskavg}
To constrain the disk-averaged excitation conditions of the detected molecules, we simultaneously fit the Keplerian deprojected, disk-averaged spectra (refer to Figure \ref{fig:avgspec}) of all transitions for a given molecule with the spectral line model-generated spectra. In addition to the error bars derived from the standard deviation of the emission, scaled by the square root of the number of independent beams within the integration area, we include an additional spectral-correlation uncertainty and a 10\% calibration uncertainty in our retrieval. Calculation of spectral correlation length is similar to \citet{Cataldi2021}. The two free parameters characterising the excitation conditions are the total column density, $N_T$, and the excitation temperature, $T_{\text{ex}}$. To mitigate the effects of artificial broadening, we represent the line shape with a Voigt profile (see Appendix~\ref{sec:drive_slab}) with 2 width parameters, the Gaussian FWHM, $v_{\text{fwhm}}$ and the Lorentzian Half width at Half maximum (HWHM), $\Gamma$. When several spectral lines are used simultaneously, lines within the same ALMA band are assumed to have the same FWHM and HWHM. Additionally, we allow the systemic velocity, $v_{\text{sys}}$, to vary within a narrow range of $5.8 \pm 0.25$ km s$^{-1}$. Depending on the molecule being fitted, the model includes a total of seven to nine free parameters (see Table \ref{tab:disk_averaged}). The parameter space is explored using the Bayesian statistical method nested sampling Monte Carlo \citep{Bunchner2016,Buchner2019}, implemented via the Python package \texttt{ultranest} \citep{Buchner2021}. A log-uniform prior is applied to $N_T$ within the range $[10^7,10^{20}]$ cm$^{-2}$, while uniform priors are used for the remaining parameters. Specifically, $T_{\text{ex}}$ is constrained to $[2.73, 100]$ K, and $\Delta v_{\text{fwhm}}$ and $\Gamma$ are restricted to $[0,10]$ km s$^{-1}$. For HCN, a narrower prior range of $[2.73, 40]$ K is adopted for $T_{\text{ex}}$ to ensure that the derived disk-averaged excitation temperatures remain within a chemically relevant regime. Each HCN transition consists of three hyperfine components, which are treated following the prescription of \citet{Guzman2021}, as described in Section \ref{sec:drive_slab}. All results are presented in Table \ref{tab:disk_averaged}, with the best-fit values taken as the 50th percentile of the posterior distributions and the error bars corresponding to the 16th and 84th percentiles. The corner plots showing the posterior distributions are presented in Appendix \ref{app:corner}. In the case of \ce{H2CO}, we were unable to obtain a strong constraint on the excitation temperature, likely due to the very close $E_u$ levels between transitions. Therefore, we restricted the excitation temperature to a narrow range of $19.42^{+1.51}_{-1.59}$ K, based on the constraints by \citet{Hernandez2024}. In Figure \ref{fig:radial}, the grey horizontal bars in the second and third columns show the disk-averaged constraints on column density and excitation temperature, respectively.

\subsection{Radial Distribution of Column Density and Excitation Temperature} \label{sec:drive-slab-radial}
We first construct a radial distribution of Keplerian-deprojected azimuthally-averaged spectra using \texttt{radial\_spectra} function in the Python package \texttt{GoFish}. For this purpose, we divide the radial extent of the disk into radial bins of width set to one-fourth of the beam major axis. In each radial bin, the velocity is then shifted according to the Keplerian rotation of the disk, and the spectra are then azimuthally averaged. To get the constraints on the radial distribution of column density and excitation temperature, we first fit the radial spectra for all the transitions for a given molecule corresponding to each radial bin independently in the same way we did the disk-averaged calculations (refer to Section \ref{sec:drive-diskavg}). The radial profiles for the constrained excitation temperature, $T_{\text{ex}}$, are presented in the third column of Figure \ref{fig:radial}. For different molecules, we have placed different outer radial cut-offs based on the radial extent of emission after visually inspecting the radial intensity profiles (leftmost column of Figure \ref{fig:radial}) and the moment 0 maps (Figure \ref{fig:m0}). Once we have determined the radial distribution of the excitation temperature ($T_{\text{ex}}$) using all available transitions, we proceed to fit the radial distribution of the total column density ($N_T$). This is done using radial spectra generated from the most spatially resolved line cube while limiting the prior range of $T_{\text{ex}}$ to the previously derived values during the fitting process. For \ce{H2CO}, we were unable to obtain a robust constraint on the radial distribution of $T_{\text{ex}}$. To derive the radial distribution of $N_T$ for \ce{H2CO}, we fixed $T_{\text{ex}}$ to $30 \times (r/75)^{-0.33} \pm 5$ K, based on the constraints provided by \citet{Hernandez2024}. For \ce{HCN}, we restricted the prior range in $T_{\text{ex}}$ between 10 K and the gas temperature referred from peak brightness temperature of $^{12}$CO 2-1 from \citet{Law2021} for a particular radius following \citet{Guzman2021}, as its emission surface is relatively flat. Radial distributions of $N_T$ for all the molecules are shown in the second column of Figure \ref{fig:radial}.

\begin{deluxetable*}{lcccccccccc}
\tablewidth{0pt}
\tablecaption{Constraints on Disk-Averaged Excitation Conditions from \textsc{DRive} LTE-SLAB Analysis \label{tab:disk_averaged}}
\tablehead{
\colhead{Molecule} &
\colhead{N$_{T}$} &
\colhead{T$_{ex}$} &
\colhead{v$_{sys}$} &
\colhead{v$_{fwhm,1}$} &
\colhead{$\Gamma_{1}$} &
\colhead{v$_{fwhm,2}$} &
\colhead{$\Gamma_{2}$} &
\colhead{v$_{fwhm,3}$} &
\colhead{$\Gamma_{3}$} &
\colhead{$\tau_{0,\text{max}}$ (transition)$^{(a)}$} \\[-1ex]
\colhead{} & \colhead{[cm$^{-2}$]} & \colhead{[K]} & \colhead{[km s$^{-1}$]} & \colhead{[km s$^{-1}$]} & \colhead{[km s$^{-1}$]} & \colhead{[km s$^{-1}$]} & \colhead{[km s$^{-1}$]} & \colhead{[km s$^{-1}$]} & \colhead{[km s$^{-1}$]} & \colhead{}
}
\startdata
\ce{NH2CHO} & $< 6.7\times 10^{11}$ & 30 (fixed) & -- & -- & -- & -- & -- & -- & -- & -- \\
\ce{HNCO} & $< 1.3\times 10^{11}$ & 30 (fixed) & -- & -- & -- & -- & -- & -- & -- & --\\
\ce{CH3OH} & $< 7.5\times 10^{12}$ & 30 (fixed) & -- & -- & -- & -- & -- & -- & -- & -- \\
\ce{H2CO} & $5.4^{+0.2}_{-0.2}\times 10^{12}$ & $19.42^{+1.59(b)}_{-1.51}$ & $5.77^{+0.01}_{-0.01}$ & $0.91^{+0.07}_{-0.08}$ & $0.06^{+0.03}_{-0.03}$ & $0.31^{+0.06}_{-0.07}$ & $0.12^{+0.02}_{-0.02}$ & -- & -- & 0.21 (3(0,3) - 2(0,2)) \\
CS & $7.10^{+2.54}_{-1.46}\times 10^{13}$ & $21.15^{+1.56}_{-1.24}$ & $5.79^{+0.02}_{-0.02}$ & $0.97^{+0.52}_{-0.59}$ & $1.05^{+0.08}_{-0.09}$ & $2.34^{+0.05}_{-0.05}$ & $0.00^{+0.00}_{-0.00}$ & -- & -- & 5.06 (3-2) \\
\ce{DCO+} & $1.3^{+0.1}_{-0.1}\times 10^{12}$ & $20.97^{+1.16}_{-1.07}$ & $5.79^{+0.02}_{-0.02}$ & $0.58^{+0.21}_{-0.27}$ & $0.37^{+0.05}_{-0.05}$ & $0.86^{+0.16}_{-0.16}$ & $0.13^{+0.06}_{-0.06}$ & -- & -- & 0.39 (5-4) \\
\ce{HCO+} & $3.0^{+6.1}_{-1.4}\times 10^{13 (c)}$ & $16.67^{+1.19}_{-1.11}$ & $5.74^{+0.01}_{-0.01}$ & $1.16^{+0.04}_{-0.04}$ & $0.00^{+0.00}_{-0.00}$ & $0.88^{+0.12}_{-0.12}$ & $0.21^{+0.03}_{-0.03}$ & -- & --  & 15.75 (1-0)\\
\ce{DCN} & $6.7^{+2.2}_{-0.9}\times 10^{11}$ & $32.74^{+23.49}_{-10.85}$ & $5.76^{+0.10}_{-0.09}$ & $2.34^{+2.35}_{-1.73}$ & $1.71^{+0.53}_{-0.82}$ & $1.06^{+0.80}_{-0.73}$ & $1.04^{+0.23}_{-0.29}$ & -- & -- & 0.07 (2-1)\\
HCN & $5.08^{+0.26}_{-0.22}\times 10^{13}$ & $31.16^{+1.57}_{-1.34}$ & $5.73^{+0.01}_{-0.01}$ & $0.78^{+0.09}_{-0.09}$ & $0.38^{+0.02}_{-0.01}$ & $0.23^{+0.23}_{-0.16}$ & $0.78^{+0.07}_{-0.07}$ & -- & -- & 2.16$^{(c)}$  (3(3)-2(2)) \\
\ce{HC3N} & $1.6^{+0.1}_{-0.1}\times 10^{13}$ & $73.86^{+3.25(d)}_{-3.04}$ & $5.80^{+0.02}_{-0.02}$ & $3.19^{+0.57}_{-0.62}$ & $0.75^{+0.22}_{-0.24}$ & $4.04^{+1.33}_{-1.55}$ & $1.30^{+0.48}_{-0.54}$ & $1.88^{+0.19}_{-0.20}$ & $0.30^{+0.06}_{-0.06}$ & 0.11 (16-15) \\
\enddata
\tablecomments{
$^{(a)}$ We report the optical depth only for the transition (indicated in parentheses) that has the largest central optical depth, $\tau_0$.
$^{(b)}$ The excitation temperature (T$_\mathrm{ex}$) for \ce{H2CO} is fixed following \citet{Hernandez2024}.
$^{(c)}$ The impact of opacity on disk-averaged constraints and their robustness is discussed in Section \ref{sec:optical_depth}.
$^{(d)}$ The temperature constraint for \ce{HC3N} may be uncertain, as discussed in Section~\ref{sec:hc3n}.
}
\end{deluxetable*}

\subsection{Optical Depth and It's Implication on Excitation Conditions} \label{sec:optical_depth}
Our spectral line model inherently computes the optical depth based on the column density and excitation temperature, assuming purely thermal broadening (see Section \ref{sec:drive_slab}). The radial distributions of the line-centre optical depths, $\tau_0$, for each transition of every molecule are shown in the fourth column of Figure \ref{fig:radial}. These values are calculated using Equation \eqref{eq:tau_0_final}, adopting the median (50th percentile) posterior values of column density and excitation temperature (see Section \ref{sec:drive-slab-radial}). We also compute the optical depth for all transitions using the derived disk-averaged column density and excitation temperature. For each species, we report the optical depth of the transition with the largest line-centre optical depth, $\tau_0$, in the last column of Table \ref{tab:disk_averaged}, and we display these same values as grey horizontal bars in the fourth column of Figure \ref{fig:radial}.

The lines from molecules \ce{DCO+}, \ce{DCN}, \ce{H2CO}, and \ce{HC3N} are found to be optically thin, while the lines from CS show moderate optical thickness ($1 < \tau_0 < 10$). In contrast, \ce{HCO+} and \ce{HCN} are highly optically thick ($\tau_0 > 10$). Because the optical depths are calculated assuming only intrinsic (thermal) line broadening, our derived radial profiles of column density and excitation temperature should remain robust even for optically thick species, provided that local turbulence in the disk is negligible \citep[e.g.,][]{Teague2018b}.

However, for the disk-averaged constraints to remain meaningful for optically thick lines, the optical depth should not vary strongly with radius across the region used to compute the disk-integrated spectrum (i.e., the radial range indicated by the grey bars in Figure \ref{fig:radial}). For CS, line opacities vary only marginally in the region where we integrate the emission $(\leq 1.3'')$, making the CS disk-averaged column density and excitation temperature values reliable. However, this is not the case for \ce{HCO+}, for which the disk-averaged optical depth,  and as a consequence, the disk-averaged column density may be underestimated. Comparison of disk-averaged values with the radial distribution of \ce{HCO+} column density in Figure \ref{fig:radial} shows that this underestimation is relatively minor. A stronger effect occurs for HCN, where opacities span two orders of magnitude, also affecting the disk-averaged excitation temperature. Fortunately, the disk-averaged column density remains relatively independent of this temperature in the range of derived values.

\subsection{Calculation of Upper Limits on Column Density} \label{sec:upper_limit}
To determine upper limits on the column density of the undetected molecules \ce{NH2CHO}, \ce{HNCO}, and \ce{CH3OH}, we assume a $3\sigma$ intensity value, $S_{\nu,3\sigma}$, integrated over a velocity range of $\Delta v = [3.75,7.5]$ km s$^{-1}$. The noise level, $\sigma$, is estimated as the standard deviation of 10,000 samples randomly drawn from the integrated spectra outside the selected velocity range. The integration area ($\Omega$) used to compute the integrated spectra is calculated with the radial range specified in the $r_{\text{range}}$ column of Table \ref{tab:line-props}. Under the assumption of LTE, the upper limit on the total column density, $N_T$, follows a Boltzmann distribution:
\begin{align}
    N_T = N_u \frac{Q(T_{\text{ex}})}{g_u} \exp{ \frac{E_u}{T_{\text{ex}}} } \\
    N_u = \frac{4 \pi}{hc A_{ul}} \frac{S_{\nu,3\sigma} \Delta v}{\Omega}
\end{align}
\noindent
Here, $N_u$ is the column density of molecules occupying the upper energy level $E_u$ of the transition. Here, $g_{\text{u}}$ is the degeneracy, $A_{ul}$ is the Einstein coefficient and $Q(T_{\text{ex}})$ is the partition function evaluated at the excitation temperature, $T_{\text{ex}}$ calculated through cubic interpolation of tabulated $Q$ values for discrete $T_{\text{ex}}$ values found in CDMS \citep{Muller2001, Muller2005, Endres2016} or JPL \citep{pickett1998} database. For these calculations, we fixed the excitation temperature, $T_{\text{ex}}$, to be 30 K.  The derived upper limits on the column densities for the undetected molecules are reported in the $N_T$ column of Table \ref{tab:disk_averaged}.

\section{PEGASIS DISK Model} \label{sec:thermo_chem_model}
In this section, we describe the thermochemical model used to analyze molecular non-detections and multi-transition detections in order to better understand the disk chemistry and constrain the elemental C/O ratio.
\subsection{Disk Physical Structure}\label{sec:phy_model}
\paragraph{Density structure}
We model an azimuthally symmetric disk structure consisting of three distinct mass components: gas, a population of small grains, and a population of large grains. Each of these components follows a parametric mass surface density distribution, $\Sigma_i(r)$, characteristic of a self-similar viscous disk, as described by \citet{Lynden1974} and \citet{Andrews2011}:

\begin{align}
\Sigma_i (r) = \Sigma_{c,i} \left(\frac{r}{r_c}\right)^{-\gamma} \exp{ \left[- \left(\frac{r}{r_c} \right)^{2-\gamma} \right]} \label{eq:sigma_i}
\end{align}

\noindent
where $\Sigma_{c,i}$ represents the characteristic surface density of the $i^{\text{th}}$ mass component at the characteristic radius $r_c$, and $\gamma$ is the surface density exponent. To determine $\Sigma_{c,i}$, we utilize previously constrained total mass values, $M_i$, for each component \citep{Zhang2021} :

\begin{align}
M_i = \int_{r_{\text{in}}}^{r_{\text{out}}} \Sigma_{i}(r)r dr d\phi \label{eq:M_i}
\end{align}

\begin{figure*}
    \centering
    \includegraphics[width=\linewidth]{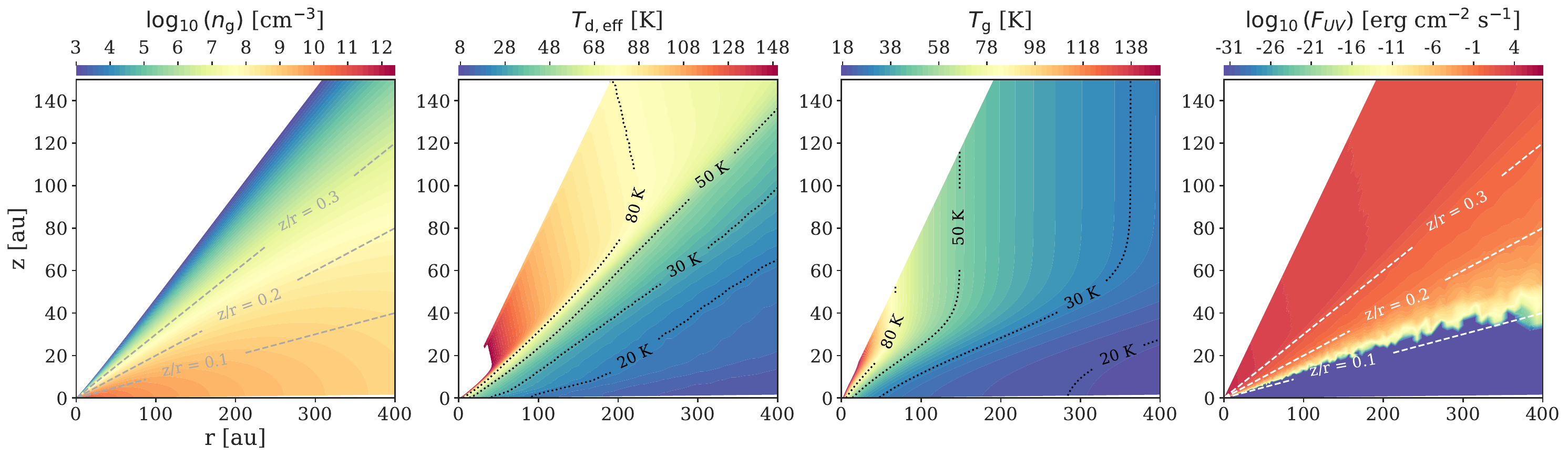}
    \caption{The physical structure of the protoplanetary disk around HD 163296. The first panel illustrates the distribution of gas number density, given by $n_g = \rho_g / (\mu m_H)$, where $\mu=2.31$ is the mean molecular weight, and $m_H$ is the mass of a hydrogen atom. The second and the third panels present the effective dust temperature (Equation \eqref{eq:td_eff}) and the gas temperature distribution (Equation \eqref{eq:Tg}), respectively, with black dotted lines representing the temperature contours. The fourth panel displays the ultraviolet flux distribution (Equation \eqref{eq:Fuv}). The grey dashed contours represent constant $z/r$ surfaces within the disk.}
    \label{fig:phy_struct}
\end{figure*}

After substituting our parametric prescription for mass surface density (Equation \ref{eq:sigma_i}) in the above relation (Equation \ref{eq:M_i}), we obatain:
\begin{align}
    \Sigma_{c,i} = \frac{(2-\gamma) M_{i}}{2\pi r_c^2} \left[ e^{-\left( \frac{r_{\text{in}}}{r_{\text{c}}} \right)^{2-\gamma}} -  e^{-\left( \frac{r_{\text{out}}}{r_{\text{c}}} \right)^{2-\gamma}}\right]^{-1}
\end{align}
\noindent
In the above calculations, our disk is confined between the inner radius, $r_{\text{in}} = 0.45$ au and the outer radius $r_{\text{out}} = 600$ au. The calculated values of $\Sigma_{c,i}$ are presented in Table \ref{tab:physical_model}. Based on the solutions of hydrostatic pressure balance for a vertically isothermal disk, the vertical density structure, $\rho_{i}$ takes a Gaussian form,
\begin{align}
    \rho_i(r,z) = \frac{\Sigma_i(r)}{\sqrt{2\pi}H_i(r)} \exp{ \left[ - \frac{1}{2} \left(\frac{z}{H_i(r)} \right)^2 \right] } \label{eq:rho_rz}\\
    H_i(r) = \chi_i H_{100} \left( \frac{r}{100 \text{ au}}\right)^\psi
\end{align}
\noindent
Here, $H_{100}$ is the scale height at 100 au, the $\psi$ is the power law index for radial distribution of scale height, and $\chi_i$ is the settling factor. We assume that the gas particles and the small grains are well coupled and share the same radial scale height, $H_i$. We keep the $\chi_i$ value at 1 for the gas and small grain population. Large grains, however, experience a significant settling effect \citep{Nakagawa1986, Dullemond2004}. To account for this, we adopt a reduced scale height by setting the $\chi_{\text{lg}}$ value to 0.2 for the large grains \citep{Andrews2011}. The gas number density structure n$_g$ throughout the disk is calculated from gas volume density, $\rho_{g}$, and is shown in the first panel of Figure \ref{fig:phy_struct}.

\paragraph{Stellar spectrum}
HD 163296 is a relatively warm Herbig Ae star, characterized by a stellar mass, $M_*$ of $2.0 M_\odot$ \citep{Teague2021}, a stellar luminosity, $L_*$ of $17.0 L_\odot$ \citep{Fairlamb2015} and an accretion rate of $\dot M$ of $10^{-7.4}$ $M_\odot$ yr$^{-1}$ \citep{Fairlamb2015}. The central star is expected to emit strong X-ray and ultraviolet (UV) radiation \citep{Hartmann2016}, which plays a crucial role in shaping the disk’s chemistry. For this study, we adopt the composite UV-X-ray stellar spectrum utilized by \citet{Zhang2021}, as shown in Figure 24 in the Appendix of their paper.

\paragraph{Dust temperature structure}
Following the description of dust properties in \citet{Zhang2021}, we assume that the grain populations adhere to the Mathis, Rumpl, \& Nordsieck (MRN) size distribution, $n(a) \propto a^{-p}$ \citep{Mathis1977}, with grain sizes ranging from $a_{\text{min}}$ to $a_{\text{max}}$, as listed in Table \ref{tab:physical_model}. Wavelength-dependent dust opacity values are directly adopted from publicly available MAPS data \citep{Oberg2021, Sierra2021}, which is based on DSHARP calculations \citep{Birnstiel2018}. Given the dust density structure, wavelength-dependent opacities, and the stellar spectrum, we compute the dust temperature distribution, $T_j$, for each dust population using the 3D Monte Carlo radiative transfer code \texttt{RADMC-3D} \citep{Dullemond2012}. Since our model includes two dust populations, the effective dust temperature, $T_{d,\text{eff}}$, at each grid cell (r,z) results from the combined contributions of both populations. We determine $T_{d,\text{eff}}$ under the assumption that dust grains behave as perfect black bodies, ensuring that the total power radiated at each (r,z) is conserved according to the Stefan-Boltzmann law:
\begin{align}
    \sigma_{\text{SB}} A_{\text{eff}}(r,z)T_{d,\text{eff}}^4(r,z) = \sum_{j} \sigma_{\text{SB}} n_j(r,z) A_j T_j^4(r,z) \label{eq:SBlaw}
\end{align}
\noindent
where the effective surface area, \( A_{\text{eff}}(r,z) \), at each grid point \((r,z)\) is given by the grain number-weighted area summed over all dust populations (\( j \)) i.e. $A_{\text{eff}} = \sum_j A_j$ with $A_j = n_ja_j^2$ representing the effective area of the \( j^{\text{th}} \) dust grain population. The number density of grains in the \( j^{\text{th}} \) population at a given grid point \((r,z)\) is expressed as $n_j = \frac{V_{\text{cell}}\rho_j(r,z)}{\frac{4}{3}\pi a_j^3 d_j}$, where \( V_{\text{cell}} \) is the volume of the grid cell, \( \rho_j(r,z) \) is the dust density, and \( d_j \) is the material density of the dust grains. For this study, we assume a constant grain material density of $d_j = 3.0$ g cm$^{-3}$ for both dust populations \citep{Birnstiel2018}. After substituting $n_j$ and $A_j$ in Equation \eqref{eq:SBlaw}, we obtain the following expression, which is used to calculate the effective dust temperature, \( T_{d,\text{eff}}(r,z) \), at each grid point:
\begin{align}
    T_{d,\text{eff}} (r,z) = \left[ \frac{\sum_j \frac{\rho_jT_j^4}{a_jd_j}}{\sum_j \frac{\rho_j}{a_jd_j}} \right]^{1/4} \label{eq:td_eff}
\end{align}
\noindent
The $T_{d, \text{eff}}$ distribution in the disk is presented in the second panel of Figure \ref{fig:phy_struct}.

\paragraph{Gas temperature structure}
We construct a 2D distribution of gas temperature similar to the one proposed by \citet{Dartois2003} and later modified by \citet{Dullemond2020}, where the gas temperature is divided into two layers, a midplane layer and an atmospheric layer. Their temperatures are characterized by radial power laws:
\begin{align}
    T_{g,\text{mid}} (r) = T_{\text{mid},0} \left( \frac{r}{100 \text{ au}} \right)^{q_{\text{mid}}} \\
    T_{g,\text{atm}} (r) = T_{\text{atm},0} \left( \frac{r}{100 \text{ au}} \right)^{q_{\text{atm}}} 
\end{align}

\noindent
where $q_{\text{mid}}$ and $q_{\text{atm}}$ are the power-law exponents for the radial distribution of midplane temperature, $T_{g,\text{mid}}$ and atmospheric temperature, $T_{g,\text{atm}}$ respectively. The two gas layers are connected through a tangent hyperbolic function:
\begin{align}
    T_g^4(r,z) = T^4_{g,\text{mid}}(r) + \frac{1}{2} \left[ 1 + \tanh{ \left( \frac{z - \alpha z_q(r)}{z_q(r)} \right)} \right] T_{g,\text{atm}}^4(r) \label{eq:Tg}
\end{align}
where $z_q(r) = z_0(r/100 \text{ au})^s$. Here, $\alpha$ controls the height at which the transition in the tanh vertical temperature profile occurs and $s$ controls how the transition height varies with radius. All the parameters describing the 2D gas temperature distribution are constrained using \ce{CO} and their isotopologue observations by \citet{Law2021} and can be found in Table \ref{tab:physical_model}. The distribution of $T_g$ in the disk is presented in the third panel of Figure \ref{fig:phy_struct}.

\paragraph{Radiation field and visual extinction}
For the aforementioned dust and stellar properties, we perform a monochromatic Monte Carlo radiative transfer using \texttt{RADMC-3D} to calculate the 2D distribution of frequency-dependent mean intensities, $J_\nu(r,z)$. For simplicity, we assume that radiation is scattered isotropically by dust particles and compute the UV flux, $F_{\text{UV}}(r,z)$, using the following expression:
\begin{align}
    F_{\text{UV}}(r,z) = \int_{\text{UV}} I_{\nu}(r,z)\Omega d\Omega = 4\pi \int_{\text{UV}} J_{\nu}(r,z) d\nu \label{eq:Fuv}
\end{align}
\noindent
The fourth panel in Figure \ref{fig:phy_struct} shows the F$_{\text{UV}}$ distribution in the disk. We calculate the visual extinction, $A_V(r,z)$ from F$_{\text{UV}}$ following the prescription by \citet{Du2014},  

\begin{align}  
    A_V(r,z) = -1.086 \ln{\left[ \frac{F_{\text{UV}}(r,z)}{F_{\text{UV},0}(r, z)} \right]} \cdot \dfrac{1}{Q_{\text{UV}}}
    \label{eq:av}
\end{align}

where $F_{\text{UV},0}(r, z)$ represents the unattenuated UV flux from the star as received at the same point, assuming the fall in flux only due to the inverse-square law and we take $Q_{\text{UV}} = 2.6$ which transforms UV extinction to visual \citep{Wild2011}.

\paragraph{Effective grain size}
We consider two distinct dust populations, each characterized by MRN distribution (refer to the \textit{Dust Temperature Structure} section). To quantify the representative grain size for each population, we first compute the mean grain size, $\langle a_j\rangle$, from the averaged surface area:

\begin{align}
    4 \pi \langle a_j\rangle^2 &= \frac{ \int_{a_{\text{min},j}}^{a_{\text{max},j}} 4 \pi a^2 a^{-p} da }{\int_{a_{\text{min},j}}^{a_{\text{max},j}} a^{-p} da} \\
\text{or, } \langle a_j \rangle &= \left[ \left( \frac{1-p}{3-p} \right) \frac{a_{\text{max},j}^{3-p} - a_{\text{min},j}^{3-p}}{a_{\text{max},j}^{1-p} - a_{\text{min},j}^{1-p}} \right]^{1/2}
\end{align}
\noindent
In our chemical model, we provide the number density weighted effective grain size at each grid point (r,z), calculated as:
\begin{align}
    \langle a \rangle (r,z) = \frac{\sum_j \frac{\rho_j(r,z) \langle a_j \rangle}{\langle a_j \rangle^3 d_j}}{\sum_j \frac{\rho_j(r,z)}{\langle a_j \rangle^3 d_j}}
\end{align}
where, $\rho_j(r,z)$ is the volume density of the $j^{\text{th}}$ dust population previously defined parametrically in Equation \ref{eq:rho_rz}.

Our physical model builds on \citet{Zhang2021}, who used the thermo-chemical code \texttt{RAC2D} \citep{Du2014} to compute chemical abundances and gas temperatures in the disk around HD 163296. However, we adopt the observationally constrained temperature profile from \citet{Law2021} and then compute chemistry using the astrochemical code \texttt{PEGASIS} (refer to Section \ref{sec:chem_mod}). To reproduce the observed radial profiles, \citet{Zhang2021} introduced a radial dependence in the CO abundance relative to \ce{H2}. Similarly, \citet{Calahan2021} extended the \citet{Zhang2021} framework to constrain the disk’s thermal structure, but modified the CO depletion profile using additional constraints from CO isotopologues, multiple higher-level CO transitions, and HD flux. Their resulting temperature structure shows good agreement with the empirically derived temperatures of \citet{Law2021}, which we also employ in this work.

\begin{deluxetable*}{lcc}
\tablecaption{Parameter prescription of the disk physical model\label{tab:physical_model}}
\tablehead{
\colhead{Parameter Description} & \colhead{Values} & \colhead{References}
}
\startdata
Characteristic radius, $r_c$ [au] & 165 & \citet{Zhang2021} \\
Surface density exponent, $\gamma$ & 0.8 & \citet{Zhang2021} \\
Total gas mass, $M_{g}$ [$M_\odot$] & 0.14 & \citet{Zhang2021} \\
Total mass of the small grain population, $M_{sg}$ [$M_\odot$] & $2.00 \times 10^{-4}$ & \citet{Zhang2021} \\
Total mass of the large grain population, $M_{lg}$ [$M_\odot$] & $2.31 \times 10^{-3}$ & \citet{Zhang2021}\\
Surface density of gas at $r_c$, $\Sigma_{c,g}$ [g cm$^{-2}$] & 8.80 & calculated \\
Surface density of the small grain population at $r_c$, $\Sigma_{c,sg}$ [g cm$^{-2}$] & $1.30 \times 10^{-2}$ & calculated \\
Surface density of the large grain population at $r_c$, $\Sigma_{c,lg}$ [g cm$^{-2}$] & $1.50 \times 10^{-1}$ & calculated \\
Scale height at 100 au, $H_{100}$ [au] & 8.4 & \citet{Zhang2021} \\
Power index for radial scale height distribution, $\psi$ & 1.08 & \citet{Zhang2021} \\
Minimum grain size for small and large grains, $a_{\text{min}}$ [$\mu$m] & 0.005 & \citet{Zhang2021} \\
Maximum grain size for the small grains, $a_{\text{max,sg}}$ [$\mu$m] & 1 & \citet{Zhang2021} \\
Maximum grain size for the large grains, $a_{\text{max,lg}}$ [mm] & 1 & \citet{Zhang2021} \\
Exponent of MRN dust distribution, $p$ & 3.5 & \citet{Zhang2021} \\
Atmospheric temperature at 100 au, $T_{\text{atm},0}$ [K] & 63 & \citet{Law2021} \\
Midplane temperature at 100 au, $T_{\text{mid},0}$ [K] & 24 & \citet{Law2021} \\
Radial power law exponent for atmospheric temperature, $q_{atm}$ & -0.61 & \citet{Law2021} \\
Radial power law exponent for midplane temperature, $q_{mid}$ & -0.18 & \citet{Law2021} \\
Tangent hyperbolic transition height at radius 100 au, $z_{0}$ [au] & 9 & \citet{Law2021} \\
$\alpha$ & 3.01 & \citet{Law2021}\\
$s$ & 0.42 & \citet{Law2021}
\enddata
\end{deluxetable*}

\subsection{Disk Chemistry} \label{sec:chem_mod}
We model the temporal evolution of molecular abundances using our gas-grain astrochemical model \texttt{PEGASIS} \citep{pegasis25}. \texttt{PEGASIS} is a versatile tool capable of simulating both two-phase (gas and grain surface) and three-phase (gas, grain surface, and grain bulk) chemistry across various astrochemical environments, such as protoplanetary disks, within a 1+1D framework. For this work, we employ the three-phase chemistry framework, as grain-surface and mantle interactions play an equally important role in determining the chemical evolution of a protoplanetary disk environment \citep{Andersson2008,Oberg2009}. Our chemical network is based on \texttt{KIDA-2024} \citep{kida2024}, and includes deuterated species containing up to six atoms \citep{Kashyap2024}, which comprises 1,117 gas-phase species, 518 grain-surface species, and 518 grain-mantle species, interconnected by 41,829 gas-phase reactions and 24,845 grain-surface and mantle reactions. Table \ref{tab:initial_abun} lists the initial elemental abundances, adopted from Table 7 in \citet{Kashyap2024}. Our chemical model incorporates a two-dimensional, species-dependent variation in cosmic-ray ionisation (CRI) rates based on \citet{Padovani2018}, along with a two-dimensional distribution of X-ray ionisation rates following the prescription of \citet{Gorti2004}. The central star's X-ray luminosity is assumed to be $7.76 \times 10^{29}$ erg s$^{-1}$ \citep{Anilkumar2024}. We also include ionisation from the decay of short-lived radionuclides (SLRs) in protoplanetary disks, following the prescription of \citet{Cleeves2013}.

\begin{deluxetable}{lcc}
\tablecaption{Initial abundances used in our astrochemical model \label{tab:initial_abun}}
\tablewidth{\columnwidth}
\tablehead{
\colhead{Element} & \colhead{Abundance relative to H} & \colhead{References}
}
\startdata
\ce{H2}  & $0.5$ &  \\
He  & $9.00 \times 10^{-2}$ & 1 \\
N   & $6.20 \times 10^{-5}$ & 2 \\
O   & $(0.85 - 3.3) \times 10^{-4}{}^{(a)}$ & 3 \\
\ce{C+}  & $1.70 \times 10^{-4}$ & 2 \\
\ce{S+}  & $8.00 \times 10^{-8}$ & 4 \\
\ce{Si+} & $8.00 \times 10^{-9}$ & 4 \\
\ce{Fe+} & $3.00 \times 10^{-9}$ & 4 \\
\ce{Na+} & $2.00 \times 10^{-9}$ & 4 \\
\ce{Mg+} & $7.00 \times 10^{-9}$ & 4 \\
\ce{P+}  & $2.00 \times 10^{-10}$ & 4 \\
\ce{Cl+} & $1.00 \times 10^{-9}$ & 4 \\
F   & $6.68 \times 10^{-9}$ & 5 \\
HD  & $1.60 \times 10^{-5}$ & 6 \\
\enddata
\tablerefs{ (1) \citet{Wakelam2008}; (2) \citet{Jenkins2009}; (3) \citet{Kashyap2024} (4) Low metal abundances from \citet{Graedel1982}; (5) Depleted value from \citet{Neufeld2005}; (6) \citet{Majumdar2017}. (a) O initial abundance is adjusted for C/O ratio ranging from 0.5 to 2.0} 
\end{deluxetable}

\subsection{Models with Different Initial C/O Ratios}
We investigated a set of chemical models with initial carbon-to-oxygen (C/O) ratios of 0.5, 0.7, 0.9, 1.1, 1.5, and 2.0 to identify the model that best explains our observations. We evaluated the models using chi-squared statistics by simultaneously fitting the modelled column densities to the observationally constrained, disk-averaged column densities of ten species targeted in our observations, including upper limits for the three undetected molecules. We focus on disk-averaged values because our fiducial model does not account for detailed radial substructures (such as multiple rings, gaps, or broken outer rings) and the resulting variations in the complex UV field structure (see Section \ref{sec:model_disc} for limitations and future outlook). Consequently, the radial variations of molecular column density profiles predicted by our disk-chemistry model may not fully reproduce the trends retrieved from our observations (Figure \ref{fig:radial}).

Our analysis primarily targets the region within 150 au of the central star, as the majority of the observed molecular emission peaks within this radius, although some lines do extend farther out. For each model, we calculated a reduced chi-squared ($\chi^2_{\text{red}}$) value with respect to disk-averaged values within 150 au, following the methodology similar to \citet{Kashyap2024}:

\begin{align}
    \chi^2_{\text{red}} = \frac{1}{\nu} \sum_{j} \left( \frac{N_{\text{obs}}(j) - N_{\text{model}}(j)}{\sigma_{\text{obs}}(j) + \sigma_{\text{model}}(j)} \right)^2
\end{align}

In this equation, $\nu$ denotes the number of degrees of freedom. $N_{\text{obs}}(j)$ and $N_{\text{model}}(j)$ refer to the observed and modelled disk-averaged column densities of the $j$-th molecule, respectively, with the terms $\sigma_{\text{obs}}(j)$ and $\sigma_{\text{model}}(j)$ representing the associated uncertainties. We assume a 20\% model uncertainty, based on the discussion in \citet{Kashyap2024}.

Additionally, we investigated the impact of the inheritance versus reset scenario on protoplanetary disk composition. In the inheritance case, we adopted initial abundances from a typical molecular cloud, characterized by a gas and dust temperature of 10 K, a total gas density of $2 \times 10^4$ cm$^{-3}$, a visual extinction of 15 mag, and a CRI rate of $1.3 \times 10^{-17}$ s$^{-1}$, evolved over $10^6$ years \citep{Kashyap2024}. For the reset case, we initialised our model directly with atomic abundances. We then computed the time-dependent molecular abundances for our static disk structure (as described in Section~\ref{sec:phy_model}) over $6 \times 10^6$ years.

\subsection{Best-fit Chemical Model}
\begin{figure*}
    \centering
    \includegraphics[width=\linewidth]{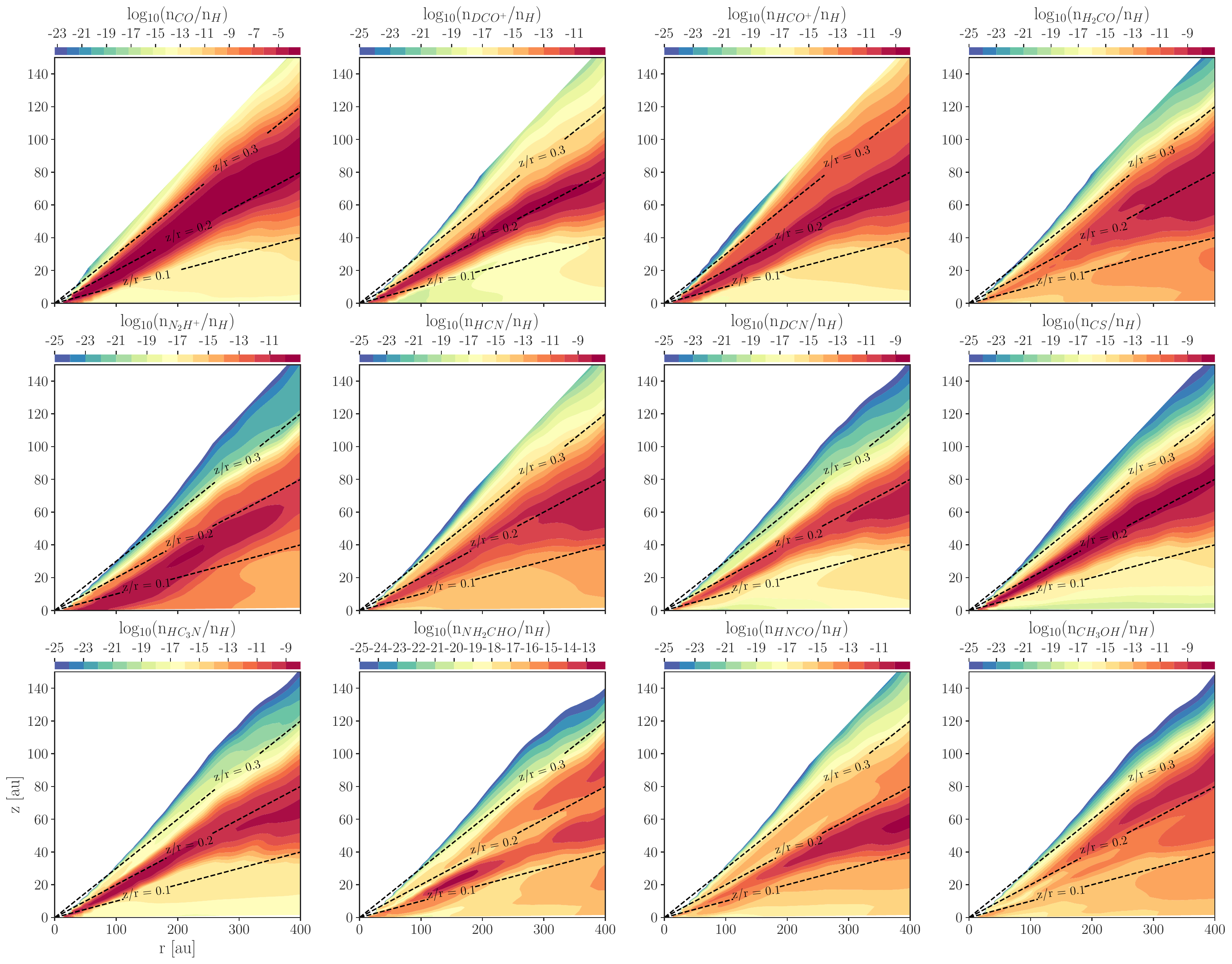}
    \caption{Simulated 2D abundance distributions for various molecules as predicted by the best model characterized by the C/O ratio of 1.1. Here, a reset scenario is shown; however, both inheritance and reset scenarios show similar abundance distributions in the case of gas phase molecules.}
    \label{fig:abun_model}
\end{figure*}

\begin{figure*}
    \centering
    \includegraphics[width=\linewidth]{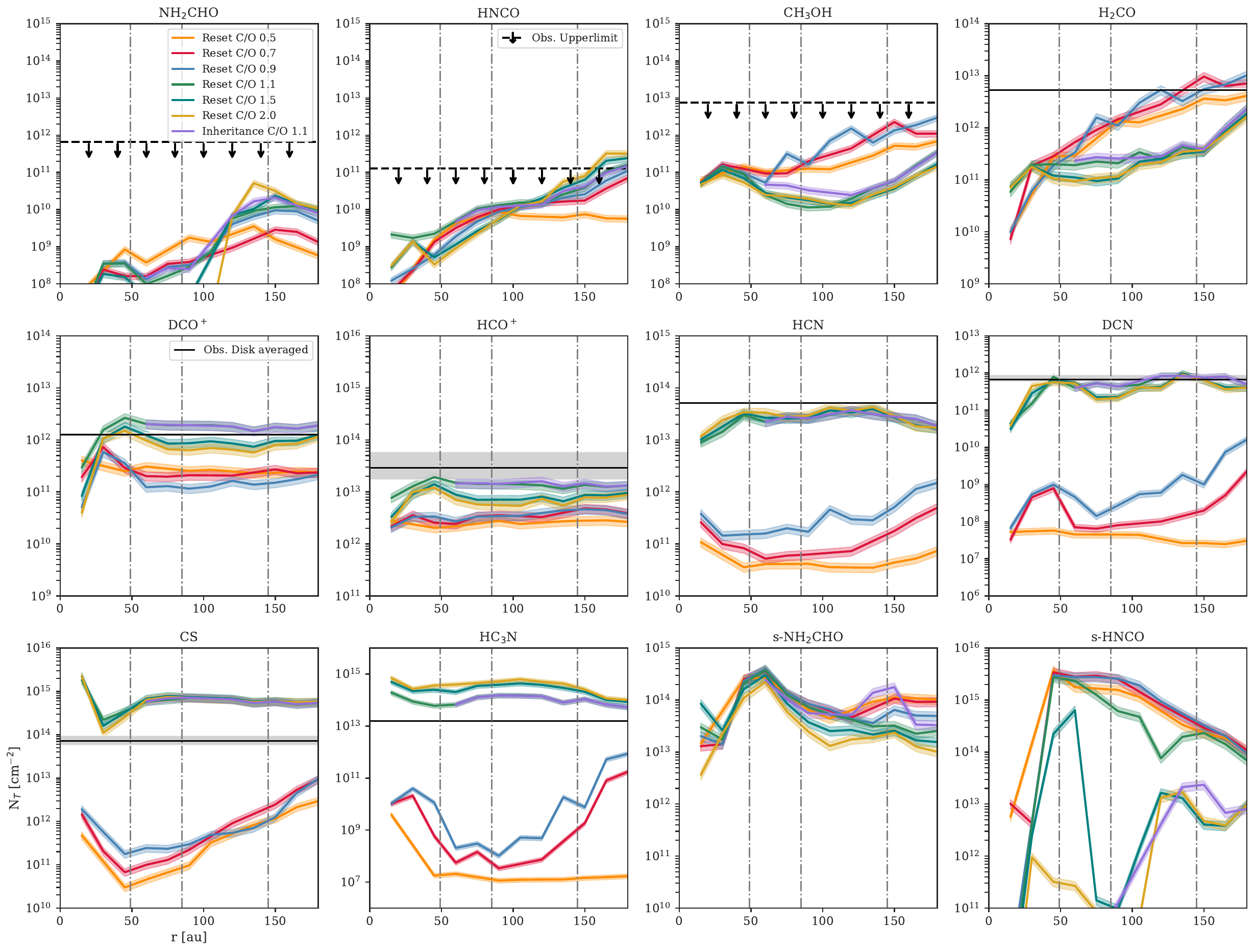}
    \caption{Comparison between observationally constrained, disk-averaged column densities and model-predicted radial column density distributions for varying initial carbon-to-oxygen (C/O) abundance ratios. A chi-squared analysis identifies the best-fit model with a C/O ratio of 1.1. Two chemical evolution scenarios are presented for this case: (1) an inheritance model, in which molecular abundances are inherited from a 1 Myr-old molecular cloud, and (2) a reset model, where the disk starts with primarily elemental abundances. A 20\% uncertainty is assumed for the model column densities when comparing with observations, following the methodology of \citet{Kashyap2024}. In the figure, dot-dashed vertical lines indicate the locations of continuum gaps at 49, 85, and 145 au. Horizontal dotted black lines with downward arrows show the observationally constrained upper limits of undetected molecules, while horizontal black bars (with grey shaded uncertainty regions) indicate the disk-averaged column densities of detected molecules determined from the DRive spectral line analysis (see section \ref{sec:drive-diskavg}).}
    \label{fig:cden_fit}
\end{figure*}

The model that best reproduces the observed disk-averaged column densities of the seven detected molecules, as well as the upper limits for the three undetected species, is characterised by a carbon-to-oxygen (C/O) ratio of 1.1 with the lowest reduced chi-squared value ($\chi^2_\text{red}$) among all tested models. Although our primary $\chi^2_\text{red}$ calculations were restricted to disk-averaged values considering radii $< 150$ au, the C/O = 1.1 model remains the best fit even when extending the analysis to the full disk (radii $< 400$ au). To further test the robustness of this result, we also performed chi-squared statistics using radially resolved column density profiles. In all cases, the C/O = 1.1 model consistently provides the best match, regardless of the chosen outer boundary. Note that for most detected molecules, the emission remains unresolved within the innermost $\sim$30 au, and thus the derived C/O ratio may not hold in this region. Simulated 2D molecular abundance distributions based on the best model are shown in Figure~\ref{fig:abun_model}. Figure~\ref{fig:cden_fit} compares the observationally constrained, disk-averaged molecular column densities with model-predicted radial distributions for a range of initial C/O ratios from 0.5 to 2.0. While the chi-squared analysis identifies C/O = 1.1 as the statistically best-fit value, models with C/O $> 1$ generally reproduce the observations well, whereas those with C/O $< 1$ show significantly poorer agreement. This trend is consistent with earlier studies such as \citet{LeGal2021} and \citet{Calahan2023}. For the best-fit case, results from the inheritance scenario are also presented in the same figure. Both the inheritance and reset scenarios yield similar outcomes for the gas-phase molecules considered here, suggesting that disk chemistry plays the dominant role in shaping their abundances. The global number density–weighted temperatures from our best-fit model are 32 K for \ce{NH2CHO}, 28 K for \ce{CH3OH}, and 29 K for HNCO. These values support the use of a 30 K excitation temperature as a reasonable assumption for estimating the upper limits on their column densities (see Section~\ref{sec:upper_limit}). A detailed discussion of the chemistry of individual molecules from the best-fit model is provided in Section~\ref{sec:discussion}.

\section{Discussion} \label{sec:discussion}

\subsection{\ce{NH2CHO}}
We observe at most a $3\sigma$ signal for \ce{NH2CHO} 7(0,7)-6(0,6) (Figures \ref{fig:matched}, \ref{fig:avgspec}), which is insufficient to confirm a detection. However, we have established the most stringent and the first ever upper limit on the \ce{NH2CHO} column density to date towards the protoplanetary disk around HD 163296, at approximately $6.73 \times 10^{11}$ cm$^{-2}$. \citet{Fadul2025} is the only study which reports a tentative detection of this molecule towards the protoplanetary disk V883 Ori and reported a column density of $6.025 \times 10^{14}$ cm$^{-2}$ at an excitation temperature of 100 K. However, V883 Ori is an outbursting source and is therefore not directly comparable to typical Class II disks such as HD 163296. In young protostellar systems, \ce{NH2CHO} column densities are usually on the order of $10^{12}$–$10^{13}$ cm$^{-2}$ \citep[e.g.,][]{Codella2017, marcelino2018}. By contrast, much higher values, ranging from $10^{14}$ to $10^{17}$ cm$^{-2}$, have been reported in hot cores \citep[e.g.,][]{ Colzi2021,  Taniguchi2023, Duan2025}.

In our modelled abundances (Figure \ref{fig:abun_model}), \ce{NH2CHO} is seen to increase predominantly peaking in the $0.1 < z/r < 0.2$ region. Although previous studies have investigated a gas-phase formation route for \ce{NH2CHO} via the reaction \ce{NH2 + H2CO -> NH2CHO + H} \citep{Barone2015, Skouteris2017}, our chemical modelling suggests that, in a Herbig Ae protoplanetary disk such as ours, \ce{NH2CHO} is predominantly formed on grain surfaces. The primary formation pathway involves the surface reaction between s-\ce{NH2CO} and s-\ce{H}, resulting in the production of s-\ce{NH2CHO} (where the prefix 's-' denotes species residing on the grain surface). Chemical desorption is the main mechanism responsible for releasing \ce{NH2CHO} into the gas phase, whereas photodesorption contributes only marginally. All our explored models predict \ce{NH2CHO} column densities significantly lower than the observationally constrained upper limit, which is consistent with the observations. However, the grain-surface \ce{s-NH2CHO} shows a column density far exceeding this upper limit (see Figure \ref{fig:cden_fit}). This indicates that while \ce{NH2CHO} is abundantly present on the grain surface, its desorption processes are inefficient in the disk environment, making its detection in the gas phase unlikely. Additionally, neither thermal desorption nor gas-phase formation is expected to be active, as most of the disk remains below 100 K. Given these factors, detecting \ce{NH2CHO} in the bulk disk gas with the current sensitivity of ALMA would likely require significantly longer integration times, but it offers a promising scope for potential detection.

\subsection{HNCO} \label{sec:hnco}
We also report the non-detection of isocyanic acid (\ce{HNCO}), the simplest peptide-bearing molecule, with a $3\sigma$ upper limit on its column density of $< 1.29 \times 10^{11}$ cm$^{-2}$. \ce{HNCO} has not yet been detected in any protoplanetary disk but has been observed in pre-stellar, protostellar, and cometary environments, where its column density ranges from $\sim 10^{15} - 10^{17}$ cm$^{-2}$ \citep[][and references therein]{Nazari2022}. Our chemical modelling indicates that \ce{HNCO} primarily forms in the disk layer between $z/r=0.1$ and $z/r=0.2$, mainly through the hydrogenation of \ce{s-OCN} on grain surfaces. In the inner disk, chemical desorption is the dominant mechanism releasing \ce{HNCO} into the gas phase. However, moving radially outward, photodesorption becomes increasingly important. Figure~\ref{fig:cden_fit} shows that \ce{HNCO} chemistry is highly sensitive to variations in the C/O ratio. The best-fit model with a C/O ratio of 1.1 predicts column densities below our observational upper limit within 150 au.

\subsection{\ce{CH3OH}} \label{sec:ch3oh}
We place an upper limit on the column density of methanol (\ce{CH3OH}), one of the simplest complex organic molecules, at $< 7.50 \times 10^{12}$ cm$^{-2}$. \ce{CH3OH} has previously been detected in the protoplanetary disk around the T Tauri star TW Hydrae \citep{Walsh2016}, with a column density of $3$–$6 \times 10^{12}$ cm$^{-2}$ peaking at a radial distance of 30 au. Methanol has also been observed in several Herbig Ae disks, including IRS 48 \citep{Vandermarel2021_IRS48}, HD 16942 \citep{Booth2023_hd169142}, HD 100546 \citep{Booth2024_100546, Evans2025}, and HD 100453 \citep{Booth2025_hd100453}. Reported column densities span $10^{13}$–$10^{16}$ cm$^{-2}$, with excitation temperatures ranging from $\sim$50 K in the outer regions of HD 100546 to 228 K in the innermost regions of HD 100453. It is worth noting that \citet{Evans2025} attributes the presence of \ce{CH3OH} in HD 100546 to thermal desorption of ice in the inner disk (0–110 au) and non-thermal desorption in the outer disk (180–260 au). For our disk, \citet{Carney2019} provided a stricter $3\sigma$ upper limit of $5.0 \times 10^{11}$ cm$^{-2}$. 

Our models suggest that \ce{CH3OH} is primarily formed on grain surfaces, with its major formation pathways involving the hydrogenation of \ce{s-CH3O} and \ce{s-CH2OH}. These intermediates originate from the successive hydrogenation of \ce{s-CO} in our model predictions.
\begin{gather}
    \ce{s-CO ->[s-H] s-HCO ->[s-H] s-H2CO} \\
    \ce{s-H2CO ->[s-H] s-CH3O/s-CH2OH ->[s-H] s-CH3OH}
\end{gather}
\noindent
The efficiency of surface reactions, particularly the conversion of \ce{s-H2CO} into \ce{s-CH3O} and \ce{s-CH2OH}, plays a crucial role in determining the abundance of methanol (\ce{s-CH3OH}) on dust grains. In our chemical model, once \ce{s-CH3OH} is formed, it enters the gas phase primarily through chemical desorption. 
Our model predictions suggest that photodesorption, which depends on UV penetration in the disk, has a minimal impact on the gas-phase abundance of \ce{CH3OH} in a protoplanetary disk like ours. This is in contrast with the explanation by \citet{Carney2019}, who attributed the non-detection of \ce{CH3OH} in the disk around HD 163296 to inhibited UV photodesorption. Our results instead point toward inefficient chemical desorption as a potential cause for the observed absence of gas-phase methanol. Additionally, the depletion of gas-phase \ce{CH3OH} through freeze-out onto grain surfaces also dominates, making gas temperature a relevant factor. However, it is important to note that HD 163296 disk is a complex source with multiple substructures that are not fully captured in our physical model. In particular, the gaps within the disk are regions where increased UV penetration is expected, leading to elevated gas and dust temperatures \citep{Marel2018, Calahan2021}. In such regions, thermal desorption and photodesorption processes may become more significant in governing the gas-phase abundance of methanol. Our best-fit model predicts a column density of \ce{CH3OH} that is below the observationally constrained upper limit. However, it remains slightly higher than the upper limit reported by \citet{Carney2019}, indicating a potential overproduction of \ce{s-CH3OH} on grain surfaces in our model. Although most reactions in the \ce{s-CH3OH} formation pathway are barrierless, the conversion of \ce{s-H2CO} to \ce{s-CH3O} and \ce{s-CH2OH} requires activation energy. In our model, we consider activation barriers of 5400~K for \ce{s-CH2OH} and 2200~K for \ce{s-CH3O}, each with a branching ratio of 33\%, based on the KIDA 2024 chemical network. However, the uncertainties associated with these values cannot be entirely ruled out. If the actual activation energies are higher than assumed, the efficiency of this conversion would decline, resulting in greater retention of \ce{H2CO} and reduced methanol production.

\subsection{\ce{H2CO}}
Formaldehyde (\ce{H2CO}) is observed throughout the disk, as evident from the radial integrated intensity distribution shown in Figure~\ref{fig:radial}. \citet{Hernandez2024} determined that \ce{H2CO} emission in HD~163296 originates from a region with an excitation temperature of approximately $T_{\text{ex}} \sim 19.4 \pm 1.6$~K. Both gas-phase and grain-surface pathways may contribute to the formation of gas-phase \ce{H2CO}, but the dominant mechanism in protoplanetary disks remains uncertain. In cold regions, \ce{H2CO} is expected to form mainly through successive hydrogenation of \ce{s-CO} on dust grains \citep[e.g.,][]{Watanabe2002, Fuchs2009}. \citet{Hernandez2024} also suggested icy origins of \ce{H2CO} in HD 163296 protoplanetary disk based on observational constraints on excitation conditions and ortho-to-para column density ratio. On the other hand, laboratory experiments \citep[e.g.,][]{Fockenberg2002, Atkinson2006} and chemical models \citep[e.g.,][]{vandermarel2014, Loomis2015} have shown that \ce{H2CO} can also form in the gas phase via neutral–neutral reactions. In our best-fit chemical model, \ce{H2CO} is produced primarily in the gas phase through the reaction \ce{CH3 + O -> H2CO + H}, because the disk is warm enough that CO ice chemistry, and thus the hydrogenation of s-CO to form \ce{s-HCO}, is suppressed. Nevertheless, all tested chemical models underpredict the \ce{H2CO} column density within 200 au and overpredict it beyond this radius. The underestimation of \ce{H2CO} within 200 au of the disk could have several possible explanations. First, our chemical network may lack high-temperature gas-phase formation pathways for \ce{H2CO}. Second, inefficient grain-surface formation due to limited CO freeze-out suppresses the hydrogenation pathway (\ce{s-CO ->[s-H] s-HCO ->[s-H] s-H2CO}). Without sufficient surface production, thermal desorption cannot replenish gas-phase \ce{H2CO} to the observed levels. This interpretation is supported by the similar inner disk column density profiles of \ce{H2CO} and \ce{CH3OH}, both of which are believed to share a common origin via grain-surface hydrogenation of CO. Third, photodissociation by FUV photons in the surface layers of the inner disk could further deplete gas-phase \ce{H2CO}, particularly if vertical shielding is weaker than assumed.

\subsection{\ce{DCO+}} \label{sec:dco+}
We report the most spatially-resolved \ce{DCO+} emission with a beam $0.31'' \times 0.27''$ characterized by triple concentric rings from the $J=5-4$ line observation in the disk around HD 163296 (see Figure \ref{fig:radial}). \citet{Flahery2017} previously observed the triple ringed structure of \ce{DCO+} ($J=3-2$) with a beam of $0.5'' \times 0.59''$. In most disks, \ce{DCO+} is resolved to show a single-ringed radial distribution \citep{Qi2008, Mathews2013, Teague2015, Huang2017}. Previously, \citet{Oberg2015a} observed a double-ringed \ce{DCO+} structure in the IM Lup disk, while \citet{Kashyap2024} reported a partially overlapping double-ringed \ce{DCO+} structure in the GG Tau A disk. Both explained that the double-ring results from differential UV penetration in the disk, owing to a stark change in surface density in the outer disk. From our modeled simulations (Figure \ref{fig:abun_model}), we observe an enhancement of \ce{DCO+} originating from the warm molecular layer of the disk, specifically within the vertical height ranges of $0.1 < z/r < 0.2$ for $r < 200$ au and around $z/r = 0.2$ for $r>200$ au. Notably, \citet{Romero-Mirza2023} observationally constrained the \ce{DCO+} emission layer to $0.05 < z/r < 0.25$ within 200 au in the low-mass disk around TW Hydrae. The dominant formation pathway for \ce{DCO+} around the region of the first gap ( $r < 60$ au), is the gas-phase deuterium exchange reaction: \ce{HCO+ + D -> DCO+ + H}, while it is primarily destroyed through grain surface charge neutralization processes. This is consistent with the radially integrated intensity distribution of \ce{DCO+} (5-4) with \ce{HCO+} (1-0) in Figure \ref{fig:radial}, where \ce{DCO+} peaks in the same region as \ce{HCO+} at around 49 au. In a small radial region around $\sim 75$ au, we also see \ce{CO + N2D+ -> N2 + DCO+} being dominant. In regions beyond 100 au, \ce{DCO+} primarily forms via reactions between \ce{CO} and deuterated ionized hydrocarbons such as \ce{CH4D+} and \ce{CH3D2+}, with a smaller contribution from the gas-phase deuterium exchange reaction: \ce{HCO+ + D -> DCO+ + H}. In this region, destruction of \ce{DCO+} occurs mainly through various dissociative recombination processes in the gas phase.
Regardless of the formation pathways, the abundance of \ce{DCO+} is ultimately governed by the gas-phase \ce{CO} abundance, which is dependent on the interplay of UV penetration and surface density in the disk. In the first column of Figure \ref{fig:radial}, the dashed and dotted vertical lines mark the ring peaks and gap depressions in the continuum distribution, respectively. Notably, the first peak in \ce{DCO+} (5-4) coincides with the continuum gap at 49 au. This suggests that in regions of lower surface density, increased UV penetration enhances photodesorption, releasing more \ce{CO} into the gas phase and driving up \ce{DCO+} production. Consequently, the first \ce{DCO+} peak aligns with the first continuum gap. The second peak, on the contrary, seems to be arising at the third continuum ring at 101 au. According to \citet{Muro2018}, these outer two \ce{DCO+} rings are under the shadow of the first continuum ring, making UV penetration shallower as compared to the first \ce{DCO+} ring region. In this case, the higher surface density of the third continuum ring directly corresponds to an increase in \ce{DCO+} column density. We did not observe any clear correlation between the continuum substructures and the third \ce{DCO+} ring. In our physical model, we have not accounted for the disk's substructures or the shadowing effect of the innermost continuum ring. As a result, the two outer peaks do not appear in the modelled column density distribution.

\subsection{\ce{HCO+}}
We observe the innermost \ce{HCO+} ring at around the first continuum gap ($\sim 49$ au) in the radial integrated intensity profile of J=1-0 transition in Figure \ref{fig:radial}, similar to the case of \ce{DCO+}. However, the second ring around the continuum peak at 101 au is not as prominent as \ce{DCO+}. There is also a third ring in the far outer region (r $\sim 300$ au). \citet{Paneque-Carreno2023} reported that \ce{HCO+} emission arises very close to the midplane at $z/r < 0.1$, consistent with our derived radial distribution of excitation temperature. In our chemical model, however, significant \ce{HCO+} emission near the midplane is found only within $\sim 60$ au; beyond this radius, \ce{HCO+} predominantly arises in the warmer molecular layer along $z/r \sim 0.2$. This discrepancy hints at the complex vertical chemical stratification at play within the disk. 

Our model suggests two major formation channels, active in the regions ($> 60$ au) :
\begin{gather}
    \ce{CO + H3+ -> HCO+ + H2}\\
    \ce{CO + N2H+ -> HCO+ + N2}
\end{gather}
Among the above two formation reactions, the relative contribution of either depends on the region. In relatively warmer regions (at smaller radial distances), the first one dominates, but as we go radially outwards towards colder regions, the second one's contribution starts increasing. Both \ce{H3+} and \ce{N2H+} directly reflect the degree of ionisation in the disk. Hence, \ce{HCO+} and in turn, \ce{DCO+}, which forms through deuterium exchange with \ce{HCO+} (see Section \ref{sec:dco+}), are very sensitive to the ionisation structure of the disk. Radial distribution of chemically modelled \ce{HCO+} column densities being slightly below the observationally constrained column densities in Figure \ref{fig:cden_fit} suggests that the ionisation structure in the disk is more complex than assumed. However, this does not affect our inferred C/O ratio, as the \ce{HCO+} column density is not strongly sensitive to C/O (Figure \ref{fig:cden_fit}). Previous studies, such as \citet{Aikawa2021}, have examined the ionisation structure of the disk surrounding HD 163296 using the \ce{HCO+} (1-0) transition. Their findings suggest that the nearly constant \ce{HCO+} column density beyond 100 au is primarily due to sustained X-ray ionisation, which dominates in the upper layers of the disk. Constant \ce{HCO+} is also evident in our case beyond 200 au. In the inner regions (r$< 60$ au), our model suggests that \ce{HCO+} primarily forms through reactions between \ce{CO} and ionized hydrocarbons such as \ce{CH5+}, \ce{CH4D+}, and \ce{CH3D2+}. The \ce{HCO+} destruction happens either through electron recombination to form \ce{H} and \ce{CO} or the grain surface charge neutralisation processes. It is worth noting that the isotope exchange with \ce{HCO+} to form \ce{DCO+} is not a major destruction pathway for \ce{HCO+} in most regions, indicating that \ce{HCO+} is significantly more abundant than \ce{DCO+} throughout the disk. This trend is also evident from the column density constraints (see Table \ref{tab:disk_averaged} and the second column of Figure \ref{fig:radial}).

\subsection{HCN}
Our \ce{HCN} (3–2) emission shows a four-ringed structure. While the two inner rings are clearly visible, the outer two are less distinct in our radial profiles. In comparison, \citet{Law2021_MAPSIII} present the outer rings with much greater contrast by restricting the averaging region to a narrow wedge along the disk’s major axis, rather than averaging over the full azimuth as done in our analysis. We do not see any clear correlation between its chemical substructures and the disk continuum substructures (see Figure \ref{fig:radial}). The observational characteristics of this transition have been extensively studied by \citet{Bergner2021}, while our focus lies in exploring its chemical aspects. Our simulations (Figure \ref{fig:abun_model}) indicate that \ce{HCN} is most abundant in the warm molecular layer ($z/r \approx 0.2$), with its abundance gradually decreasing toward the midplane. This is consistent with observational constraints from \citet{Paneque-Carreno2023}, who located the HCN emission layer at $z/r \approx 0.2$ within 350 au in the disk around HD 163296, as well as  with the radial constraints on \ce{HCN} excitation temperature (third column in Figure \ref{fig:radial}). The column densities predicted by our best-fit model fit the disk-averaged observed values really well.
Multiple gas-phase formation pathways are active in different regions of the disk ($r<300$ au), including \ce{CH3CNH+ + e- -> HCN + CH3}, \ce{CN + C2H4 -> HCN + C2H3}, etc. In outer regions ($r \sim 300$ au), s-HCN is formed on the grain surface, following the reaction: \ce{s-H2 + s-CN -> s-HCN + s-H} and then released into the gas phase through photodesorption.

\subsection{DCN}
Our analysis of the \ce{DCN} (2-1) transition shows radially decreasing emission. High-resolution observations of the 3-2 transition have a very low signal-to-noise emission. However, we do see a peak at the same location as \ce{HCN} inner ring, while there is a suppressed ring-like feature at the location of the outer \ce{HCN} ring (third row in Figure \ref{fig:radial}). Previous MAPS studies have assumed that \ce{HCN} and \ce{DCN} originate from the same molecular layers to constrain their column densities and excitation temperatures \citep{Bergner2021}. To refine these constraints, we analyse two \ce{DCN} transitions using a slab model (see Table \ref{tab:line-props}). Our simulated abundance distribution indicates that DCN forms closer to the midplane ($0.1 < z/r < 0.2$) in the inner disk regions. Here, it is primarily produced on grain surfaces via the reaction \ce{s-D + s-CN -> s-DCN}, and subsequently released into the gas phase through photodesorption and chemical desorption. In the outer disk ($r > 300$ au), where DCN peaks in the molecular layer ($z/r \sim 0.2$), the dominant formation pathway shifts to the gas-phase reaction \ce{D + C2N -> C + DCN}. Toward the midplane in these outer regions, DCN increasingly freezes out onto dust grains.

\subsection{CS}
We observe a ring-like radial distribution in both \ce{CS} transitions, with emission peaking in the gap region at approximately 49~au (see Figure~\ref{fig:radial}). Previous studies by \citet{LeGal2021} and \citet{Ma2024} have extensively examined sulfur chemistry in HD~163296. Our study utilises a different \ce{CS} transition (J=3–2) to establish observational constraints (see Table~\ref{tab:line-props}). Our disk-averaged estimates for the \ce{CS} column density and excitation temperature are approximately $7 \times 10^{13}$~cm$^{-2}$ and $21 \pm 2$ K, respectively. \citet{Law2025} performed the most comprehensive multi-line \ce{CS} study in HD~163296, covering a wide range of $E_u$ values from 7.1~K to 129.3~K, and reported a disk-averaged column density of $(5.53 \pm 0.33) \times 10^{12}$~cm$^{-2}$ at an excitation temperature of $24.4 \pm 1.3$~K. \citet{LeGal2021} reported a column density of $6.2 \times 10^{12}$~cm$^{-2}$ at $22.7 \pm 2.2$~K, while \citet{Ma2024} estimated a radial column density of around $10^{13}$~cm$^{-2}$. Both \citet{LeGal2021} and \citet{Ma2024} modelled the disk chemically, but their physical models involved several simplifications, including assumptions about grain size, UV treatment, visual extinction, and vertical settling, as discussed in Section~\ref{sec:model_disc}. Our best-fit model overpredicts the \ce{CS} column density roughly by a factor of 10 (Figure~\ref{fig:cden_fit}). This discrepancy may be related to the long-standing sulfur depletion problem in protoplanetary disks, which is a common finding in Herbig disks \citep{ Keyte2024_b, Keyte2024} and remains an open issue in astrochemistry \citep{semenov18, LeGal2021}. In fact, \citet{Ma2024} emphasised the need to consider a highly depleted initial sulfur abundance—at least a factor of 10 lower than that used in this study and in other traditional disk chemistry models—in order to reconcile the model predictions with the observed \ce{CS} abundances. 

In our chemical model, \ce{CS} peaks within the vertical layer $0.1<z/r<0.2$ in the innermost disk region ($r < 100 $ au), where several formation pathways are active, including: \ce{HOCS+ + e- -> CS + OH} and \ce{H + HCS -> CS + H2}, etc. Beyond 100 au, \ce{CS} primarily originates from higher layers, between $z/r \sim 0.2$ and 0.3, with the dominant formation pathway being: \ce{S + C2 -> C + CS}. Closer to the midplane, \ce{CS} progressively depletes from the gas phase as it freezes onto grain surfaces. The peak of \ce{CS} emission in the gap region suggests that enhanced UV penetration, resulting from lower surface density, reduces the extent of \ce{CS} freeze-out in this region.

\subsection{\ce{HC3N}} \label{sec:hc3n}
\ce{HC3N} exhibits a single-ringed radial distribution, with emission concentrated within 200 au (see Figure \ref{fig:radial}), with no clear correspondence with dust continuum structure. As a key nitrogen-bearing molecule, \ce{HC3N} has been previously studied in HD 163296 disk by \citet{Bergner2018} and \citet{Ilee2021}, though without chemical modelling. \citet{Ilee2021} analyzed two \ce{HC3N} transitions with upper energy levels ($E_u$) of 28.8 K and 189.9 K, constraining the disk-averaged column density and excitation temperature to approximately $7.3 \times 10^{13}$ cm$^{-2}$ and 36.9 K, respectively, using the rotational diagram method. In our study, we incorporated an additional transition with $E_u = 59.38$ K from our dataset alongside the transitions used by \citet{Ilee2021} (see Table \ref{tab:line-props}). Our analysis yielded a disk-averaged column density of $1.57 \times 10^{13}$ cm$^{-2}$ and an excitation temperature of $74 \pm 3$ K. While this excitation temperature differs from that reported by \citet{Ilee2021}, when we constrained our analysis to only their selected transitions, our results were consistent. We also note that the disk-averaged excitation temperature is higher than suggested by the radial profile, indicating that it may not be reliable. However, the disk-averaged column density is largely insensitive to temperature variations across the disk, so our column-density constraint remains robust. Our model predicted \ce{HC3N} column density falls within a factor of 5 above the disk-averaged observational constraint.
 In our simulations, \ce{HC3N} primarily forms in the molecular layer at $z/r \approx 0.2$, driven by several bimolecular gas-phase reactions:
\begin{gather}
    \ce{C3H3N+ + e- -> H2 + HC3N} \label{reac:hc3n_1}\\
    \ce{ N + CH2C2H -> H2 + HC3N}\\
    \ce{CN + C2H2 -> H + HC3N}
\end{gather}
The dominant formation pathway among the above reactions varies with radial distance from the star, while the primary destruction mechanisms are photodissociation of \ce{HC3N} into \ce{CN} and \ce{C2H} away from the midplane and adsorption onto grain surfaces close to the midplane. At the region where \ce{HC3N} ring is peaking ( refer to Figure \ref{fig:radial}), reaction \eqref{reac:hc3n_1} is the dominant channel for \ce{HC3N} formation. Similar to \ce{CS} and \ce{HCN}, \ce{HC3N} is highly sensitive to UV penetration in the disk. Higher gas-phase abundances of hydrocarbons and nitriles lead to increased production of \ce{HC3N}, emphasising the importance of accurately accounting for nitrogen abundance, alongside carbon and oxygen, when modelling disk chemistry. \ce{HC3N} is completely depleted in the midplane ($z/r < 0.1$) due to insufficient UV penetration in this region. \citet{Bergner2018}, \citet{Ilee2021} and \citet{Calahan2023} also report that \ce{HC3N} primarily originates from relatively warm regions of the disk.

\subsection{New constraints on the C/O ratio from multi-line grid-fitting and its comparison with earlier works}

In our grid exploration, where we aim to simultaneously fit the observed column densities of all the targeted molecules, the C/O ratio was varied by fixing the initial \ce{C+} abundance at $1.70 \times 10^{-4}$ and varying the \ce{O} abundance. This is motivated by the fact that O budget is uncertain due to the weak or non-detection of water vapour (\ce{H2O}) lines in observations from the \textit{Herschel Space Observatory}, suggesting that volatile oxygen is sequestered in larger, midplane solids that do not migrate to disk surface layers where photodesorption could release water vapour \citep{Oberg2021review}. Among the molecules considered, CS, HCN, DCN, \ce{HC3N}, and \ce{HNCO} are found to be particularly sensitive to changes in the C/O ratio. The best-fit C/O ratio from our grid exploration is a super-solar value of 1.1 for the full disk.

There are also other studies in the literature that have attempted to constrain the C/O ratio in this disk. For example, sulfur-bearing molecules have been proposed as effective probes of the elemental C/O ratio. \citet{LeGal2021} showed that the ratio N$_\mathrm{CS}$/N$_\mathrm{CO}$ is a promising diagnostic for the elemental C/O ratio in protoplanetary disks, and reported a super-solar C/O ratio toward another Herbig Ae disk, MWC 480 ($>1$). \ce{SO} has not yet been detected in HD 163296 disk. However, \citet{LeGal2021} and \citet{Ma2024} proposed a super-solar C/O ratio ($>0.9$) in HD 163296 disk based on CS observations combined with \ce{SO} upper limits. In another study, \citet{Bosman2021} also reported a super-solar C/O ratio, though substantially higher than ours ($\sim2.0$). Their analysis emphasised \ce{C2H} as a tracer of the C/O ratio due to its photo-sensitive chemistry. Like our work, their study adopted the physical structure from \citet{Zhang2021}. However, they initialised elemental carbon and oxygen in the forms of \ce{H2O}, CO, and \ce{CH4}, whereas our model assumes initial abundances of \ce{C+} and \ce{O}.

A carbon-to-oxygen (C/O) ratio greater than 1 can indicate depletion of volatile elements. Based on extensive studies of the interstellar medium, carbon is believed to be primarily carried by CO and carbonaceous grains, while oxygen resides in both \ce{H2O} and silicate grains. HD 163296 is a luminous star ($17 L_{\odot}$) hosting a warm disk, where most of the disk remains above the CO freeze-out temperature ($\sim$20 K). Hence, we expect the majority of CO to be in the gas phase in this environment. Given that CO, one of the primary reservoirs of carbon, is abundantly available in the gas phase, the disk’s C/O ratio is likely governed by the abundance of oxygen rather than carbon. This is consistent with several previous studies \citep{Kastner2015, Kama2016, Bergin2016, Miotello2019}, which inferred oxygen depletion in protoplanetary disks based on thermochemical modelling of hydrocarbon species such as \ce{C2H}, while fitting with ALMA molecular line observations. \citet{Bergin2016} also demonstrated that a strong UV irradiation is necessary, along with elevated C/O ratios, to explain hydrocarbon observations. \citet{Bosman2021} reported strong \ce{C2H} emission and a super-solar C/O ratio in HD 163296 disk, where they suggested small dust depletion in the innermost regions, allowing higher UV penetration into the disk. \citet{Calahan2023} has built on \citet{Bergin2016} and \citet{Bosman2021} to explain simultaneously four molecules \ce{CH3CN}, \ce{HCN}, \ce{HC3N}, and \ce{C2H} observed towards HD 163296 disk, making it the most comprehensive molecular constraint on C/O prior to this work. They concluded that an elevated C/O ratio is required to provide the elemental reservoir for both hydrocarbons and nitriles, while the depletion of small grains enables deeper UV penetration, facilitating their release into the gas phase. It is worth noting that HD 163296 is a UV-bright, mature (age $\sim$ 6 Myr), well-settled disk. While we have not explicitly modelled small grain depletion, at most height points of our physical model, the gas-to-dust mass ratio is $\sim 1000$, as opposed to the standard interstellar ratio of 100. This high ratio allows for enhanced UV penetration from the central star, likely promoting the release of refractory materials into the gas phase.

\subsection{Strengths, Limitations, and Future Outlook} \label{sec:model_disc}
While previous studies have provided valuable insights into specific chemical processes, they are generally limited to narrow molecular subsets. Our study is the first to employ a comprehensive chemical network with a detailed treatment of deuterium fractionation (see Section~\ref{sec:chem_mod}), enabling a holistic exploration of the chemistry in HD~163296 by considering a larger set of molecules. By simultaneously analyzing high-angular-resolution observations of a diverse molecular inventory—including organics, nitriles, sulfur-bearing species, and tracers of ionization and deuteration—our study provides a more comprehensive view of the chemical complexity within this disk, while also highlighting the challenges of fitting multiple molecular line observations simultaneously using thermo-chemical models.

HD 163296 disk exhibits multiple substructures, with rings and gaps, with evidence of ongoing planet formation. Gaps with reduced surface densities can result in an increase in gas and dust temperatures due to larger UV penetration \citep{Marel2018}. On one hand, an increase in gas temperature can lead to a decrease in frozen-out molecules; on the other hand, higher UV penetration impacts the photochemistry. Observations of scattered light in the disk around HD 163296 using the SPHERE instrument reveal that the outer two thermal continuum rings, visible in ALMA observations, are absent, suggesting that the outer disk is relatively flat and lies in the shadow of the innermost continuum ring \citep{Muro2018}. Such shadowing likely suppresses UV penetration in the disk. \citet{Rich2020} report the presence of ansae features, suggesting a fourth broken ring in the disk with a semi-major axis of 330 au. However, our adopted density structure, and consequently the UV field modelling, does not account for these substructures. Although understanding their impact on disk chemistry is important, a detailed treatment is beyond the scope of this paper and will be addressed in a future study.


\section{Conclusions} \label{sec:conclusions}
In this study, we investigate the chemistry of the Herbig Ae protoplanetary disk around HD 163296. Our analysis is based on a diverse and extended set of molecular species, including both simple and complex organics (\ce{NH2CHO}, \ce{HNCO}, \ce{CH3OH}, \ce{H2CO}), nitriles (\ce{HCN}, \ce{DCN}, \ce{HC3N}), a sulfur-bearing species (\ce{CS}), and molecules tracing both warm (\ce{HCO+}, \ce{HCN}, \ce{DCN}) and cold (\ce{DCO+}) molecular layers. Several of these molecules are highly sensitive to the elemental C/O ratio, while others are primarily influenced by ionization or grain-surface chemistry. For our analysis, we employ high-resolution ALMA observations from Bands 3, 4, 6, and 7, encompassing multiple transitions of the species mentioned above.

Our key findings are summarized below.

\begin{enumerate}
    \item We report a non-detection of \ce{NH2CHO} in the HD 163296 protoplanetary disk, with at most a 3$\sigma$ signal from the 7(0,7)-6(0,6) transition. This study establishes the first and most stringent $3\sigma$ upper limit on \ce{NH2CHO} column density in this protoplanetary disk, estimated at approximately $7 \times 10^{11}$ cm$^{-2}$.

    \item We report the non-detection of \ce{HNCO} emission in the HD 163296 disk, setting the first $3\sigma$ upper limit on its column density in a protoplanetary disk at approximately $1.3 \times 10^{11}$ cm$^{-2}$.

    \item We report the most spatially-resolved \ce{DCO+} emission in the disk around HD 163296 with a beam $0.31'' \times 0.27''$ from the $J=5-4$ line observation, revealing triple-ringed chemical substructure with strong correspondence with continuum substructure.

   \item Our study is the first to employ an extensive chemical reaction network with a detailed treatment of deuterium fractionation to model the chemistry of the disk around HD 163296, while simultaneously comparing the observed disk-averaged column densities of seven detected molecules with the observationally constrained upper limits of three undetected molecules. Our results show no significant variation in the abundance distributions of the gas-phase molecules considered between the inheritance and reset models, suggesting that the chemistry of the molecules observed in this study is likely dominated by processes occurring within the disk.

    \item Our chemical model indicates that both \ce{NH2CHO} and \ce{HNCO} form primarily through grain-surface reactions and are abundantly present on grain surfaces in this disk environment. However, they are not detected in the gas phase due to inefficient chemical desorption processes. All explored models predict \ce{NH2CHO} column densities well below the observational upper limits, suggesting that ALMA’s current sensitivity would require much longer integration times for detection in HD~163296. In contrast, the modelled \ce{HNCO} column densities approach the upper limit in the outer regions of the disk. 
    
    \item Our grid exploration shows that among the molecules studied, CS, HCN, DCN, and \ce{HC3N} are particularly sensitive to variations in the C/O ratio. Our best-fit model, representing the global C/O ratio of the entire disk, is derived using chi-squared statistics by simultaneously fitting the model predictions to the observationally constrained molecular column densities, and yields a super-solar C/O ratio of 1.1 with the lowest reduced chi-squared value ($\chi^2_{\text{red}}$). Models with C/O $> 1$ generally reproduce the observations well, whereas those with C/O $< 1$ show significantly poorer agreement.

\end{enumerate}

\section*{Acknowledgments}
This paper makes use of the following sets of ALMA data: ADS/JAO ALMA$\#$2019.1.00206.S; 2018.1.01055.L; 2018.1.01055.L; 2011.0.000010.SV; and 2021.1.00138.S. ALMA is a partnership of ESO (representing its member states), NSF (USA) and NINS (Japan), together with NRC (Canada), MOST and ASIAA (Taiwan), and KASI (Republic of Korea), in cooperation with the Republic of Chile. The Joint ALMA Observatory is operated by ESO, AUI/NRAO and NAOJ. The National Radio Astronomy Observatory is a facility of the National Science Foundation operated under a cooperative agreement by Associated Universities, Inc. L.M. acknowledges financial support from DAE, Government of India, for this work. L.M. also gratefully acknowledges support from Breakthrough Listen at the University of Oxford through a sub-award at NISER under Agreement R82799/CN013, provided as part of a global collaboration under the Breakthrough Listen project funded by the Breakthrough Prize Foundation. This research was carried out in part at the Jet Propulsion Laboratory, which is operated for NASA by the California Institute of Technology. K.W. acknowledges the financial support from the NASA Emerging Worlds grant 18-EW-18\_2-0083. We would like to thank the anonymous referee for constructive comments that helped improve the manuscript.

\facility{ALMA}   

\software{\texttt{CASA} \citep{CASATeam2022}, \texttt{keplerian\_mask} \citep{kepmask}, \texttt{GoFish} \citep{GoFish}, \texttt{VISIBLE} \citep{loomis2018visible}, \texttt{PEGASIS}(Maitrey2025), \texttt{galario} \citep{galario}, \texttt{ultranest} \citep{Bunchner2016, Buchner2019, Buchner2021}, \texttt{numpy} \citep{numpy}, astropy \citep{astropy:2013, astropy:2018, astropy:2022}, \texttt{matplotlib} \citep{matplotlib}}

\appendix

\section{DRive Retrieval Framework}\label{sec:drive}
\counterwithin{figure}{section}
\setcounter{figure}{0}

\begin{figure*}
    \centering
    \includegraphics[width=0.95\linewidth]{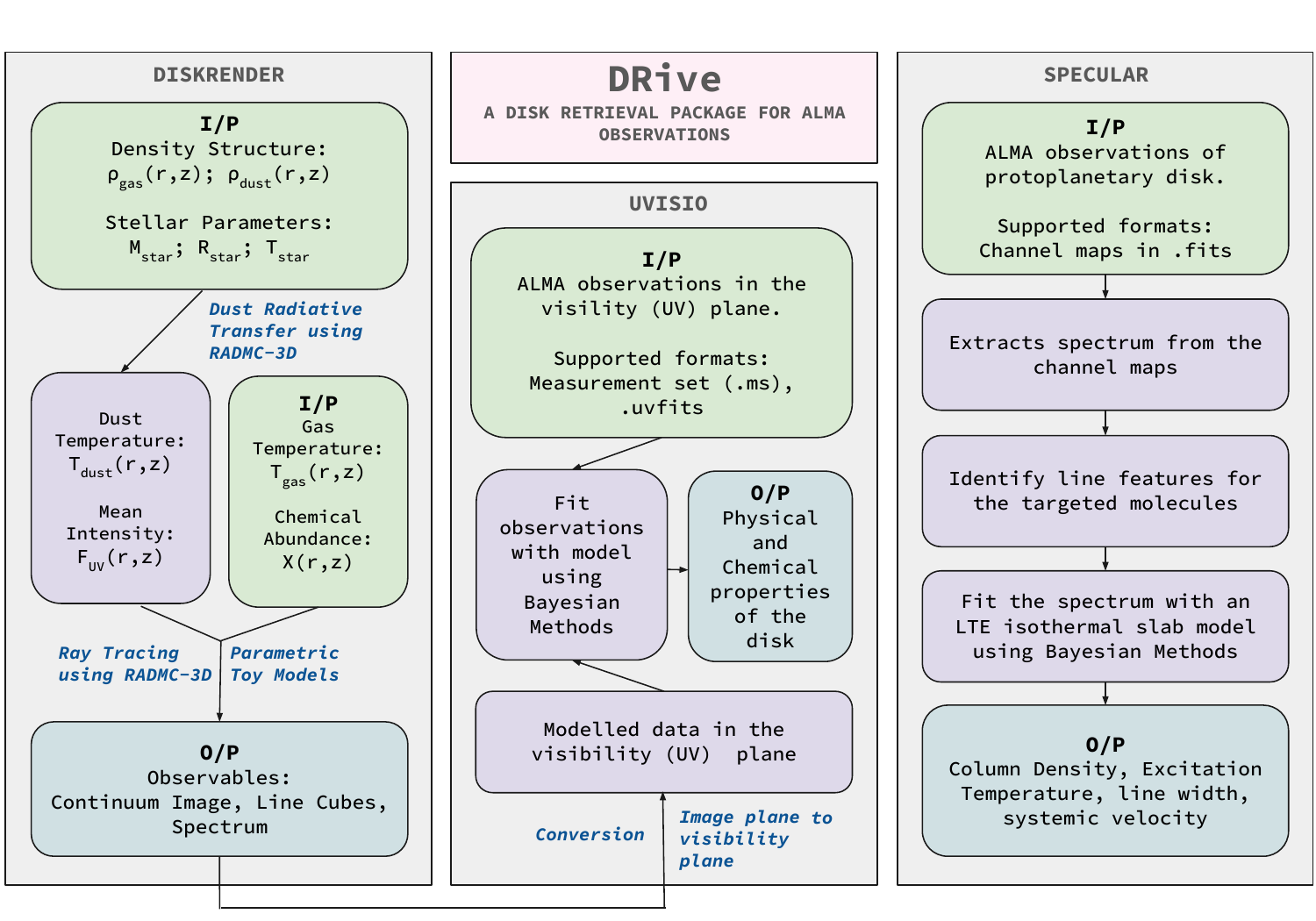}
    \caption{The workflow of our Disk Retrieval module, \textsc{DRive} comprising of three sub-modules (refer to Section \ref{sec:drive}). In the above schematic, I/P and O/P represent inputs and outputs of the model.}
    \label{fig:drive}
\end{figure*}

Here, we introduce \textsc{DRive}, our in-house, state-of-the-art, Python-based disk retrieval package designed to interpret ALMA observations of protoplanetary disks. To fully leverage the wealth of available continuum and molecular line data, and to extract the underlying physical and chemical properties of these systems, robust retrieval models are essential. The core philosophy behind \textsc{DRive} is to offer a comprehensive suite of tools and utilities that streamline the retrieval, processing, and visualization of data from protoplanetary disk observations, in both the image and visibility domains. The workflow of \textsc{DRive} is illustrated in Figure~\ref{fig:drive}.

Currently, \textsc{DRive} comprises three primary sub-modules: \texttt{diskrender}, \texttt{uvisio}, and \texttt{specular}. The \texttt{diskrender} sub-module generates observables in the imaging plane using either full 3D Monte Carlo radiative transfer and ray-tracing with \textsc{RADMC-3D} \citep{Dullemond2012} or simplified toy models. In contrast, \texttt{uvisio} focuses on data in the visibility plane, facilitating tasks such as reading observed visibility data, converting modelled imaging-plane data into the visibility plane, and fitting observed data using modelled representations while varying free parameters. Under the hood, \texttt{uvisio} uses \texttt{casatools}\footnote{\url{https://casadocs.readthedocs.io/en/v6.2.0/api/casatools.html}} to handle observed visibility data and \texttt{galario} \citep{galario} to transform imaging-plane data into the visibility plane. In Section \ref{sec:data_redcn}, we make use of the \texttt{diskrender.toy\_models.PowerLawDisk} function to generate a continuum image and employ the \texttt{uvisio} submodule to fit geometric parameters against observations in the visibility plane. While \texttt{uvisio} specializes in directly fitting observables in the visibility plane, \textsc{DRive}'s third module, \texttt{specular}, is designed to fit disk-averaged spectra. It achieves this by comparing spectral data generated from observed line images against parametric spectral line models, as detailed in Section \ref{sec:drive_slab}.

\textsc{DRive} has the ability to explore parameter space using both Markov Chain Monte Carlo (MCMC) \citep{Goodman2010} and nested sampling Monte Carlo Bayesian statistical methods \citep{Bunchner2016,Buchner2019}, implemented via the Python packages \texttt{emcee} \citep{emcee2013} and \texttt{ultranest} \citep{Buchner2021}. Through extensive testing, we have determined that \texttt{ultranest} outperforms \texttt{emcee} in terms of both convergence time and computational efficiency, requiring fewer live points in nested sampling compared to the number of walkers in MCMC.

Several other radiative transfer retrieval tools exist, such as \texttt{pdspy} \citep{pdspy} and \texttt{DiskFit} \citep{Pietu2007}. Similarly, spectral analysis tools like the eXtended CASA Line Analysis Software Suite \citep[XCLASS,][]{Moller2017} and the Centre d'Analyse Scientifique de Spectres Infrarouges et Submillimetriques \citep[CASSIS,][]{Vastel2015} are also available. While all the aforementioned tools have their strengths, we opted for our own implementation, \textsc{DRIve}, to accommodate greater flexibility in incorporating custom radiative transfer models and refining the slab model-generated spectra.

\subsection{DRive- Spectral Line Model}\label{sec:drive_slab}
We constrained the excitation conditions of the detected molecules using the \texttt{specular} sub-module in \textsc{DRive}. In this sub-module, synthetic spectra are generated under the assumption of Local Thermodynamic Equilibrium (LTE) in the emitting region, with all transitions of a given molecule sharing a common excitation temperature. LTE is a reasonable approximation for modelling the gas spectra in the protoplanetary disk around HD 163296 within a radius of 400 au, given its characteristic gas density \citep{Cataldi2021}. By solving the fundamental radiative transfer equation for an isothermal and uniform slab, we get the intrinsic emission spectra, $I_{\nu}$ as:
\begin{align}
    I_{\nu} = \left( B_{\nu}(T_{ex}) - B_{\nu}(T_{\text{CMB}}) \right) \left( 1 - e^{-\tau_{\nu}} \right) \label{eq:I_nu}
\end{align}
\noindent
Here, $B_{\nu}$ is the Planck function for blackbody radiation, and the background radiation is contributed by the cosmic microwave background (CMB) temperature, $T_{\text{CMB}} = 2.73$ K. The frequency-dependent optical depth, $\tau_{\nu}$, for each transition is calculated assuming a Gaussian line profile centred around Doppler shifted the Doppler-shifted rest frequency, $\nu_c$, of the transition as follows:
\begin{align}
    \tau_{\nu} = \tau_{0} \exp \left( - \frac{\left( \nu - \nu_{c} \right)^2}{2 \sigma^2_{\nu}} \right) \label{eq:tau_nu}
\end{align}
where $\nu$ is the frequency axis, $\sigma_{\nu}$ is the standard deviation in frequency of the Gaussian line profile, and $\tau_0$ is the optical depth at the line centre. $\tau_0$ is calculated as follows:
\begin{align}
    \tau_{0} = \frac{hc}{4\pi} \left( x_l B_{lu} - x_u B_{ul} \right) N_T \frac{\nu_{0}}{\sqrt{2\pi} \sigma_{\nu} c } \label{eq:tau0}
\end{align}
\noindent
Here, $N_T$ is the total column density of the molecule, $B_{lu}$ and $B_{ul}$ are Einstein coefficients, $x_u$ and $x_l$ are the fraction of molecules in the upper and lower energy levels, respectively, and $\nu_{0}$ is the rest frequency. We calculate the central frequency of the Gaussian profile, $\nu_{c}$  in terms of the rest frequency of the transition, $\nu_{0}$ and the systemic velocity of the source, $v_{\text{sys}}$ as follows:
\begin{align}
    \nu_{c} = \nu_{0} \left( 1 - \frac{v_{\text{sys}}}{c} \right) \label{eq:nu_c}
\end{align}

For transitions with multiple hyperfine components (i.e., identical J quantum numbers but different F), we follow the prescription of \citet{Guzman2021} to compute the frequency-dependent optical depth, $\tau_{\nu}$. 
\begin{align}
    \tau_{\nu} = \sum_{i} \tau_{0,i} r_i \exp \left( - \frac{\left( \nu - \nu_{c,i} \right)^2}{2 \sigma^2_{\nu,i}} \right) \label{eq:tau_nu}
\end{align}
\noindent
where $\tau_{0, i}$ denotes the total optical depth at the line center of the \textit{i}$^{\text{th}}$ hyperfine component calculated using equation \eqref{eq:tau0}, and $\sigma_{\nu,i}$ and $\nu_{c,i}$ represent its frequency linewidth and central frequency, respectively. Here, $\nu_{c,i}$ is calculated in terms of the rest frequency of the hyperfine component, $\nu_{0,i}$ and the systemic velocity of the source, $v_{\text{sys}}$, using equation \eqref{eq:nu_c}. The contribution of the hyperfine components towards the optical depth is calculated by relative intensities of the hyperfine components, $r_{i}$, with respect to the brightest hyperfine component in the group:
\begin{align}
    r_i = \frac{g_{u,i}A_{ul,i}}{\max( g_{u,j}A_{ul,j})} \label{eq:ri}
\end{align}
\noindent
In Equation \eqref{eq:ri}, the denominator corresponds to the brightest transition among all available hyperfine components, not only the detected ones \citep{Guzman2021}. \par

In the case of LTE, the level population follows the Maxwell-Boltzmann distribution. Hence, we have:
\begin{align}
    g_u B_{lu} = g_u B_{ul} \quad \quad
    B_{ul} = \frac{c^3}{8\pi h \nu^3} A_{ul} \label{eq:Bul_Blu}\\
    x_{u} = \frac{g_u}{Q(T_{ex})} \exp{\left(-\frac{E_u}{k_B T_{\text{ex}}}\right)} \label{eq:xu} \\
    x_{l} = \frac{g_l}{Q(T_{ex})} \exp{\left(-\frac{{E_l}}{k_B T_{\text{ex}}}\right)} \label{eq:xl}
\end{align}
\noindent
where $k_B$ is the Boltzmann constant and $Q(T_{\text{ex}})$ is the partition function evaluated at the excitation temperature, $T_{\text{ex}}$ calculated by interpolating tabulated $Q$ values for discrete $T_{\text{ex}}$ values found in CDMS or JPL database.
Using Equations \eqref{eq:Bul_Blu}, \eqref{eq:xu}, \eqref{eq:xl} and $E_{u} - E_{l} = h\nu_{0}$ in Equation \eqref{eq:tau0}, we get the following expression for $\tau_0$ for LTE conditions:
\begin{align}
    \tau_{0} = \frac{c^2 A_{ul}}{8\pi \nu^2_0 \sqrt{2\pi} \sigma_{\nu}} \left( \exp{ \left(\frac{h\nu_0}{k_B T_{\text{ex}}}\right)}  - 1\right) N_u \label{eq:tau_0_final} \\
    N_u = N_T \frac{g_u}{Q(T_{\text{ex}})}\exp{\left( -\frac{E_u}{k_B T_{\text{ex}}} \right)}
\end{align}

The optical depth is fundamentally determined by the intrinsic line width, which is primarily governed by thermal broadening \citep[e.g.][]{Teague2016,Flaherty2020}. However, in observations, the spectral lines appear artificially broadened beyond their thermally expected width. This additional broadening can be attributed to several factors: (i) beam smearing, where multiple Keplerian velocity components are blended within the same beam due to finite spectral resolution;
(ii) emission from different disk surfaces, where contributions from both the front and back sides of an elevated disk layer introduce distinct projected velocities. We first adopt the thermal line width, $\sigma_{\nu,\text{thermal}}$, in the optical depth calculation of Equation \eqref{eq:tau_nu}:
\begin{align} 
\sigma_{\nu,\text{thermal}} = \sqrt{ \frac{k_B T_{\text{kin}}}{\mu m_H} } 
\end{align}
\noindent
where $T_{\text{kin}}$ is the kinetic temperature of the emitting region, $\mu$ is the mass of the species in \textit{amu}, and $m_H$ is the mass of a hydrogen atom. Under the assumption of LTE, $T_{\text{kin}}$ is equivalent to $T_{ex}$.

To account for artificial broadening seen in observations, the modelled intensity, computed using thermal broadening in Equation \eqref{eq:I_nu}, is convolved with a Voigt profile, the convolution of a Gaussian and a Lorentzian component, with freely constrained FWHM and HWHM for the Gaussian and Lorentzian terms, respectively. The Voigt profile is necessary because it naturally captures the combined effects of Gaussian-like processes (e.g., instrumental broadening) and Lorentzian-like processes (e.g., turbulence, pressure/natural broadening, or line wings introduced by velocity gradients) \citep{yamato24}. This redistributes the modelled spectra to a broadened line profile while preserving the total velocity-integrated intensity. The measured line widths, expressed as the FWHM of velocity, $\Delta v_{\text{fwhm}}$, are related to frequency line width $\sigma_{\nu}$ by:

\begin{align} 
\sigma_{\nu,i} = \frac{\Delta v_{\text{fwhm}}}{2\sqrt{\ln{2}}} \frac{\nu_{0,i}}{c} 
\end{align}

This approach effectively accounts for artificial broadening while preserving the physical accuracy of the modelled spectra. Similar methods have been employed by \citet{Cataldi2021}, \citet{Guzman2021}, and \citet{Bergner2021} to incorporate line broadening. Additionally, to counteract the effects of finite channel widths, the spectra are initially simulated at a much higher frequency resolution, uniform across all transitions and then interpolated to match the observed resolution. We generate the observed images for all transitions of a particular molecule with the smallest possible common beam size. The modelled emission (Equation \ref{eq:I_nu}), computed in Jy/sr, is converted to Jy/beam by multiplying it by the common beam size in sr. In cases where observed spectra are presented along the velocity axis, $v_{\text{obs}}$, we convert the velocity axis to the frequency axis, $\nu_{\text{obs}}$, using
$\nu_{\text{obs}} = \nu_{0} \left( 1 - \frac{v_{\text{obs}}}{c} \right)$.

\section{Matched Filter Responses for the weak lines}
\begin{figure*}
    \centering
    \includegraphics[width=0.9\linewidth]{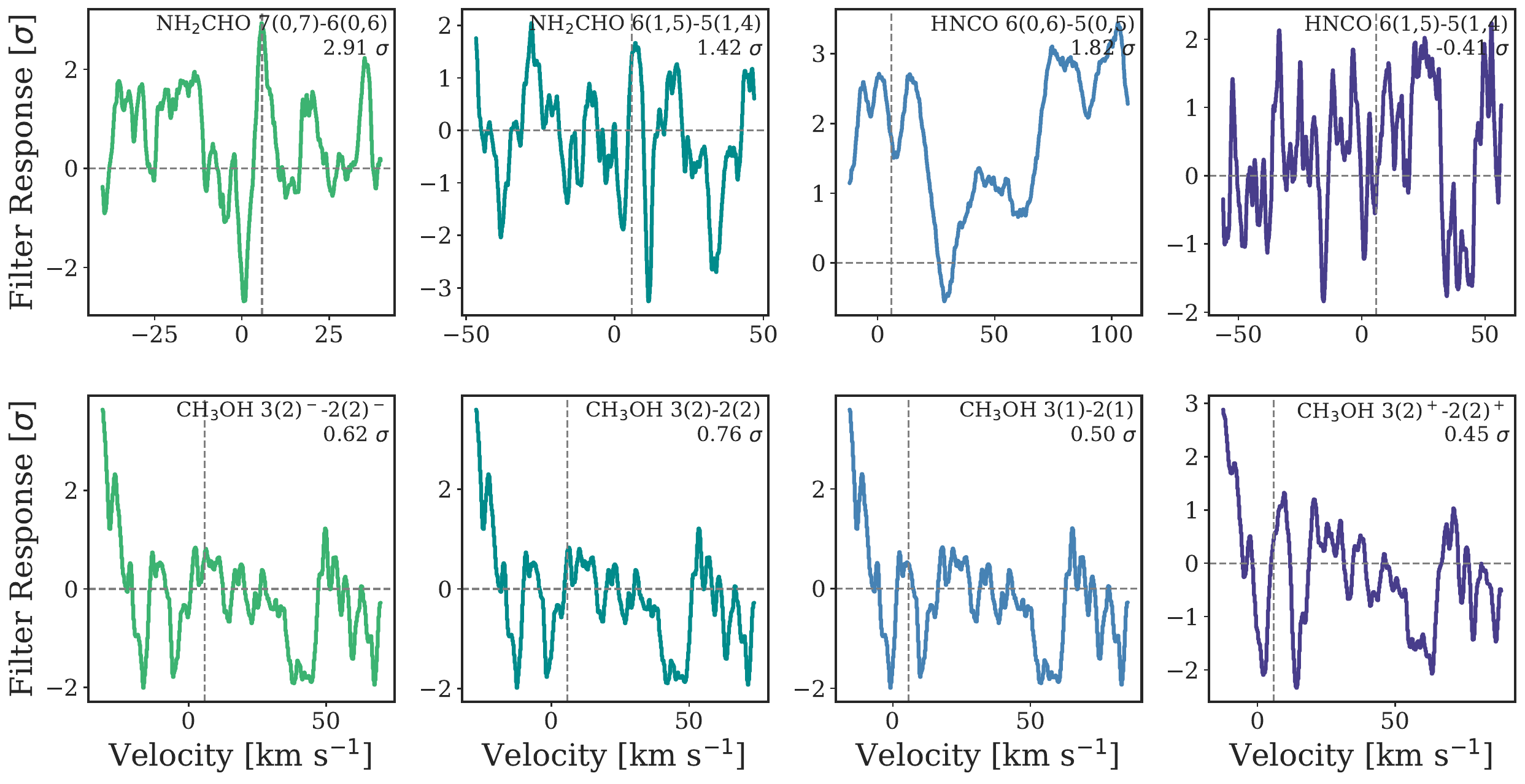}
    \caption{Matched filter response for the lines with low signal-to-noise ratio.}
    \label{fig:matched}
\end{figure*}

We also generate the matched filter responses for our molecular line observations, taking Keplerian masks as the filters, as explained in section \ref{sec:obs_results}. The filter responses for observations with low signal-to-noise ratio are presented in the figure \ref{fig:matched}.

\section{Channel Maps}
\label{app:ch_maps}

\begin{figure*}[t!]
    \centering
    \includegraphics[width=\linewidth]{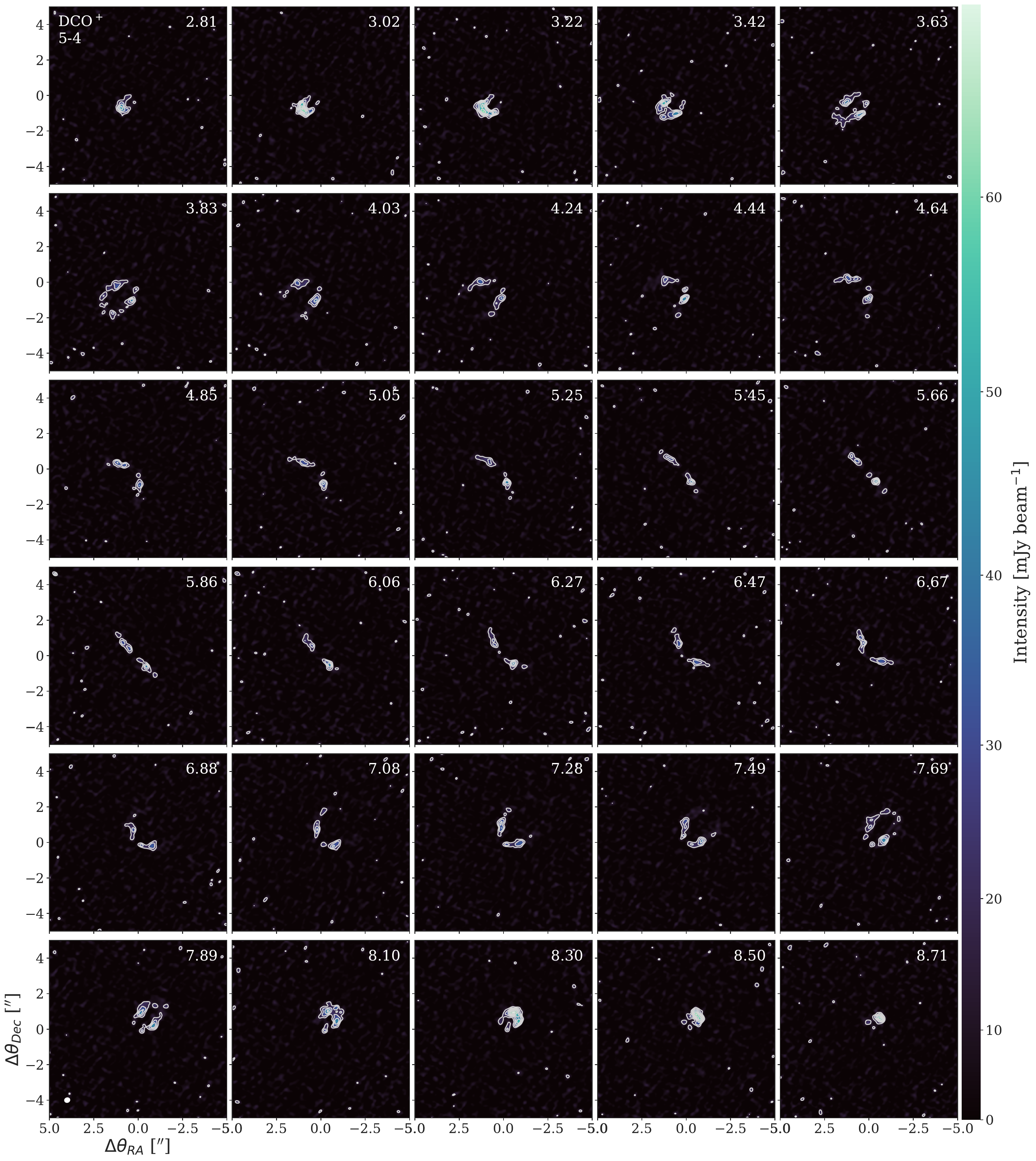}
    \caption{Channel maps of the observed DCO$^+$ 5-4 transition. Channel LSRK velocities are shown in the top-right corner of each panel in km s$^{-1}$. Solid contours represent [3, 5, 7, 9, ...]~$\times~\sigma$ levels, where $\sigma = 5.05$ mJy\,beam$^{-1}$. The synthesized beam is indicated by the ellipse in the lower-left corner, with a size of $0.31''\times0.27''$ and a position angle of 108.94$^\circ$. The colour scale is non-linearly stretched to better highlight faint line emission features.}
    \label{fig:ch_map_dco+_54}
\end{figure*}

\begin{figure*}
    \centering
    \includegraphics[width=\linewidth]{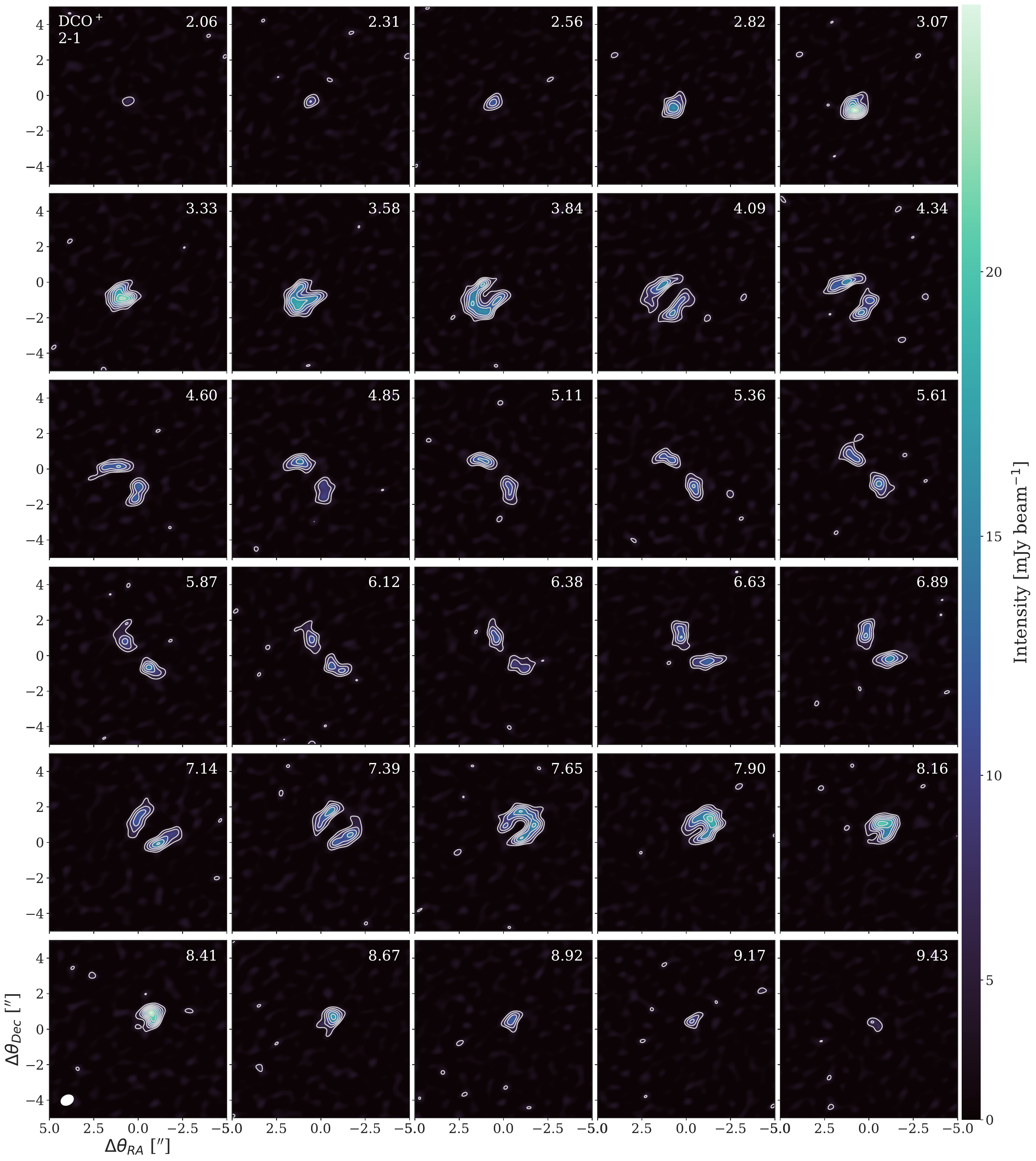}
    \caption{Channel maps of the observed DCO$^+$ 2-1 transition. Channel LSRK velocities are shown in the top-right corner of each panel in km s$^{-1}$. Solid contours represent [3, 5, 7, 9, ...]~$\times~\sigma$ levels, where $\sigma = 1.46$ mJy\,beam$^{-1}$. The synthesized beam is indicated by the ellipse in the lower-left corner, with a size of $0.74''\times0.56''$ and a position angle of 117.39$^\circ$. The colour scale is non-linearly stretched to better highlight faint line emission features.}
    \label{fig:ch_map_dco+_21}
\end{figure*}

\begin{figure*}
    \centering
    \includegraphics[width=\linewidth]{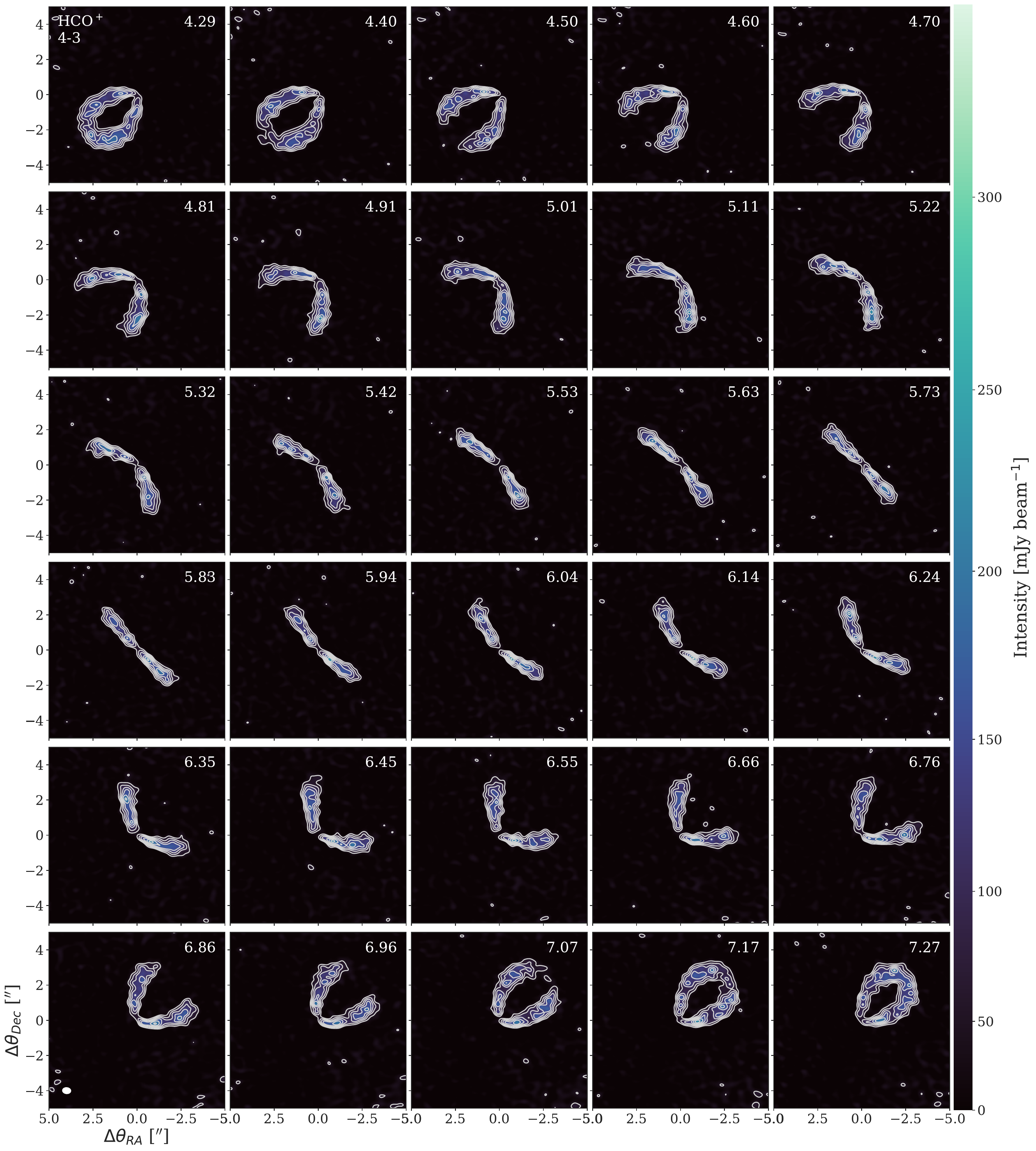}
    \caption{Channel maps of the observed HCO$^+$ 4-3 transition. Channel LSRK velocities are shown in the top-right corner of each panel in km s$^{-1}$. Solid contours represent [3, 5, 7, 9, ...]~$\times~\sigma$ levels, where $\sigma = 15.73$ mJy\,beam$^{-1}$. The synthesized beam is indicated by the ellipse in the lower-left corner, with a size of $0.48''\times0.36''$ and a position angle of 84.82$^\circ$. The colour scale is non-linearly stretched to better highlight faint line emission features.}
    \label{fig:ch_map_hco+43}
\end{figure*}

\begin{figure*}
    \centering
    \includegraphics[width=\linewidth]{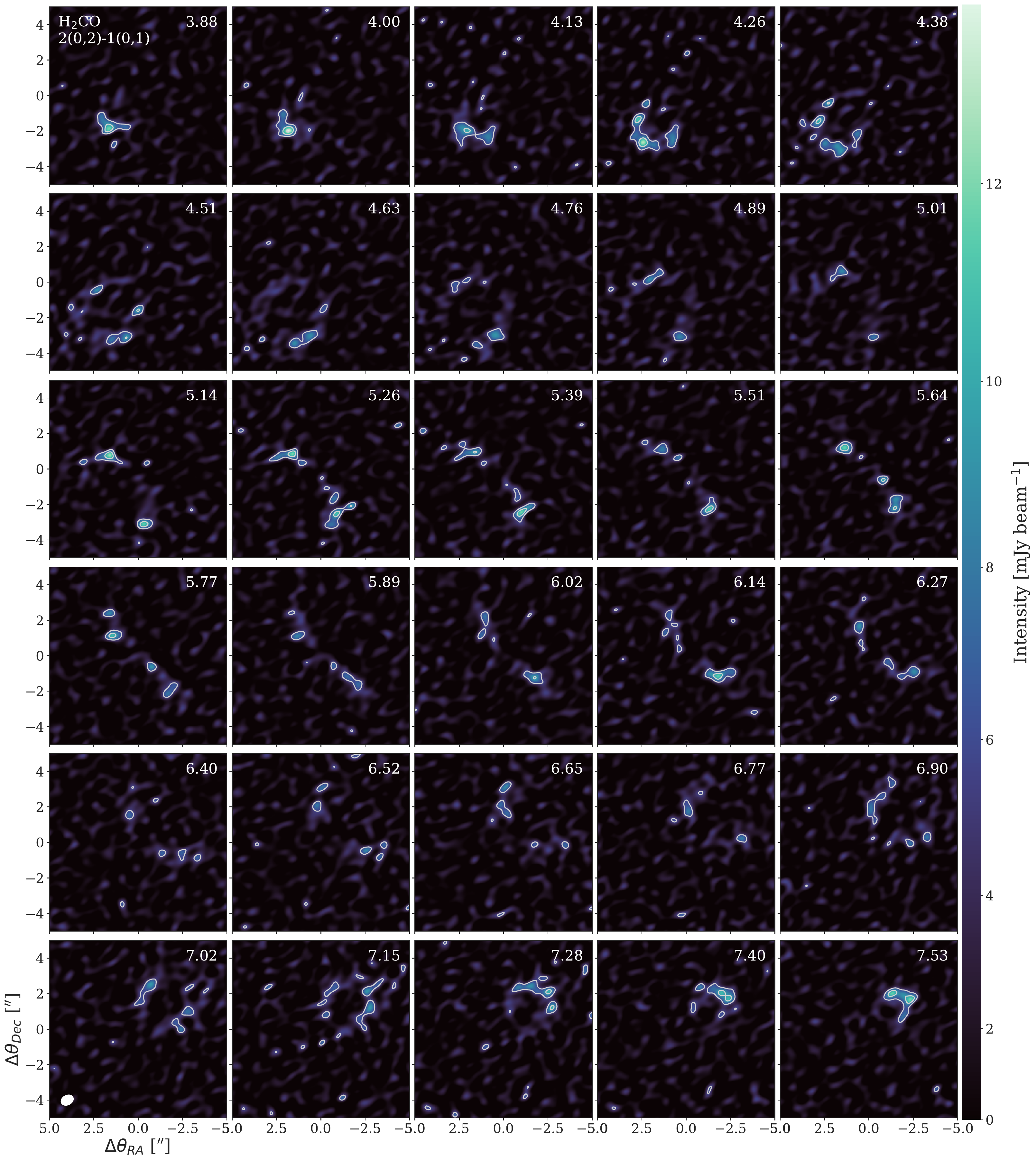}
    \caption{Channel maps of the observed H$_2$CO 2(0,2)-1(0,1) transition. Channel LSRK velocities are shown in the top-right corner of each panel in km s$^{-1}$. Solid contours represent [3, 5]~$\times~\sigma$ levels, where $\sigma = 1.87$ mJy\,beam$^{-1}$. The synthesized beam is indicated by the ellipse in the lower-left corner, with a size of $0.73''\times0.55''$ and a position angle of 117.46$^\circ$. The colour scale is non-linearly stretched to better highlight faint line emission features.}
    \label{fig:ch_map_h2co21}
\end{figure*}

\begin{figure*}
    \centering
    \includegraphics[width=\linewidth]{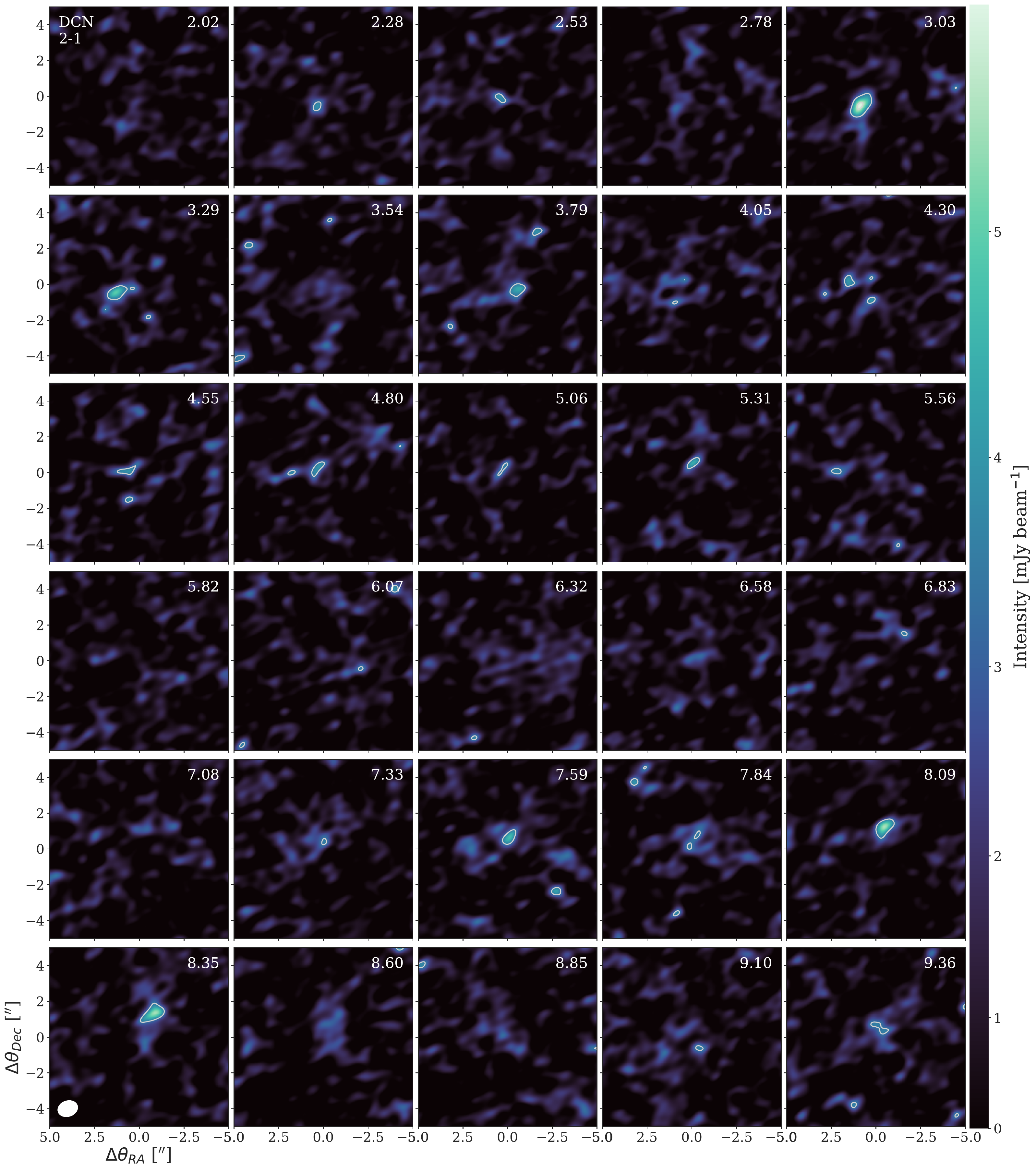}
    \caption{Channel maps of the observed DCN 2-1 transition. Channel LSRK velocities are shown in the top-right corner of each panel in km s$^{-1}$. Solid contours represent [3, 5]~$\times~\sigma$ levels, where $\sigma = 1.13$ mJy\,beam$^{-1}$. The synthesized beam is indicated by the ellipse in the lower-left corner, with a size of $1.13''\times0.87''$ and a position angle of 108.21$^\circ$. The colour scale is non-linearly stretched to better highlight faint line emission features.}
    \label{fig:ch_map_dcn21}
\end{figure*}

\begin{figure*}
    \centering
    \includegraphics[width=\linewidth]{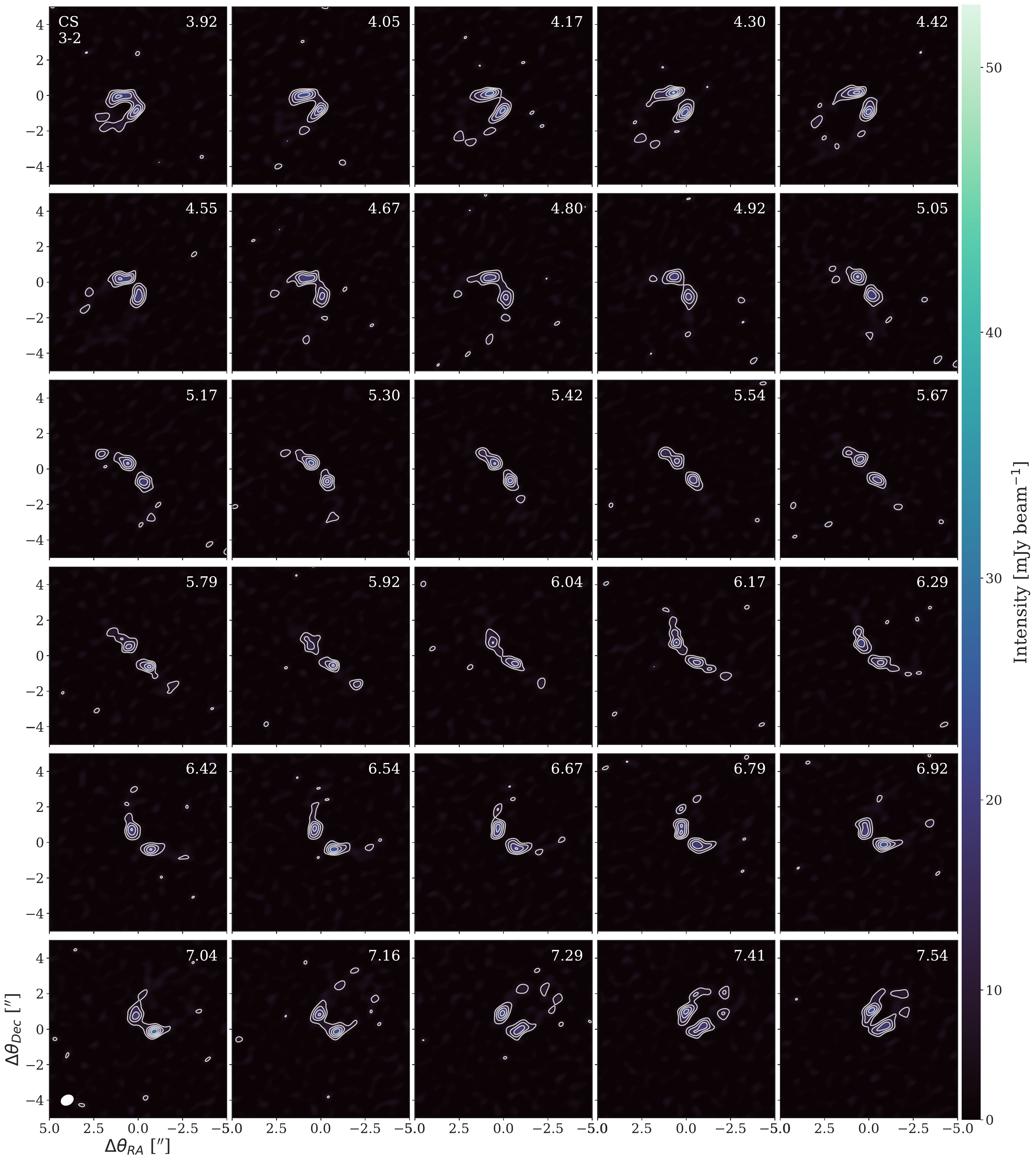}
    \caption{Channel maps of the observed CS 3-2 transition. Channel LSRK velocities are shown in the top-right corner of each panel in km s$^{-1}$. Solid contours represent [3, 5, 7, 9, ...]~$\times~\sigma$ levels, where $\sigma = 2.33$ mJy\,beam$^{-1}$. The synthesized beam is indicated by the ellipse in the lower-left corner, with a size of $0.72''\times0.54''$ and a position angle of 117.95$^\circ$. The colour scale is non-linearly stretched to better highlight faint line emission features.}
    \label{fig:ch_map_cs32}
\end{figure*}

\newpage
\section{Corner Plots}
\label{app:corner}
The corner plots present the posterior distributions of the disk-averaged column densities and excitation temperatures for the detected molecules (see Section \ref{sec:drive-diskavg} and Table \ref{tab:disk_averaged}). Along the diagonal, histograms with overlaid kernel density estimates show the one-dimensional marginalized posterior distributions for each parameter, annotated with vertical lines marking the median (50th percentile) and 1$\sigma$ uncertainties (16th and 84th percentiles). The lower triangle features two-dimensional joint posterior distributions visualized as filled contour plots, where increasingly darker shades indicate higher probability densities for specific parameter combinations.
Together, these interlinked visualizations comprehensively characterize both the individual parameter constraints and their mutual correlations or degeneracies in the fitting results.

\begin{figure}[t!]
    \centering
    \includegraphics[width=\linewidth]{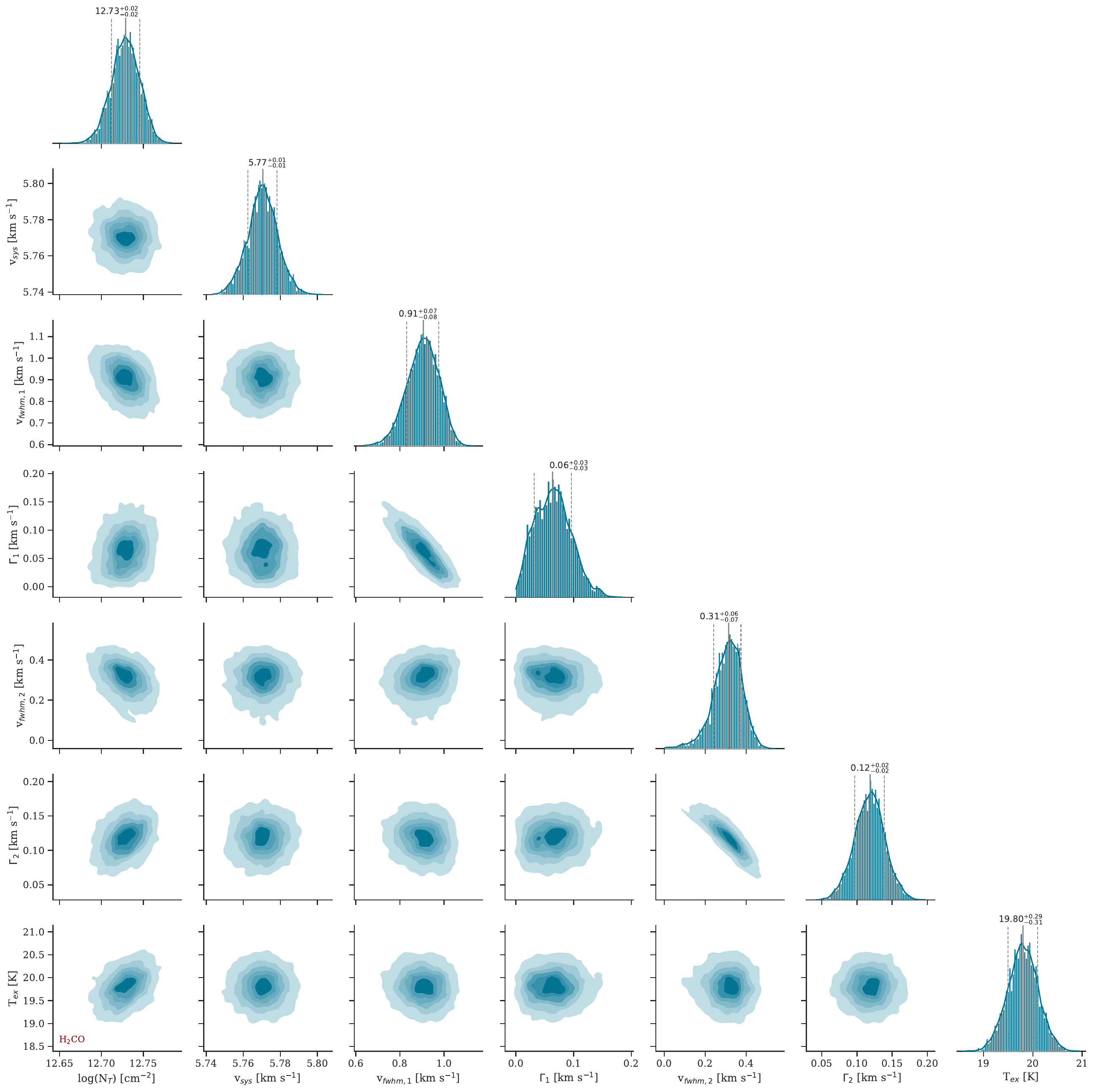}
    \caption{Corner plot of the nested-sampled posterior distributions for \ce{H2CO} disk-averaged column density and excitation temperature, showing parameter constraints and degeneracies.}
    \label{fig:corner_h2co}
\end{figure}

\begin{figure}
    \centering
    \includegraphics[width=\linewidth]{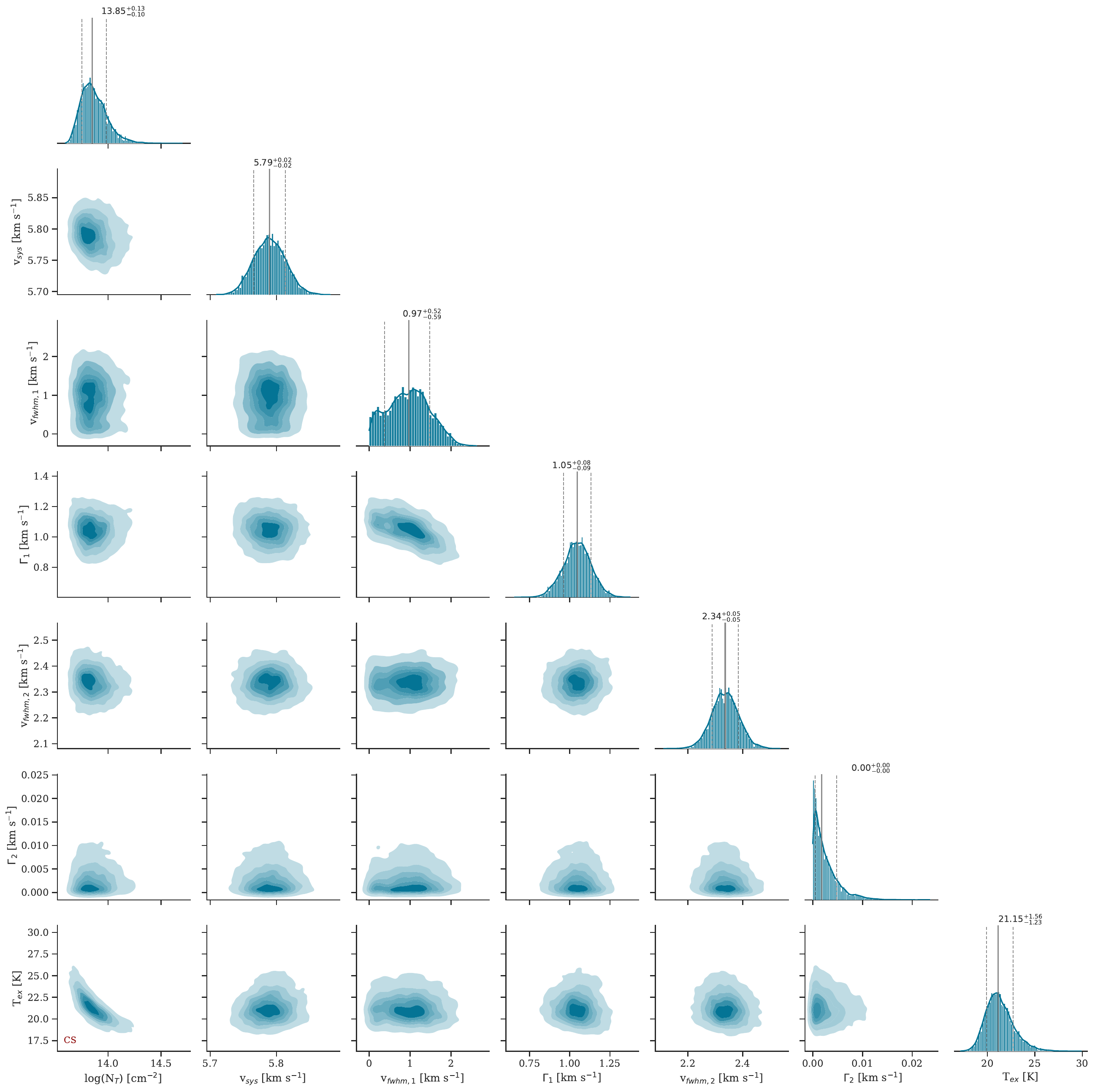}
    \caption{Corner plot of the nested-sampled posterior distributions for \ce{CS} disk-averaged column density and excitation temperature, showing parameter constraints and degeneracies.}
    \label{fig:corner_CS}
\end{figure}

\begin{figure}
    \centering
    \includegraphics[width=\linewidth]{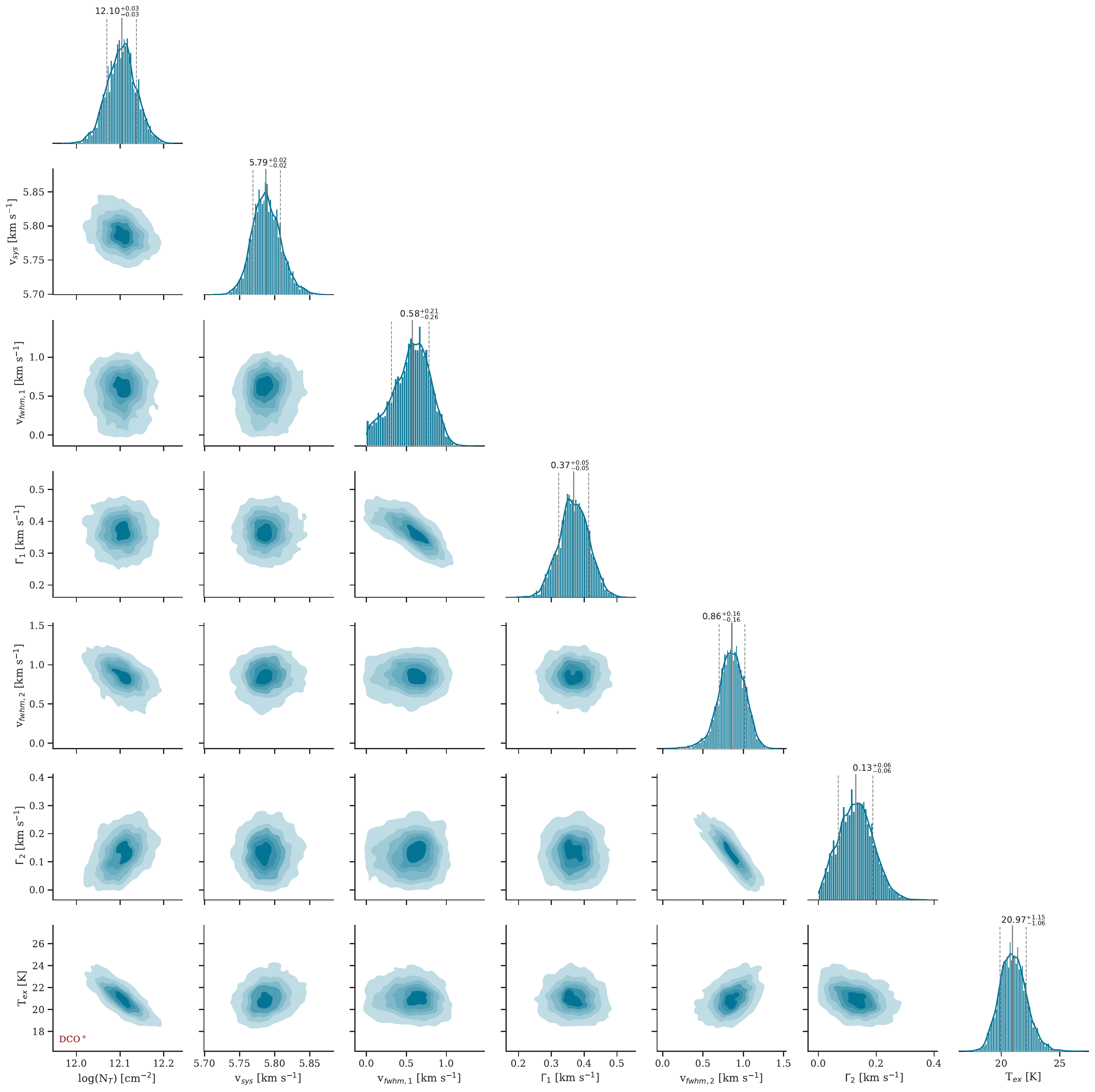}
    \caption{Corner plot of the nested-sampled posterior distributions for \ce{DCO+} disk-averaged column density and excitation temperature, showing parameter constraints and degeneracies.}
    \label{fig:corner_dcop}
\end{figure}

\begin{figure}
    \centering
    \includegraphics[width=\linewidth]{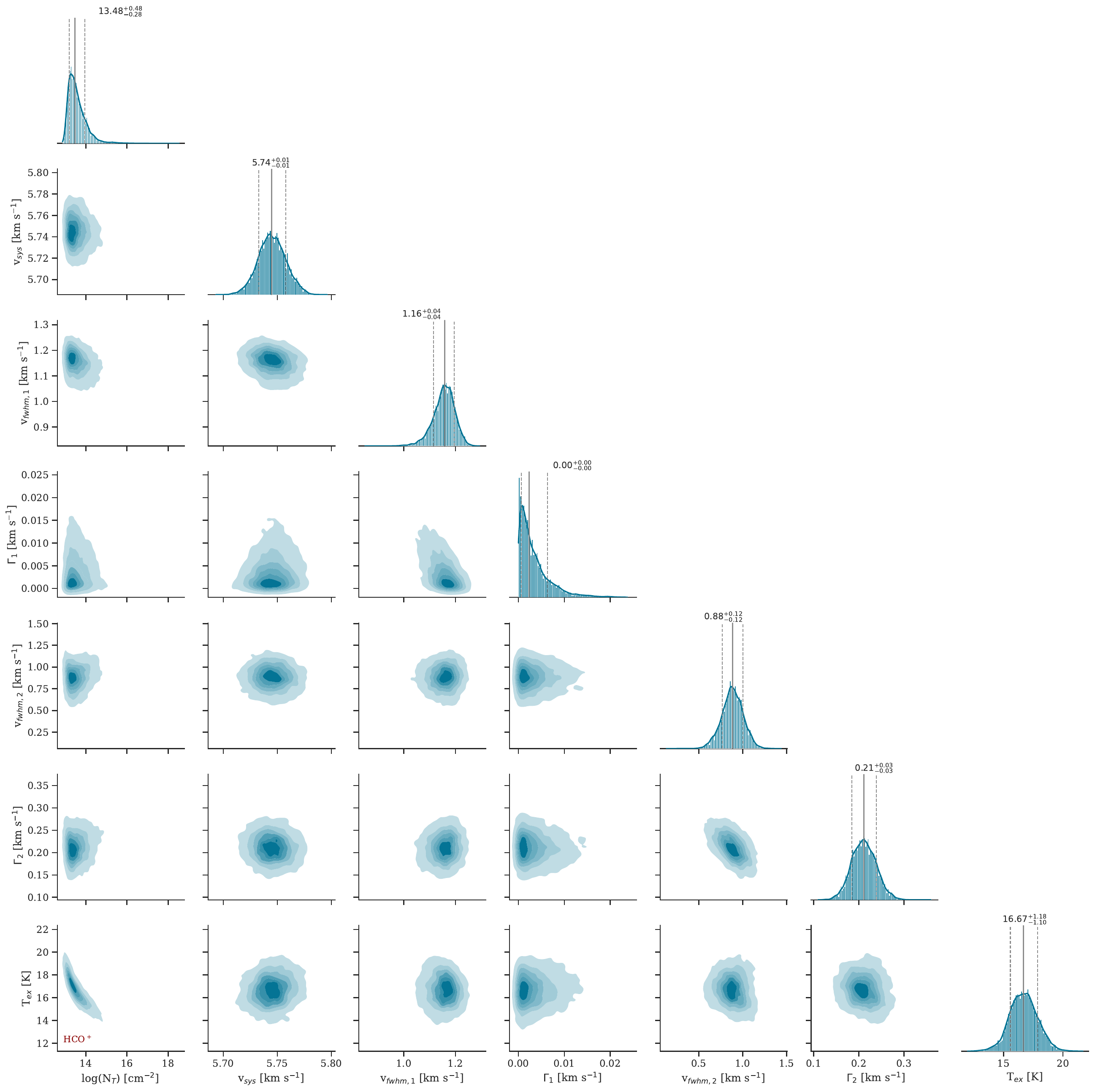}
    \caption{Corner plot of the nested-sampled posterior distributions for \ce{HCO+} disk-averaged column density and excitation temperature, showing parameter constraints and degeneracies.}
    \label{fig:corner_hcop}
\end{figure}

\begin{figure}
    \centering
    \includegraphics[width=\linewidth]{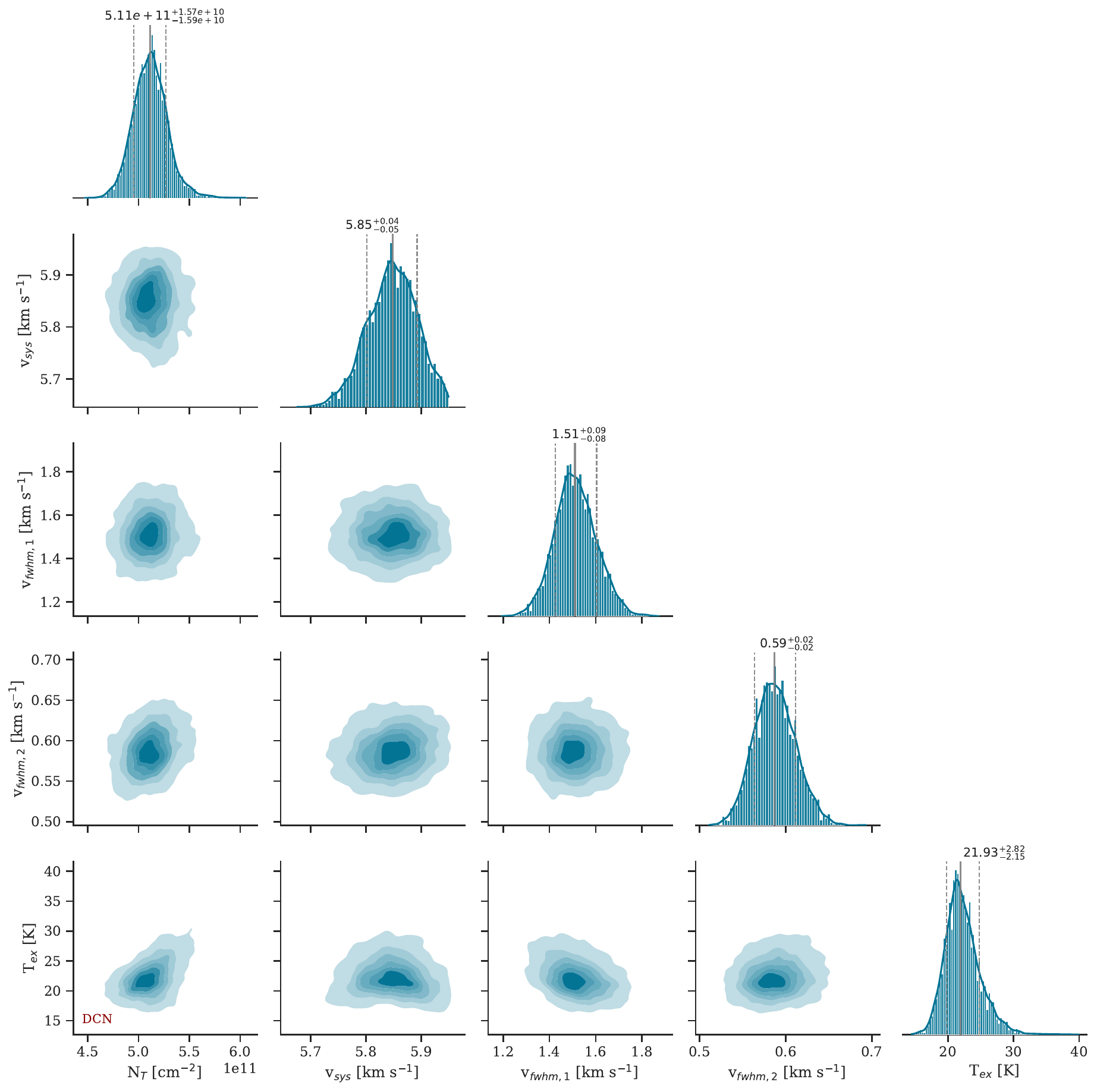}
    \caption{Corner plot of the nested-sampled posterior distributions for \ce{DCN} disk-averaged column density and excitation temperature, showing parameter constraints and degeneracies.}
    \label{fig:corner_dcn}
\end{figure}

\begin{figure}
    \centering
    \includegraphics[width=\linewidth]{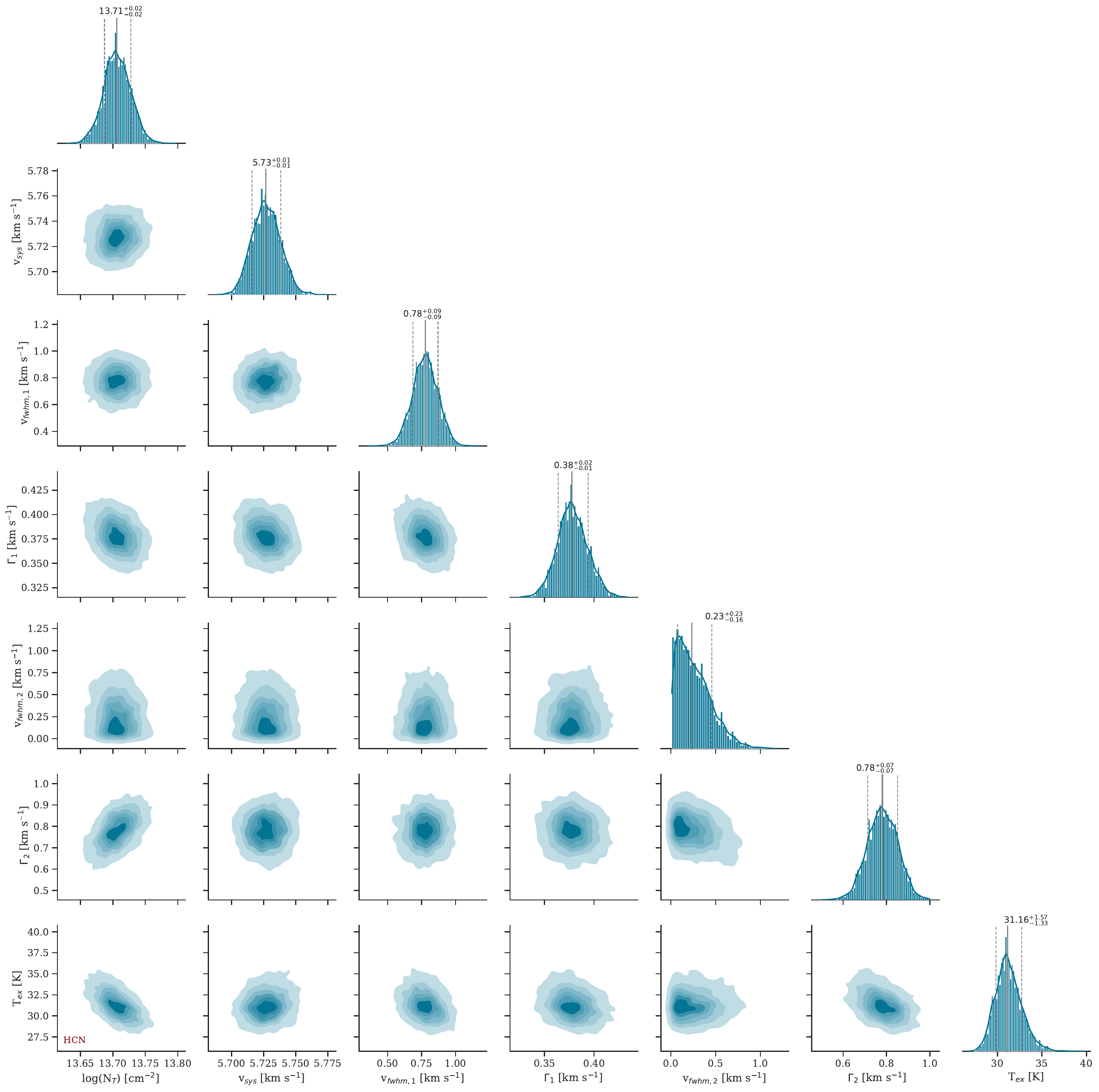}
    \caption{Corner plot of the nested-sampled posterior distributions for \ce{HCN} disk-averaged column density and excitation temperature, showing parameter constraints and degeneracies.}
    \label{fig:corner_hcn}
\end{figure}

\begin{figure}
    \centering
    \includegraphics[width=\linewidth]{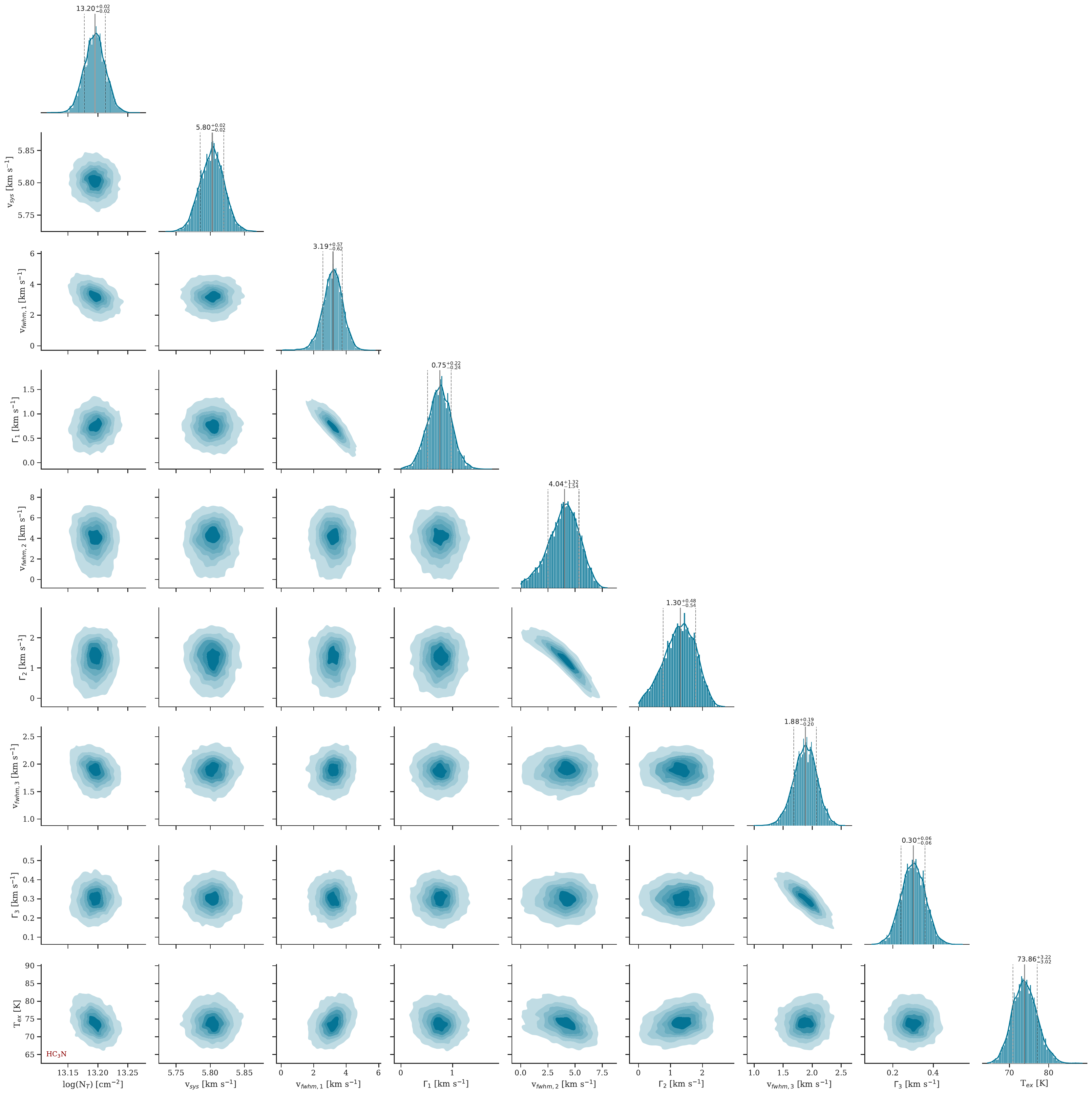}
    \caption{Corner plot of the nested-sampled posterior distributions for \ce{HC3N} disk-averaged column density and excitation temperature, showing parameter constraints and degeneracies.}
    \label{fig:corner_hc3n}
\end{figure}

\clearpage
\bibliography{main}{}

@ARTICLE{semenov18,
       author = {{Semenov}, D. and {Favre}, C. and {Fedele}, D. and {Guilloteau}, S. and {Teague}, R. and {Henning}, Th. and {Dutrey}, A. and {Chapillon}, E. and {Hersant}, F. and {Pi{\'e}tu}, V.},
        title = "{Chemistry in disks. XI. Sulfur-bearing species as tracers of protoplanetary disk physics and chemistry: the DM Tau case}",
      journal = {\aap},
         year = 2018,
        month = sep,
       volume = {617},
          eid = {A28},
        pages = {A28},
          doi = {10.1051/0004-6361/201832980},
archivePrefix = {arXiv},
       eprint = {1806.07707},
 primaryClass = {astro-ph.GA},
       adsurl = {https://ui.adsabs.harvard.edu/abs/2018A&A...617A..28S}
}

@ARTICLE{Booth21,
       author = {{Booth}, Alice S. and {Walsh}, Catherine and {Terwisscha van Scheltinga}, Jeroen and {van Dishoeck}, Ewine F. and {Ilee}, John D. and {Hogerheijde}, Michiel R. and {Kama}, Mihkel and {Nomura}, Hideko},
        title = "{An inherited complex organic molecule reservoir in a warm planet-hosting disk}",
      journal = {Nature Astronomy},
         year = 2021,
        month = jan,
       volume = {5},
        pages = {684-690},
          doi = {10.1038/s41550-021-01352-w},
archivePrefix = {arXiv},
       eprint = {2104.08348},
 primaryClass = {astro-ph.EP},
       adsurl = {https://ui.adsabs.harvard.edu/abs/2021NatAs...5..684B}
}

@ARTICLE{Henning13,
       author = {{Henning}, Thomas and {Semenov}, Dmitry},
        title = "{Chemistry in Protoplanetary Disks}",
      journal = {Chemical Reviews},
         year = 2013,
        month = dec,
       volume = {113},
       number = {12},
        pages = {9016-9042},
          doi = {10.1021/cr400128p},
archivePrefix = {arXiv},
       eprint = {1310.3151},
 primaryClass = {astro-ph.GA},
       adsurl = {https://ui.adsabs.harvard.edu/abs/2013ChRv..113.9016H}
}

@ARTICLE{pegasis25,
       author = {{Maitrey}, S. and {Majumdar}, L. and {Manilal}, V. and {Srivastava}, B. and {Rayalacheruvu}, P. and {Willacy}, K. and {Herbst}, E.},
        title = "{Role of diffusive and nondiffusive grain-surface processes in cold cores: Insights from the PEGASIS three-phase astrochemical model}",
      journal = {\aap},
         year = 2025,
        month = jul,
       volume = {699},
          eid = {A332},
        pages = {A332},
          doi = {10.1051/0004-6361/202554717},
archivePrefix = {arXiv},
       eprint = {2504.18138},
 primaryClass = {astro-ph.GA},
       adsurl = {https://ui.adsabs.harvard.edu/abs/2025A&A...699A.332M}
}

@ARTICLE{Marois08,
       author = {{Marois}, Christian and {Macintosh}, Bruce and {Barman}, Travis and {Zuckerman}, B. and {Song}, Inseok and {Patience}, Jennifer and {Lafreni{\`e}re}, David and {Doyon}, Ren{\'e}},
        title = "{Direct Imaging of Multiple Planets Orbiting the Star HR 8799}",
      journal = {Science},
         year = 2008,
        month = nov,
       volume = {322},
       number = {5906},
        pages = {1348},
          doi = {10.1126/science.1166585},
archivePrefix = {arXiv},
       eprint = {0811.2606},
 primaryClass = {astro-ph},
       adsurl = {https://ui.adsabs.harvard.edu/abs/2008Sci...322.1348M}
}

@ARTICLE{Marois10,
       author = {{Marois}, Christian and {Zuckerman}, B. and {Konopacky}, Quinn M. and {Macintosh}, Bruce and {Barman}, Travis},
        title = "{Images of a fourth planet orbiting HR 8799}",
      journal = {\nat},
         year = 2010,
        month = dec,
       volume = {468},
       number = {7327},
        pages = {1080-1083},
          doi = {10.1038/nature09684},
archivePrefix = {arXiv},
       eprint = {1011.4918},
 primaryClass = {astro-ph.EP},
       adsurl = {https://ui.adsabs.harvard.edu/abs/2010Natur.468.1080M}
}

@ARTICLE{Molliere20,
       author = {{Molli{\`e}re}, P. and {Stolker}, T. and {Lacour}, S. and {Otten}, G.~P.~P.~L. and {Shangguan}, J. and {Charnay}, B. and {Molyarova}, T. and {Nowak}, M. and {Henning}, Th. and {Marleau}, G. -D. and {Semenov}, D.~A. and {van Dishoeck}, E. and {Eisenhauer}, F. and {Garcia}, P. and {Garcia Lopez}, R. and {Girard}, J.~H. and {Greenbaum}, A.~Z. and {Hinkley}, S. and {Kervella}, P. and {Kreidberg}, L. and {Maire}, A. -L. and {Nasedkin}, E. and {Pueyo}, L. and {Snellen}, I.~A.~G. and {Vigan}, A. and {Wang}, J. and {de Zeeuw}, P.~T. and {Zurlo}, A.},
        title = "{Retrieving scattering clouds and disequilibrium chemistry in the atmosphere of HR 8799e}",
      journal = {\aap},
         year = 2020,
        month = aug,
       volume = {640},
          eid = {A131},
        pages = {A131},
          doi = {10.1051/0004-6361/202038325},
archivePrefix = {arXiv},
       eprint = {2006.09394},
 primaryClass = {astro-ph.EP},
       adsurl = {https://ui.adsabs.harvard.edu/abs/2020A&A...640A.131M}
}

@ARTICLE{Lis97,
       author = {{Lis}, D.~C. and {Mehringer}, D.~M. and {Benford}, D. and {Gardner}, M. and {Phillips}, T.~G. and {Bockel{\'e}e-Morvan}, D. and {Biver}, N. and {Colom}, P. and {Crovisier}, J. and {Despois}, D. and {Rauer}, H.},
        title = "{New Molecular Species in Comet C/1995 O1 (Hale-Bopp) Observed with the Caltech Ssubmillimeter Observatory}",
      journal = {Earth Moon and Planets},
         year = 1997,
        month = jul,
       volume = {78},
        pages = {13-20},
          doi = {10.1023/A:1006281802554},
       adsurl = {https://ui.adsabs.harvard.edu/abs/1997EM&P...78...13L}
}

@ARTICLE{Biver24,
       author = {{Biver}, N. and {Bockel{\'e}e-Morvan}, D. and {Handzlik}, B. and {Sandqvist}, Aa. and {Boissier}, J. and {Drozdovskaya}, M.~N. and {Moreno}, R. and {Crovisier}, J. and {Lis}, D.~C. and {Cordiner}, M. and {Milam}, S. and {Roth}, N.~X. and {Bonev}, B.~P. and {Dello Russo}, N. and {Vervack}, R. and {Opitom}, C. and {Kawakita}, H.},
        title = "{Chemical composition of comets C/2021 A1 (Leonard) and C/2022 E3 (ZTF) from radio spectroscopy and the abundance of HCOOH and HNCO in comets}",
      journal = {\aap},
         year = 2024,
        month = oct,
       volume = {690},
          eid = {A271},
        pages = {A271},
          doi = {10.1051/0004-6361/202450921},
archivePrefix = {arXiv},
       eprint = {2408.10759},
 primaryClass = {astro-ph.EP},
       adsurl = {https://ui.adsabs.harvard.edu/abs/2024A&A...690A.271B}
}

@ARTICLE{Morvan2000,
       author = {{Bockel{\'e}e-Morvan}, D. and {Lis}, D.~C. and {Wink}, J.~E. and {Despois}, D. and {Crovisier}, J. and {Bachiller}, R. and {Benford}, D.~J. and {Biver}, N. and {Colom}, P. and {Davies}, J.~K. and {G{\'e}rard}, E. and {Germain}, B. and {Houde}, M. and {Mehringer}, D. and {Moreno}, R. and {Paubert}, G. and {Phillips}, T.~G. and {Rauer}, H.},
        title = "{New molecules found in comet C/1995 O1 (Hale-Bopp). Investigating the link between cometary and interstellar material}",
      journal = {\aap},
         year = 2000,
        month = jan,
       volume = {353},
        pages = {1101-1114},
       adsurl = {https://ui.adsabs.harvard.edu/abs/2000A&A...353.1101B}
}

@ARTICLE{yamato24,
       author = {{Yamato}, Yoshihide and {Notsu}, Shota and {Aikawa}, Yuri and {Okoda}, Yuki and {Nomura}, Hideko and {Sakai}, Nami},
        title = "{Chemistry of Complex Organic Molecules in the V883 Ori Disk Revealed by ALMA Band 3 Observations}",
      journal = {\aj},
         year = 2024,
        month = feb,
       volume = {167},
       number = {2},
          eid = {66},
        pages = {66},
          doi = {10.3847/1538-3881/ad11d9},
archivePrefix = {arXiv},
       eprint = {2312.01300},
 primaryClass = {astro-ph.EP},
       adsurl = {https://ui.adsabs.harvard.edu/abs/2024AJ....167...66Y}
}

@ARTICLE{oberg23,
       author = {{{\"O}berg}, Karin I. and {Facchini}, Stefano and {Anderson}, Dana E.},
        title = "{Protoplanetary Disk Chemistry}",
      journal = {\araa},
         year = 2023,
        month = aug,
       volume = {61},
        pages = {287-328},
          doi = {10.1146/annurev-astro-022823-040820},
archivePrefix = {arXiv},
       eprint = {2309.05685},
 primaryClass = {astro-ph.EP},
       adsurl = {https://ui.adsabs.harvard.edu/abs/2023ARA&A..61..287O}
}

@ARTICLE{Hernandez2024,
       author = {{Hern{\'a}ndez-Vera}, Claudio and {Guzm{\'a}n}, Viviana V. and {Artur de la Villarmois}, Elizabeth and {{\"O}berg}, Karin I. and {Cleeves}, L. Ilsedore and {Hogerheijde}, Michiel R. and {Qi}, Chunhua and {Carpenter}, John and {Fayolle}, Edith C.},
        title = "{Radial and Vertical Constraints on the Icy Origin of H$_{2}$CO in the HD 163296 Protoplanetary Disk}",
      journal = {\apj},
         year = 2024,
        month = may,
       volume = {967},
       number = {1},
          eid = {68},
        pages = {68},
          doi = {10.3847/1538-4357/ad3cdb},
archivePrefix = {arXiv},
       eprint = {2404.06133},
 primaryClass = {astro-ph.EP},
       adsurl = {https://ui.adsabs.harvard.edu/abs/2024ApJ...967...68H}
}

@ARTICLE{Oberg2021,
       author = {{{\"O}berg}, Karin I. and {Guzm{\'a}n}, Viviana V. and {Walsh}, Catherine and {Aikawa}, Yuri and {Bergin}, Edwin A. and {Law}, Charles J. and {Loomis}, Ryan A. and {Alarc{\'o}n}, Felipe and {Andrews}, Sean M. and {Bae}, Jaehan and {Bergner}, Jennifer B. and {Boehler}, Yann and {Booth}, Alice S. and {Bosman}, Arthur D. and {Calahan}, Jenny K. and {Cataldi}, Gianni and {Cleeves}, L. Ilsedore and {Czekala}, Ian and {Furuya}, Kenji and {Huang}, Jane and {Ilee}, John D. and {Kurtovic}, Nicolas T. and {Le Gal}, Romane and {Liu}, Yao and {Long}, Feng and {M{\'e}nard}, Fran{\c{c}}ois and {Nomura}, Hideko and {P{\'e}rez}, Laura M. and {Qi}, Chunhua and {Schwarz}, Kamber R. and {Sierra}, Anibal and {Teague}, Richard and {Tsukagoshi}, Takashi and {Yamato}, Yoshihide and {van't Hoff}, Merel L.~R. and {Waggoner}, Abygail R. and {Wilner}, David J. and {Zhang}, Ke},
        title = "{Molecules with ALMA at Planet-forming Scales (MAPS). I. Program Overview and Highlights}",
      journal = {\apjs},
         year = 2021,
        month = nov,
       volume = {257},
       number = {1},
          eid = {1},
        pages = {1},
          doi = {10.3847/1538-4365/ac1432},
archivePrefix = {arXiv},
       eprint = {2109.06268},
 primaryClass = {astro-ph.EP},
       adsurl = {https://ui.adsabs.harvard.edu/abs/2021ApJS..257....1O}
}

@ARTICLE{galario,
   author = {{Tazzari}, M. and {Beaujean}, F. and {Testi}, L.},
    title = "{GALARIO: a GPU accelerated library for analysing radio interferometer observations}",
  journal = {\mnras},
archivePrefix = "arXiv",
   eprint = {1709.06999},
 primaryClass = "astro-ph.IM",
     year = 2018,
    month = jun,
   volume = 476,
    pages = {4527-4542},
      doi = {10.1093/mnras/sty409},
   adsurl = {http://adsabs.harvard.edu/abs/2018MNRAS.476.4527T}
}

@ARTICLE{Bunchner2016,
       author = {{Buchner}, Johannes},
        title = "{A statistical test for Nested Sampling algorithms}",
      journal = {Statistics and Computing},
         year = 2016,
        month = jan,
       volume = {26},
       number = {1-2},
        pages = {383-392},
          doi = {10.1007/s11222-014-9512-y},
archivePrefix = {arXiv},
       eprint = {1407.5459},
 primaryClass = {stat.CO},
       adsurl = {https://ui.adsabs.harvard.edu/abs/2016S&C....26..383B}
}

@ARTICLE{Buchner2019,
       author = {{Buchner}, Johannes},
        title = "{Collaborative Nested Sampling: Big Data versus Complex Physical Models}",
      journal = {\pasp},
         year = 2019,
        month = oct,
       volume = {131},
       number = {1004},
        pages = {108005},
          doi = {10.1088/1538-3873/aae7fc},
archivePrefix = {arXiv},
       eprint = {1707.04476},
 primaryClass = {stat.CO},
       adsurl = {https://ui.adsabs.harvard.edu/abs/2019PASP..131j8005B}
}

@ARTICLE{Buchner2021,
       author = {{Buchner}, Johannes},
        title = "{UltraNest - a robust, general purpose Bayesian inference engine}",
      journal = {The Journal of Open Source Software},
         year = 2021,
        month = apr,
       volume = {6},
       number = {60},
          eid = {3001},
        pages = {3001},
          doi = {10.21105/joss.03001},
archivePrefix = {arXiv},
       eprint = {2101.09604},
 primaryClass = {stat.CO},
       adsurl = {https://ui.adsabs.harvard.edu/abs/2021JOSS....6.3001B}
}

@ARTICLE{loomis2018detecting,
       author = {{Loomis}, Ryan A. and {{\"O}berg}, Karin I. and {Andrews}, Sean M. and {Walsh}, Catherine and {Czekala}, Ian and {Huang}, Jane and {Rosenfeld}, Katherine A.},
        title = "{Detecting Weak Spectral Lines in Interferometric Data through Matched Filtering}",
      journal = {\aj},
         year = 2018,
        month = apr,
       volume = {155},
       number = {4},
          eid = {182},
        pages = {182},
          doi = {10.3847/1538-3881/aab604},
archivePrefix = {arXiv},
       eprint = {1803.04987},
 primaryClass = {astro-ph.IM},
       adsurl = {https://ui.adsabs.harvard.edu/abs/2018AJ....155..182L}
}

@software{loomis2018visible,
       author = {{Loomis}, Ryan A. and {Oberg}, Karin I. and {Andrews}, Sean M. and {Walsh}, Catherine and {Czekala}, Ian and {Huang}, Jane and {Rosenfeld}, Katherine A.},
        title = "{VISIBLE: VISIbility Based Line Extraction}",
 howpublished = {Astrophysics Source Code Library, record ascl:1802.006},
         year = 2018,
        month = feb,
          eid = {ascl:1802.006},
       adsurl = {https://ui.adsabs.harvard.edu/abs/2018ascl.soft02006L}
}

@ARTICLE{bettermoments2018,
       author = {{Teague}, Richard and {Foreman-Mackey}, Daniel},
        title = "{A Robust Method to Measure Centroids of Spectral Lines}",
      journal = {Research Notes of the American Astronomical Society},
         year = 2018,
        month = Sep,
       volume = {2},
          eid = {173},
        pages = {173},
          doi = {10.3847/2515-5172/aae265},
       adsurl = {https://ui.adsabs.harvard.edu/abs/2018RNAAS...2c.173T}
}

@article{GoFish,
    doi = {10.21105/joss.01632},
    url = {https://doi.org/10.21105/joss.01632},
    year = {2019},
    month = {sep},
    publisher = {The Open Journal},
    volume = {4},
    number = {41},
    pages = {1632},
    author = {Richard Teague},
    title = {GoFish: Fishing for Line Observations in Protoplanetary Disks},
    journal = {The Journal of Open Source Software}
}

@ARTICLE{Muller2001,
       author = {{M{\"u}ller}, H.~S.~P. and {Thorwirth}, S. and {Roth}, D.~A. and {Winnewisser}, G.},
        title = "{The Cologne Database for Molecular Spectroscopy, CDMS}",
      journal = {\aap},
         year = 2001,
        month = apr,
       volume = {370},
        pages = {L49-L52},
          doi = {10.1051/0004-6361:20010367},
       adsurl = {https://ui.adsabs.harvard.edu/abs/2001A&A...370L..49M}
}

@ARTICLE{Muller2005,
       author = {{M{\"u}ller}, Holger S.~P. and {Schl{\"o}der}, Frank and {Stutzki}, J{\"u}rgen and {Winnewisser}, Gisbert},
        title = "{The Cologne Database for Molecular Spectroscopy, CDMS: a useful tool for astronomers and spectroscopists}",
      journal = {Journal of Molecular Structure},
         year = 2005,
        month = may,
       volume = {742},
       number = {1-3},
        pages = {215-227},
          doi = {10.1016/j.molstruc.2005.01.027},
       adsurl = {https://ui.adsabs.harvard.edu/abs/2005JMoSt.742..215M}
}

@ARTICLE{Endres2016,
       author = {{Endres}, Christian P. and {Schlemmer}, Stephan and {Schilke}, Peter and {Stutzki}, J{\"u}rgen and {M{\"u}ller}, Holger S.~P.},
        title = "{The Cologne Database for Molecular Spectroscopy, CDMS, in the Virtual Atomic and Molecular Data Centre, VAMDC}",
      journal = {Journal of Molecular Spectroscopy},
         year = 2016,
        month = sep,
       volume = {327},
        pages = {95-104},
          doi = {10.1016/j.jms.2016.03.005},
       adsurl = {https://ui.adsabs.harvard.edu/abs/2016JMoSp.327...95E}
}

@article{pickett1998,
title = {SUBMILLIMETER, MILLIMETER, AND MICROWAVE SPECTRAL LINE CATALOG},
journal = {Journal of Quantitative Spectroscopy and Radiative Transfer},
volume = {60},
number = {5},
pages = {883-890},
year = {1998},
issn = {0022-4073},
doi = {https://doi.org/10.1016/S0022-4073(98)00091-0},
url = {https://www.sciencedirect.com/science/article/pii/S0022407398000910},
author = {H.M. Pickett and R.L. Poynter and E.A. Cohen and M.L. Delitsky and J.C. Pearson and H.S.P. {M{\"u}ller}}
}

@article{Cataldi2021,
doi = {10.3847/1538-4365/ac143d},
url = {https://dx.doi.org/10.3847/1538-4365/ac143d},
year = {2021},
month = {nov},
publisher = {The American Astronomical Society},
volume = {257},
number = {1},
pages = {10},
author = {Cataldi, Gianni and Yamato, Yoshihide and Aikawa, Yuri and Bergner, Jennifer B. and Furuya, Kenji and Guzmán, Viviana V. and Huang, Jane and Loomis, Ryan A. and Qi, Chunhua and Andrews, Sean M. and Bergin, Edwin A. and Booth, Alice S. and Bosman, Arthur D. and Cleeves, L. Ilsedore and Czekala, Ian and Ilee, John D. and Law, Charles J. and Le Gal, Romane and Liu, Yao and Long, Feng and Ménard, François and Nomura, Hideko and Öberg, Karin I. and Schwarz, Kamber R. and Teague, Richard and Tsukagoshi, Takashi and Walsh, Catherine and Wilner, David J. and Zhang, Ke},
title = {Molecules with ALMA at Planet-forming Scales (MAPS). X. Studying Deuteration at High Angular Resolution toward Protoplanetary Disks},
journal = {The Astrophysical Journal Supplement Series}
}

@ARTICLE{Teague2016,
       author = {{Teague}, R. and {Guilloteau}, S. and {Semenov}, D. and {Henning}, Th. and {Dutrey}, A. and {Pi{\'e}tu}, V. and {Birnstiel}, T. and {Chapillon}, E. and {Hollenbach}, D. and {Gorti}, U.},
        title = "{Measuring turbulence in TW Hydrae with ALMA: methods and limitations}",
      journal = {\aap},
         year = 2016,
        month = jul,
       volume = {592},
          eid = {A49},
        pages = {A49},
          doi = {10.1051/0004-6361/201628550},
archivePrefix = {arXiv},
       eprint = {1606.00005},
 primaryClass = {astro-ph.SR},
       adsurl = {https://ui.adsabs.harvard.edu/abs/2016A&A...592A..49T}
}

@ARTICLE{Flaherty2020,
       author = {{Flaherty}, Kevin and {Hughes}, A. Meredith and {Simon}, Jacob B. and {Qi}, Chunhua and {Bai}, Xue-Ning and {Bulatek}, Alyssa and {Andrews}, Sean M. and {Wilner}, David J. and {K{\'o}sp{\'a}l}, {\'A}gnes},
        title = "{Measuring Turbulent Motion in Planet-forming Disks with ALMA: A Detection around DM Tau and Nondetections around MWC 480 and V4046 Sgr}",
      journal = {\apj},
         year = 2020,
        month = jun,
       volume = {895},
       number = {2},
          eid = {109},
        pages = {109},
          doi = {10.3847/1538-4357/ab8cc5},
archivePrefix = {arXiv},
       eprint = {2004.12176},
 primaryClass = {astro-ph.SR},
       adsurl = {https://ui.adsabs.harvard.edu/abs/2020ApJ...895..109F}
}

@ARTICLE{Guzman2021,
       author = {{Guzm{\'a}n}, Viviana V. and {Bergner}, Jennifer B. and {Law}, Charles J. and {{\"O}berg}, Karin I. and {Walsh}, Catherine and {Cataldi}, Gianni and {Aikawa}, Yuri and {Bergin}, Edwin A. and {Czekala}, Ian and {Huang}, Jane and {Andrews}, Sean M. and {Loomis}, Ryan A. and {Zhang}, Ke and {Le Gal}, Romane and {Alarc{\'o}n}, Felipe and {Ilee}, John D. and {Teague}, Richard and {Cleeves}, L. Ilsedore and {Wilner}, David J. and {Long}, Feng and {Schwarz}, Kamber R. and {Bosman}, Arthur D. and {P{\'e}rez}, Laura M. and {M{\'e}nard}, Fran{\c{c}}ois and {Liu}, Yao},
        title = "{Molecules with ALMA at Planet-forming Scales (MAPS). VI. Distribution of the Small Organics HCN, C$_{2}$H, and H$_{2}$CO}",
      journal = {\apjs},
         year = 2021,
        month = nov,
       volume = {257},
       number = {1},
          eid = {6},
        pages = {6},
          doi = {10.3847/1538-4365/ac1440},
archivePrefix = {arXiv},
       eprint = {2109.06391},
 primaryClass = {astro-ph.EP},
       adsurl = {https://ui.adsabs.harvard.edu/abs/2021ApJS..257....6G}
}

@ARTICLE{Bergner2021,
       author = {{Bergner}, Jennifer B. and {{\"O}berg}, Karin I. and {Guzm{\'a}n}, Viviana V. and {Law}, Charles J. and {Loomis}, Ryan A. and {Cataldi}, Gianni and {Bosman}, Arthur D. and {Aikawa}, Yuri and {Andrews}, Sean M. and {Bergin}, Edwin A. and {Booth}, Alice S. and {Cleeves}, L. Ilsedore and {Czekala}, Ian and {Huang}, Jane and {Ilee}, John D. and {Le Gal}, Romane and {Long}, Feng and {Nomura}, Hideko and {M{\'e}nard}, Fran{\c{c}}ois and {Qi}, Chunhua and {Schwarz}, Kamber R. and {Teague}, Richard and {Tsukagoshi}, Takashi and {Walsh}, Catherine and {Wilner}, David J. and {Yamato}, Yoshihide},
        title = "{Molecules with ALMA at Planet-forming Scales (MAPS). XI. CN and HCN as Tracers of Photochemistry in Disks}",
      journal = {\apjs},
         year = 2021,
        month = nov,
       volume = {257},
       number = {1},
          eid = {11},
        pages = {11},
          doi = {10.3847/1538-4365/ac143a},
archivePrefix = {arXiv},
       eprint = {2109.06694},
 primaryClass = {astro-ph.SR},
       adsurl = {https://ui.adsabs.harvard.edu/abs/2021ApJS..257...11B}
}

@ARTICLE{emcee2013,
       author = {{Foreman-Mackey}, Daniel and {Hogg}, David W. and {Lang}, Dustin and {Goodman}, Jonathan},
        title = "{emcee: The MCMC Hammer}",
      journal = {\pasp},
         year = 2013,
        month = mar,
       volume = {125},
       number = {925},
        pages = {306},
          doi = {10.1086/670067},
archivePrefix = {arXiv},
       eprint = {1202.3665},
 primaryClass = {astro-ph.IM},
       adsurl = {https://ui.adsabs.harvard.edu/abs/2013PASP..125..306F}
}

@ARTICLE{Colzi2021,
       author = {{Colzi}, L. and {Rivilla}, V.~M. and {Beltr{\'a}n}, M.~T. and {Jim{\'e}nez-Serra}, I. and {Mininni}, C. and {Melosso}, M. and {Cesaroni}, R. and {Fontani}, F. and {Lorenzani}, A. and {S{\'a}nchez-Monge}, A. and {Viti}, S. and {Schilke}, P. and {Testi}, L. and {Alonso}, E.~R. and {Kolesnikov{\'a}}, L.},
        title = "{The GUAPOS project. II. A comprehensive study of peptide-like bond molecules}",
      journal = {\aap},
         year = 2021,
        month = sep,
       volume = {653},
          eid = {A129},
        pages = {A129},
          doi = {10.1051/0004-6361/202141573},
archivePrefix = {arXiv},
       eprint = {2107.11258},
 primaryClass = {astro-ph.GA},
       adsurl = {https://ui.adsabs.harvard.edu/abs/2021A&A...653A.129C}
}

@ARTICLE{marcelino2018,
       author = {{Marcelino}, N. and {Gerin}, M. and {Cernicharo}, J. and {Fuente}, A. and {Wootten}, H.~A. and {Chapillon}, E. and {Pety}, J. and {Lis}, D.~C. and {Roueff}, E. and {Commer{\c{c}}on}, B. and {Ciardi}, A.},
        title = "{ALMA observations of the young protostellar system Barnard 1b: Signatures of an incipient hot corino in B1b-S}",
      journal = {\aap},
         year = 2018,
        month = nov,
       volume = {620},
          eid = {A80},
        pages = {A80},
          doi = {10.1051/0004-6361/201731955},
archivePrefix = {arXiv},
       eprint = {1809.08014},
 primaryClass = {astro-ph.GA},
       adsurl = {https://ui.adsabs.harvard.edu/abs/2018A&A...620A..80M}
}

@ARTICLE{Barone2015,
       author = {{Barone}, V. and {Latouche}, C. and {Skouteris}, D. and {Vazart}, F. and {Balucani}, N. and {Ceccarelli}, C. and {Lefloch}, B.},
        title = "{Gas-phase formation of the prebiotic molecule formamide: insights from new quantum computations.}",
      journal = {\mnras},
         year = 2015,
        month = oct,
       volume = {453},
        pages = {L31-L35},
          doi = {10.1093/mnrasl/slv094},
archivePrefix = {arXiv},
       eprint = {1507.03741},
 primaryClass = {astro-ph.GA},
       adsurl = {https://ui.adsabs.harvard.edu/abs/2015MNRAS.453L..31B}
}

@ARTICLE{Skouteris2017,
       author = {{Skouteris}, D. and {Vazart}, F. and {Ceccarelli}, C. and {Balucani}, N. and {Puzzarini}, C. and {Barone}, V.},
        title = "{New quantum chemical computations of formamide deuteration support gas-phase formation of this prebiotic molecule}",
      journal = {\mnras},
         year = 2017,
        month = jun,
       volume = {468},
       number = {1},
        pages = {L1-L5},
          doi = {10.1093/mnrasl/slx012},
archivePrefix = {arXiv},
       eprint = {1701.06138},
 primaryClass = {astro-ph.SR},
       adsurl = {https://ui.adsabs.harvard.edu/abs/2017MNRAS.468L...1S}
}

@ARTICLE{kida2024,
       author = {{Wakelam}, V. and {Gratier}, P. and {Loison}, J. -C. and {Hickson}, K.~M. and {Penguen}, J. and {Mechineau}, A.},
        title = "{The 2024 KIDA network for interstellar chemistry}",
      journal = {\aap},
         year = 2024,
        month = sep,
       volume = {689},
          eid = {A63},
        pages = {A63},
          doi = {10.1051/0004-6361/202450606},
archivePrefix = {arXiv},
       eprint = {2407.15958},
 primaryClass = {astro-ph.GA},
       adsurl = {https://ui.adsabs.harvard.edu/abs/2024A&A...689A..63W}
}

@ARTICLE{Carney2019,
       author = {{Carney}, M.~T. and {Hogerheijde}, M.~R. and {Guzm{\'a}n}, V.~V. and {Walsh}, C. and {{\"O}berg}, K.~I. and {Fayolle}, E.~C. and {Cleeves}, L.~I. and {Carpenter}, J.~M. and {Qi}, C.},
        title = "{Upper limits on CH$_{3}$OH in the HD 163296 protoplanetary disk. Evidence for a low gas-phase CH$_{3}$OH-to-H$_{2}$CO ratio}",
      journal = {\aap},
         year = 2019,
        month = mar,
       volume = {623},
          eid = {A124},
        pages = {A124},
          doi = {10.1051/0004-6361/201834353},
archivePrefix = {arXiv},
       eprint = {1901.02689},
 primaryClass = {astro-ph.SR},
       adsurl = {https://ui.adsabs.harvard.edu/abs/2019A&A...623A.124C}
}

@ARTICLE{Muro2018,
       author = {{Muro-Arena}, G.~A. and {Dominik}, C. and {Waters}, L.~B.~F.~M. and {Min}, M. and {Klarmann}, L. and {Ginski}, C. and {Isella}, A. and {Benisty}, M. and {Pohl}, A. and {Garufi}, A. and {Hagelberg}, J. and {Langlois}, M. and {Menard}, F. and {Pinte}, C. and {Sezestre}, E. and {van der Plas}, G. and {Villenave}, M. and {Delboulb{\'e}}, A. and {Magnard}, Y. and {M{\"o}ller-Nilsson}, O. and {Pragt}, J. and {Rabou}, P. and {Roelfsema}, R.},
        title = "{Dust modeling of the combined ALMA and SPHERE datasets of HD 163296. Is HD 163296 really a Meeus group II disk?}",
      journal = {\aap},
         year = 2018,
        month = jun,
       volume = {614},
          eid = {A24},
        pages = {A24},
          doi = {10.1051/0004-6361/201732299},
archivePrefix = {arXiv},
       eprint = {1802.03328},
 primaryClass = {astro-ph.EP},
       adsurl = {https://ui.adsabs.harvard.edu/abs/2018A&A...614A..24M}
}

@ARTICLE{Nazari2022,
       author = {{Nazari}, P. and {Meijerhof}, J.~D. and {van Gelder}, M.~L. and {Ahmadi}, A. and {van Dishoeck}, E.~F. and {Tabone}, B. and {Langeroodi}, D. and {Ligterink}, N.~F.~W. and {Jaspers}, J. and {Beltr{\'a}n}, M.~T. and {Fuller}, G.~A. and {S{\'a}nchez-Monge}, {\'A}. and {Schilke}, P.},
        title = "{N-bearing complex organics toward high-mass protostars. Constant ratios pointing to formation in similar pre-stellar conditions across a large mass range}",
      journal = {\aap},
         year = 2022,
        month = dec,
       volume = {668},
          eid = {A109},
        pages = {A109},
          doi = {10.1051/0004-6361/202243788},
archivePrefix = {arXiv},
       eprint = {2208.11128},
 primaryClass = {astro-ph.GA},
       adsurl = {https://ui.adsabs.harvard.edu/abs/2022A&A...668A.109N}
}

@ARTICLE{Walsh2016,
       author = {{Walsh}, Catherine and {Loomis}, Ryan A. and {{\"O}berg}, Karin I. and {Kama}, Mihkel and {van 't Hoff}, Merel L.~R. and {Millar}, Tom J. and {Aikawa}, Yuri and {Herbst}, Eric and {Widicus Weaver}, Susanna L. and {Nomura}, Hideko},
        title = "{First Detection of Gas-phase Methanol in a Protoplanetary Disk}",
      journal = {\apjl},
         year = 2016,
        month = may,
       volume = {823},
       number = {1},
          eid = {L10},
        pages = {L10},
          doi = {10.3847/2041-8205/823/1/L10},
archivePrefix = {arXiv},
       eprint = {1606.06492},
 primaryClass = {astro-ph.EP},
       adsurl = {https://ui.adsabs.harvard.edu/abs/2016ApJ...823L..10W}
}

@ARTICLE{Goodman2010,
       author = {{Goodman}, Jonathan and {Weare}, Jonathan},
        title = "{Ensemble samplers with affine invariance}",
      journal = {Communications in Applied Mathematics and Computational Science},
         year = 2010,
        month = jan,
       volume = {5},
       number = {1},
        pages = {65-80},
          doi = {10.2140/camcos.2010.5.65},
       adsurl = {https://ui.adsabs.harvard.edu/abs/2010CAMCS...5...65G}
}

@ARTICLE{Pietu2007,
       author = {{Pi{\'e}tu}, V. and {Dutrey}, A. and {Guilloteau}, S.},
        title = "{Probing the structure of protoplanetary disks: a comparative study of DM Tau, LkCa 15, and MWC 480}",
      journal = {\aap},
         year = 2007,
        month = may,
       volume = {467},
       number = {1},
        pages = {163-178},
          doi = {10.1051/0004-6361:20066537},
archivePrefix = {arXiv},
       eprint = {astro-ph/0701425},
 primaryClass = {astro-ph},
       adsurl = {https://ui.adsabs.harvard.edu/abs/2007A&A...467..163P}
}

@misc{pdspy,
       author = {{Sheehan}, Patrick},
        title = "{pdspy: MCMC tool for continuum and spectral line radiative transfer modeling}",
 howpublished = {Astrophysics Source Code Library, record ascl:2207.026},
         year = 2022,
        month = jul,
          eid = {ascl:2207.026},
       adsurl = {https://ui.adsabs.harvard.edu/abs/2022ascl.soft07026S}
}

@ARTICLE{Moller2017,
       author = {{M{\"o}ller}, T. and {Endres}, C. and {Schilke}, P.},
        title = "{eXtended CASA Line Analysis Software Suite (XCLASS)}",
      journal = {\aap},
         year = 2017,
        month = feb,
       volume = {598},
          eid = {A7},
        pages = {A7},
          doi = {10.1051/0004-6361/201527203},
archivePrefix = {arXiv},
       eprint = {1508.04114},
 primaryClass = {astro-ph.IM},
       adsurl = {https://ui.adsabs.harvard.edu/abs/2017A&A...598A...7M}
}

@INPROCEEDINGS{Vastel2015,
       author = {{Vastel}, C. and {Bottinelli}, S. and {Caux}, E. and {Glorian}, J. -M. and {Boiziot}, M.},
        title = "{CASSIS: a tool to visualize and analyse instrumental and synthetic spectra.}",
    booktitle = {SF2A-2015: Proceedings of the Annual meeting of the French Society of Astronomy and Astrophysics},
         year = 2015,
       editor = {{Martins}, F. and {Boissier}, S. and {Buat}, V. and {Cambr{\'e}sy}, L. and {Petit}, P.},
        month = dec,
        pages = {313-316},
       adsurl = {https://ui.adsabs.harvard.edu/abs/2015sf2a.conf..313V}
}

@article{McGuire2022,
doi = {10.3847/1538-4365/ac2a48},
url = {https://dx.doi.org/10.3847/1538-4365/ac2a48},
year = {2022},
month = {mar},
publisher = {The American Astronomical Society},
volume = {259},
number = {2},
pages = {30},
author = {McGuire, Brett A.},
title = {2021 Census of Interstellar, Circumstellar, Extragalactic, Protoplanetary Disk, and Exoplanetary Molecules},
journal = {The Astrophysical Journal Supplement Series}
}

@ARTICLE{Lynden1974,
       author = {{Lynden-Bell}, D. and {Pringle}, J.~E.},
        title = "{The evolution of viscous discs and the origin of the nebular variables.}",
      journal = {\mnras},
         year = 1974,
        month = sep,
       volume = {168},
        pages = {603-637},
          doi = {10.1093/mnras/168.3.603},
       adsurl = {https://ui.adsabs.harvard.edu/abs/1974MNRAS.168..603L}
}

@article{Andrews2011,
doi = {10.1088/0004-637X/732/1/42},
url = {https://dx.doi.org/10.1088/0004-637X/732/1/42},
year = {2011},
month = {apr},
publisher = {The American Astronomical Society},
volume = {732},
number = {1},
pages = {42},
author = {Andrews, Sean M. and Wilner, David J. and Espaillat, Catherine and Hughes, A. M. and Dullemond, C. P. and McClure, M. K. and Qi, Chunhua and Brown, J. M.},
title = {RESOLVED IMAGES OF LARGE CAVITIES IN PROTOPLANETARY TRANSITION DISKS},
journal = {The Astrophysical Journal}
}

@ARTICLE{Zhang2021,
       author = {{Zhang}, Ke and {Booth}, Alice S. and {Law}, Charles J. and {Bosman}, Arthur D. and {Schwarz}, Kamber R. and {Bergin}, Edwin A. and {{\"O}berg}, Karin I. and {Andrews}, Sean M. and {Guzm{\'a}n}, Viviana V. and {Walsh}, Catherine and {Qi}, Chunhua and {van't Hoff}, Merel L.~R. and {Long}, Feng and {Wilner}, David J. and {Huang}, Jane and {Czekala}, Ian and {Ilee}, John D. and {Cataldi}, Gianni and {Bergner}, Jennifer B. and {Aikawa}, Yuri and {Teague}, Richard and {Bae}, Jaehan and {Loomis}, Ryan A. and {Calahan}, Jenny K. and {Alarc{\'o}n}, Felipe and {M{\'e}nard}, Fran{\c{c}}ois and {Le Gal}, Romane and {Sierra}, Anibal and {Yamato}, Yoshihide and {Nomura}, Hideko and {Tsukagoshi}, Takashi and {P{\'e}rez}, Laura M. and {Trapman}, Leon and {Liu}, Yao and {Furuya}, Kenji},
        title = "{Molecules with ALMA at Planet-forming Scales (MAPS). V. CO Gas Distributions}",
      journal = {\apjs},
         year = 2021,
        month = nov,
       volume = {257},
       number = {1},
          eid = {5},
        pages = {5},
          doi = {10.3847/1538-4365/ac1580},
archivePrefix = {arXiv},
       eprint = {2109.06233},
 primaryClass = {astro-ph.EP},
       adsurl = {https://ui.adsabs.harvard.edu/abs/2021ApJS..257....5Z}
}

@article{Nakagawa1986,
title = {Settling and growth of dust particles in a laminar phase of a low-mass solar nebula},
journal = {Icarus},
volume = {67},
number = {3},
pages = {375-390},
year = {1986},
issn = {0019-1035},
doi = {https://doi.org/10.1016/0019-1035(86)90121-1},
url = {https://www.sciencedirect.com/science/article/pii/0019103586901211},
author = {Yoshitsugu Nakagawa and Minoru Sekiya and Chushiro Hayashi}
}

@article{Dullemond2004,
	author = {{Dullemond, C. P.} and {Dominik, C.}},
	title = {The effect of dust settling on the appearance   of protoplanetary disks},
	DOI= "10.1051/0004-6361:20040284",
	url= "https://doi.org/10.1051/0004-6361:20040284",
	journal = {A\&A},
	year = 2004,
	volume = 421,
	number = 3,
	pages = "1075-1086",
}

@ARTICLE{Mathis1977,
       author = {{Mathis}, J.~S. and {Rumpl}, W. and {Nordsieck}, K.~H.},
        title = "{The size distribution of interstellar grains.}",
      journal = {\apj},
         year = 1977,
        month = oct,
       volume = {217},
        pages = {425-433},
          doi = {10.1086/155591},
       adsurl = {https://ui.adsabs.harvard.edu/abs/1977ApJ...217..425M}
}

@article{Hartmann2016,
   author = "Hartmann, Lee and Herczeg, Gregory and Calvet, Nuria",
   title = "Accretion onto Pre-Main-Sequence Stars", 
   journal= "Annual Review of Astronomy and Astrophysics",
   year = "2016",
   volume = "54",
   number = "Volume 54, 2016",
   pages = "135-180",
   doi = "https://doi.org/10.1146/annurev-astro-081915-023347",
   url = "https://www.annualreviews.org/content/journals/10.1146/annurev-astro-081915-023347",
   publisher = "Annual Reviews",
   issn = "1545-4282",
   type = "Journal Article"
  }

@misc{Dullemond2012,
       author = {{Dullemond}, C.~P. and {Juhasz}, A. and {Pohl}, A. and {Sereshti}, F. and {Shetty}, R. and {Peters}, T. and {Commercon}, B. and {Flock}, M.},
        title = "{RADMC-3D: A multi-purpose radiative transfer tool}",
 howpublished = {Astrophysics Source Code Library, record ascl:1202.015},
         year = 2012,
        month = feb,
          eid = {ascl:1202.015},
       adsurl = {https://ui.adsabs.harvard.edu/abs/2012ascl.soft02015D}
}

@article{Dartois2003,
	author = {{Dartois, E.} and {Dutrey, A.} and {Guilloteau, S.}},
	title = {Structure of the DM Tau Outer Disk: Probing the vertical
kinetic temperature gradient},
	DOI= "10.1051/0004-6361:20021638",
	url= "https://doi.org/10.1051/0004-6361:20021638",
	journal = {A\&A},
	year = 2003,
	volume = 399,
	number = 2,
	pages = "773-787",
}

@article{Dullemond2020,
	author = {{Dullemond, C. P.} and {Isella, A.} and {Andrews, S. M.} and {Skobleva, I.} and {Dzyurkevich, N.}},
	title = {Midplane temperature and outer edge of the protoplanetary disk around HD 163296},
	DOI= "10.1051/0004-6361/201936438",
	url= "https://doi.org/10.1051/0004-6361/201936438",
	journal = {A\&A},
	year = 2020,
	volume = 633,
	pages = "A137",
}

@ARTICLE{Law2021,
       author = {{Law}, Charles J. and {Teague}, Richard and {Loomis}, Ryan A. and {Bae}, Jaehan and {{\"O}berg}, Karin I. and {Czekala}, Ian and {Andrews}, Sean M. and {Aikawa}, Yuri and {Alarc{\'o}n}, Felipe and {Bergin}, Edwin A. and {Bergner}, Jennifer B. and {Booth}, Alice S. and {Bosman}, Arthur D. and {Calahan}, Jenny K. and {Cataldi}, Gianni and {Cleeves}, L. Ilsedore and {Furuya}, Kenji and {Guzm{\'a}n}, Viviana V. and {Huang}, Jane and {Ilee}, John D. and {Le Gal}, Romane and {Liu}, Yao and {Long}, Feng and {M{\'e}nard}, Fran{\c{c}}ois and {Nomura}, Hideko and {P{\'e}rez}, Laura M. and {Qi}, Chunhua and {Schwarz}, Kamber R. and {Soto}, Daniela and {Tsukagoshi}, Takashi and {Yamato}, Yoshihide and {van't Hoff}, Merel L.~R. and {Walsh}, Catherine and {Wilner}, David J. and {Zhang}, Ke},
        title = "{Molecules with ALMA at Planet-forming Scales (MAPS). IV. Emission Surfaces and Vertical Distribution of Molecules}",
      journal = {\apjs},
         year = 2021,
        month = nov,
       volume = {257},
       number = {1},
          eid = {4},
        pages = {4},
          doi = {10.3847/1538-4365/ac1439},
archivePrefix = {arXiv},
       eprint = {2109.06217},
 primaryClass = {astro-ph.GA},
       adsurl = {https://ui.adsabs.harvard.edu/abs/2021ApJS..257....4L}
}

@ARTICLE{Du2014,
       author = {{Du}, Fujun and {Bergin}, Edwin A.},
        title = "{Water Vapor Distribution in Protoplanetary Disks}",
      journal = {\apj},
         year = 2014,
        month = sep,
       volume = {792},
       number = {1},
          eid = {2},
        pages = {2},
          doi = {10.1088/0004-637X/792/1/2},
archivePrefix = {arXiv},
       eprint = {1408.2026},
 primaryClass = {astro-ph.SR},
       adsurl = {https://ui.adsabs.harvard.edu/abs/2014ApJ...792....2D}
}

@ARTICLE{Andersson2008,
       author = {{Andersson}, S. and {van Dishoeck}, E.~F.},
        title = "{Photodesorption of water ice. A molecular dynamics study}",
      journal = {\aap},
         year = 2008,
        month = dec,
       volume = {491},
       number = {3},
        pages = {907-916},
          doi = {10.1051/0004-6361:200810374},
archivePrefix = {arXiv},
       eprint = {0810.1916},
 primaryClass = {astro-ph},
       adsurl = {https://ui.adsabs.harvard.edu/abs/2008A&A...491..907A}
}

@ARTICLE{Oberg2009,
       author = {{{\"O}berg}, K.~I. and {Garrod}, R.~T. and {van Dishoeck}, E.~F. and {Linnartz}, H.},
        title = "{Formation rates of complex organics in UV irradiated CH\_3OH-rich ices. I. Experiments}",
      journal = {\aap},
         year = 2009,
        month = sep,
       volume = {504},
       number = {3},
        pages = {891-913},
          doi = {10.1051/0004-6361/200912559},
archivePrefix = {arXiv},
       eprint = {0908.1169},
 primaryClass = {astro-ph.GA},
       adsurl = {https://ui.adsabs.harvard.edu/abs/2009A&A...504..891O}
}

@ARTICLE{Kashyap2024,
       author = {{Kashyap}, Parashmoni and {Majumdar}, Liton and {Dutrey}, Anne and {Guilloteau}, St{\'e}phane and {Willacy}, Karen and {Chapillon}, Edwige and {Teague}, Richard and {Semenov}, Dmitry and {Henning}, Thomas and {Turner}, Neal and {Sahai}, Raghvendra and {K{\'o}sp{\'a}l}, {\'A}gnes and {Coutens}, Audrey and {Pi{\'e}tu}, V. and {Gratier}, Pierre and {Ruaud}, Maxime and {Phuong}, N.~T. and {Di Folco}, E. and {Lee}, Chin-Fei and {Tang}, Y. -W.},
        title = "{Chemistry in the GG Tau A Disk: Constraints from H$_{2}$D$^{+}$, N$_{2}$H$^{+}$, and DCO$^{+}$ High Angular Resolution ALMA Observations}",
      journal = {\apj},
         year = 2024,
        month = dec,
       volume = {976},
       number = {2},
          eid = {258},
        pages = {258},
          doi = {10.3847/1538-4357/ad815c},
archivePrefix = {arXiv},
       eprint = {2407.07238},
 primaryClass = {astro-ph.EP},
       adsurl = {https://ui.adsabs.harvard.edu/abs/2024ApJ...976..258K}
}

@ARTICLE{Padovani2018,
       author = {{Padovani}, Marco and {Ivlev}, Alexei V. and {Galli}, Daniele and {Caselli}, Paola},
        title = "{Cosmic-ray ionisation in circumstellar discs}",
      journal = {\aap},
         year = 2018,
        month = jun,
       volume = {614},
          eid = {A111},
        pages = {A111},
          doi = {10.1051/0004-6361/201732202},
archivePrefix = {arXiv},
       eprint = {1803.09348},
 primaryClass = {astro-ph.HE},
       adsurl = {https://ui.adsabs.harvard.edu/abs/2018A&A...614A.111P}
}

@ARTICLE{Evans2025,
       author = {{Evans}, Lucy and {Booth}, Alice S. and {Walsh}, Catherine and {Ilee}, John D. and {Keyte}, Luke and {Law}, Charles J. and {Leemker}, Margot and {Notsu}, Shota and {{\"O}berg}, Karin and {Temmink}, Milou and {van der Marel}, Nienke},
        title = "{ALMA reveals thermal and non-thermal desorption of methanol ice in the HD 100546 protoplanetary disk}",
      journal = {arXiv e-prints},
         year = 2025,
        month = feb,
          eid = {arXiv:2502.04957},
        pages = {arXiv:2502.04957},
          doi = {10.48550/arXiv.2502.04957},
archivePrefix = {arXiv},
       eprint = {2502.04957},
 primaryClass = {astro-ph.EP},
       adsurl = {https://ui.adsabs.harvard.edu/abs/2025arXiv250204957E}
}

@article{Watanabe2002,
doi = {10.1086/341412},
url = {https://dx.doi.org/10.1086/341412},
year = {2002},
month = {may},
publisher = {},
volume = {571},
number = {2},
pages = {L173},
author = {Watanabe, Naoki and Kouchi, Akira},
title = {Efficient Formation of Formaldehyde and Methanol by the Addition of Hydrogen Atoms to CO in H2O-CO Ice at 10 K},
journal = {The Astrophysical Journal}
}

@ARTICLE{Fuchs2009,
       author = {{Fuchs}, G.~W. and {Cuppen}, H.~M. and {Ioppolo}, S. and {Romanzin}, C. and {Bisschop}, S.~E. and {Andersson}, S. and {van Dishoeck}, E.~F. and {Linnartz}, H.},
        title = "{Hydrogenation reactions in interstellar CO ice analogues. A combined experimental/theoretical approach}",
      journal = {\aap},
         year = 2009,
        month = oct,
       volume = {505},
       number = {2},
        pages = {629-639},
          doi = {10.1051/0004-6361/200810784},
       adsurl = {https://ui.adsabs.harvard.edu/abs/2009A&A...505..629F}
}

@ARTICLE{Anilkumar2024,
       author = {{Anilkumar}, Hema and {Mathew}, Blesson and {Jithesh}, V. and {Kartha}, Sreeja S. and {Manoj}, P. and {Narang}, Mayank and {Chavali}, Mahathi},
        title = "{Chandra X-ray analysis of Herbig Ae/Be stars}",
      journal = {\mnras},
         year = 2024,
        month = may,
       volume = {530},
       number = {3},
        pages = {3020-3037},
          doi = {10.1093/mnras/stae938},
archivePrefix = {arXiv},
       eprint = {2404.03403},
 primaryClass = {astro-ph.SR},
       adsurl = {https://ui.adsabs.harvard.edu/abs/2024MNRAS.530.3020A}
}

@ARTICLE{Gorti2004,
       author = {{Gorti}, U. and {Hollenbach}, D.},
        title = "{Models of Chemistry, Thermal Balance, and Infrared Spectra from Intermediate-Aged Disks around G and K Stars}",
      journal = {\apj},
         year = 2004,
        month = sep,
       volume = {613},
       number = {1},
        pages = {424-447},
          doi = {10.1086/422406},
archivePrefix = {arXiv},
       eprint = {astro-ph/0405244},
 primaryClass = {astro-ph},
       adsurl = {https://ui.adsabs.harvard.edu/abs/2004ApJ...613..424G}
}

@ARTICLE{Qi2008,
       author = {{Qi}, Chunhua and {Wilner}, David J. and {Aikawa}, Yuri and {Blake}, Geoffrey A. and {Hogerheijde}, Michiel R.},
        title = "{Resolving the Chemistry in the Disk of TW Hydrae. I. Deuterated Species}",
      journal = {\apj},
         year = 2008,
        month = jul,
       volume = {681},
       number = {2},
        pages = {1396-1407},
          doi = {10.1086/588516},
archivePrefix = {arXiv},
       eprint = {0803.2753},
 primaryClass = {astro-ph},
       adsurl = {https://ui.adsabs.harvard.edu/abs/2008ApJ...681.1396Q}
}

@ARTICLE{Mathews2013,
       author = {{Mathews}, G.~S. and {Klaassen}, P.~D. and {Juh{\'a}sz}, A. and {Harsono}, D. and {Chapillon}, E. and {van Dishoeck}, E.~F. and {Espada}, D. and {de Gregorio-Monsalvo}, I. and {Hales}, A. and {Hogerheijde}, M.~R. and {Mottram}, J.~C. and {Rawlings}, M.~G. and {Takahashi}, S. and {Testi}, L.},
        title = "{ALMA imaging of the CO snowline of the HD 163296 disk with DCO$^{+}$}",
      journal = {\aap},
         year = 2013,
        month = sep,
       volume = {557},
          eid = {A132},
        pages = {A132},
          doi = {10.1051/0004-6361/201321600},
archivePrefix = {arXiv},
       eprint = {1307.3420},
 primaryClass = {astro-ph.SR},
       adsurl = {https://ui.adsabs.harvard.edu/abs/2013A&A...557A.132M}
}

@ARTICLE{Teague2015,
       author = {{Teague}, R. and {Semenov}, D. and {Guilloteau}, S. and {Henning}, Th. and {Dutrey}, A. and {Wakelam}, V. and {Chapillon}, E. and {Pietu}, V.},
        title = "{Chemistry in disks. IX. Observations and modelling of HCO$^{+}$ and DCO$^{+}$ in DM Tauri}",
      journal = {\aap},
         year = 2015,
        month = feb,
       volume = {574},
          eid = {A137},
        pages = {A137},
          doi = {10.1051/0004-6361/201425268},
archivePrefix = {arXiv},
       eprint = {1501.00984},
 primaryClass = {astro-ph.EP},
       adsurl = {https://ui.adsabs.harvard.edu/abs/2015A&A...574A.137T}
}

@ARTICLE{Huang2017,
       author = {{Huang}, Jane and {{\"O}berg}, Karin I. and {Qi}, Chunhua and {Aikawa}, Yuri and {Andrews}, Sean M. and {Furuya}, Kenji and {Guzm{\'a}n}, Viviana V. and {Loomis}, Ryan A. and {van Dishoeck}, Ewine F. and {Wilner}, David J.},
        title = "{An ALMA Survey of DCN/H$^{13}$CN and DCO$^{+}$/H$^{13}$CO$^{+}$ in Protoplanetary Disks}",
      journal = {\apj},
         year = 2017,
        month = feb,
       volume = {835},
       number = {2},
          eid = {231},
        pages = {231},
          doi = {10.3847/1538-4357/835/2/231},
archivePrefix = {arXiv},
       eprint = {1701.01735},
 primaryClass = {astro-ph.SR},
       adsurl = {https://ui.adsabs.harvard.edu/abs/2017ApJ...835..231H}
}

@ARTICLE{Oberg2015a,
       author = {{{\"O}berg}, Karin I. and {Furuya}, Kenji and {Loomis}, Ryan and {Aikawa}, Yuri and {Andrews}, Sean M. and {Qi}, Chunhua and {van Dishoeck}, Ewine F. and {Wilner}, David J.},
        title = "{Double DCO$^{+}$ Rings Reveal CO Ice Desorption in the Outer Disk Around IM Lup}",
      journal = {\apj},
         year = 2015,
        month = sep,
       volume = {810},
       number = {2},
          eid = {112},
        pages = {112},
          doi = {10.1088/0004-637X/810/2/112},
archivePrefix = {arXiv},
       eprint = {1508.07296},
 primaryClass = {astro-ph.GA},
       adsurl = {https://ui.adsabs.harvard.edu/abs/2015ApJ...810..112O}
}

@ARTICLE{Huang2018a,
       author = {{Huang}, Jane and {Andrews}, Sean M. and {Dullemond}, Cornelis P. and {Isella}, Andrea and {P{\'e}rez}, Laura M. and {Guzm{\'a}n}, Viviana V. and {{\"O}berg}, Karin I. and {Zhu}, Zhaohuan and {Zhang}, Shangjia and {Bai}, Xue-Ning and {Benisty}, Myriam and {Birnstiel}, Tilman and {Carpenter}, John M. and {Hughes}, A. Meredith and {Ricci}, Luca and {Weaver}, Erik and {Wilner}, David J.},
        title = "{The Disk Substructures at High Angular Resolution Project (DSHARP). II. Characteristics of Annular Substructures}",
      journal = {\apjl},
         year = 2018,
        month = dec,
       volume = {869},
       number = {2},
          eid = {L42},
        pages = {L42},
          doi = {10.3847/2041-8213/aaf740},
archivePrefix = {arXiv},
       eprint = {1812.04041},
 primaryClass = {astro-ph.EP},
       adsurl = {https://ui.adsabs.harvard.edu/abs/2018ApJ...869L..42H}
}

@ARTICLE{Aikawa2021,
       author = {{Aikawa}, Yuri and {Cataldi}, Gianni and {Yamato}, Yoshihide and {Zhang}, Ke and {Booth}, Alice S. and {Furuya}, Kenji and {Andrews}, Sean M. and {Bae}, Jaehan and {Bergin}, Edwin A. and {Bergner}, Jennifer B. and {Bosman}, Arthur D. and {Cleeves}, L. Ilsedore and {Czekala}, Ian and {Guzm{\'a}n}, Viviana V. and {Huang}, Jane and {Ilee}, John D. and {Law}, Charles J. and {Le Gal}, Romane and {Loomis}, Ryan A. and {M{\'e}nard}, Fran{\c{c}}ois and {Nomura}, Hideko and {{\"O}berg}, Karin I. and {Qi}, Chunhua and {Schwarz}, Kamber R. and {Teague}, Richard and {Tsukagoshi}, Takashi and {Walsh}, Catherine and {Wilner}, David J.},
        title = "{Molecules with ALMA at Planet-forming Scales (MAPS). XIII. HCO$^{+}$ and Disk Ionization Structure}",
      journal = {\apjs},
         year = 2021,
        month = nov,
       volume = {257},
       number = {1},
          eid = {13},
        pages = {13},
          doi = {10.3847/1538-4365/ac143c},
archivePrefix = {arXiv},
       eprint = {2109.06419},
 primaryClass = {astro-ph.SR},
       adsurl = {https://ui.adsabs.harvard.edu/abs/2021ApJS..257...13A}
}

@ARTICLE{Flahery2017,
       author = {{Flaherty}, Kevin M. and {Hughes}, A. Meredith and {Rose}, Sanaea C. and {Simon}, Jacob B. and {Qi}, Chunhua and {Andrews}, Sean M. and {K{\'o}sp{\'a}l}, {\'A}gnes and {Wilner}, David J. and {Chiang}, Eugene and {Armitage}, Philip J. and {Bai}, Xue-ning},
        title = "{A Three-dimensional View of Turbulence: Constraints on Turbulent Motions in the HD 163296 Protoplanetary Disk Using DCO$^{+}$}",
      journal = {\apj},
         year = 2017,
        month = jul,
       volume = {843},
       number = {2},
          eid = {150},
        pages = {150},
          doi = {10.3847/1538-4357/aa79f9},
archivePrefix = {arXiv},
       eprint = {1706.04504},
 primaryClass = {astro-ph.EP},
       adsurl = {https://ui.adsabs.harvard.edu/abs/2017ApJ...843..150F}
}

@ARTICLE{LeGal2021,
       author = {{Le Gal}, Romane and {{\"O}berg}, Karin I. and {Teague}, Richard and {Loomis}, Ryan A. and {Law}, Charles J. and {Walsh}, Catherine and {Bergin}, Edwin A. and {M{\'e}nard}, Fran{\c{c}}ois and {Wilner}, David J. and {Andrews}, Sean M. and {Aikawa}, Yuri and {Booth}, Alice S. and {Cataldi}, Gianni and {Bergner}, Jennifer B. and {Bosman}, Arthur D. and {Cleeves}, L. Ilse and {Czekala}, Ian and {Furuya}, Kenji and {Guzm{\'a}n}, Viviana V. and {Huang}, Jane and {Ilee}, John D. and {Nomura}, Hideko and {Qi}, Chunhua and {Schwarz}, Kamber R. and {Tsukagoshi}, Takashi and {Yamato}, Yoshihide and {Zhang}, Ke},
        title = "{Molecules with ALMA at Planet-forming Scales (MAPS). XII. Inferring the C/O and S/H Ratios in Protoplanetary Disks with Sulfur Molecules}",
      journal = {\apjs},
         year = 2021,
        month = nov,
       volume = {257},
       number = {1},
          eid = {12},
        pages = {12},
          doi = {10.3847/1538-4365/ac2583},
archivePrefix = {arXiv},
       eprint = {2109.06286},
 primaryClass = {astro-ph.GA},
       adsurl = {https://ui.adsabs.harvard.edu/abs/2021ApJS..257...12L}
}

@ARTICLE{Ma2024,
       author = {{Ma}, Rong and {Quan}, Donghui and {Zhou}, Yan and {Esimbek}, Jarken and {Li}, Dalei and {Li}, Xiaohu and {Zhang}, Xia and {Tuo}, Juan and {Feng}, Yanan},
        title = "{Investigating Sulfur Chemistry in the HD163296 Disk}",
      journal = {Research in Astronomy and Astrophysics},
         year = 2024,
        month = jul,
       volume = {24},
       number = {7},
          eid = {075017},
        pages = {075017},
          doi = {10.1088/1674-4527/ad5771},
archivePrefix = {arXiv},
       eprint = {2406.07896},
 primaryClass = {astro-ph.EP},
       adsurl = {https://ui.adsabs.harvard.edu/abs/2024RAA....24g5017M}
}

@ARTICLE{Ilee2021,
       author = {{Ilee}, John D. and {Walsh}, Catherine and {Booth}, Alice S. and {Aikawa}, Yuri and {Andrews}, Sean M. and {Bae}, Jaehan and {Bergin}, Edwin A. and {Bergner}, Jennifer B. and {Bosman}, Arthur D. and {Cataldi}, Gianni and {Cleeves}, L. Ilsedore and {Czekala}, Ian and {Guzm{\'a}n}, Viviana V. and {Huang}, Jane and {Law}, Charles J. and {Le Gal}, Romane and {Loomis}, Ryan A. and {M{\'e}nard}, Fran{\c{c}}ois and {Nomura}, Hideko and {{\"O}berg}, Karin I. and {Qi}, Chunhua and {Schwarz}, Kamber R. and {Teague}, Richard and {Tsukagoshi}, Takashi and {Wilner}, David J. and {Yamato}, Yoshihide and {Zhang}, Ke},
        title = "{Molecules with ALMA at Planet-forming Scales (MAPS). IX. Distribution and Properties of the Large Organic Molecules HC$_{3}$N, CH$_{3}$CN, and c-C$_{3}$H$_{2}$}",
      journal = {\apjs},
         year = 2021,
        month = nov,
       volume = {257},
       number = {1},
          eid = {9},
        pages = {9},
          doi = {10.3847/1538-4365/ac1441},
archivePrefix = {arXiv},
       eprint = {2109.06319},
 primaryClass = {astro-ph.EP},
       adsurl = {https://ui.adsabs.harvard.edu/abs/2021ApJS..257....9I}
}

@ARTICLE{Bergner2018,
       author = {{Bergner}, Jennifer B. and {Guzm{\'a}n}, Viviana G. and {{\"O}berg}, Karin I. and {Loomis}, Ryan A. and {Pegues}, Jamila},
        title = "{A Survey of CH$_{3}$CN and HC$_{3}$N in Protoplanetary Disks}",
      journal = {\apj},
         year = 2018,
        month = apr,
       volume = {857},
       number = {1},
          eid = {69},
        pages = {69},
          doi = {10.3847/1538-4357/aab664},
archivePrefix = {arXiv},
       eprint = {1803.04986},
 primaryClass = {astro-ph.EP},
       adsurl = {https://ui.adsabs.harvard.edu/abs/2018ApJ...857...69B}
}

@ARTICLE{Cleeves2013,
       author = {{Cleeves}, L. Ilsedore and {Adams}, Fred C. and {Bergin}, Edwin A. and {Visser}, Ruud},
        title = "{Radionuclide Ionization in Protoplanetary Disks: Calculations of Decay Product Radiative Transfer}",
      journal = {\apj},
         year = 2013,
        month = nov,
       volume = {777},
       number = {1},
          eid = {28},
        pages = {28},
          doi = {10.1088/0004-637X/777/1/28},
archivePrefix = {arXiv},
       eprint = {1309.0018},
 primaryClass = {astro-ph.SR},
       adsurl = {https://ui.adsabs.harvard.edu/abs/2013ApJ...777...28C}
}

@ARTICLE{Bosman2021,
       author = {{Bosman}, Arthur D. and {Alarc{\'o}n}, Felipe and {Bergin}, Edwin A. and {Zhang}, Ke and {van't Hoff}, Merel L.~R. and {{\"O}berg}, Karin I. and {Guzm{\'a}n}, Viviana V. and {Walsh}, Catherine and {Aikawa}, Yuri and {Andrews}, Sean M. and {Bergner}, Jennifer B. and {Booth}, Alice S. and {Cataldi}, Gianni and {Cleeves}, L. Ilsedore and {Czekala}, Ian and {Furuya}, Kenji and {Huang}, Jane and {Ilee}, John D. and {Law}, Charles J. and {Le Gal}, Romane and {Liu}, Yao and {Long}, Feng and {Loomis}, Ryan A. and {M{\'e}nard}, Fran{\c{c}}ois and {Nomura}, Hideko and {Qi}, Chunhua and {Schwarz}, Kamber R. and {Teague}, Richard and {Tsukagoshi}, Takashi and {Yamato}, Yoshihide and {Wilner}, David J.},
        title = "{Molecules with ALMA at Planet-forming Scales (MAPS). VII. Substellar O/H and C/H and Superstellar C/O in Planet-feeding Gas}",
      journal = {\apjs},
         year = 2021,
        month = nov,
       volume = {257},
       number = {1},
          eid = {7},
        pages = {7},
          doi = {10.3847/1538-4365/ac1435},
archivePrefix = {arXiv},
       eprint = {2109.06221},
 primaryClass = {astro-ph.EP},
       adsurl = {https://ui.adsabs.harvard.edu/abs/2021ApJS..257....7B}
}

@ARTICLE{Snyder1972,
       author = {{Snyder}, Lewis E. and {Buhl}, David},
        title = "{Interstellar Isocyanic Acid}",
      journal = {\apj},
         year = 1972,
        month = nov,
       volume = {177},
        pages = {619},
          doi = {10.1086/151739},
       adsurl = {https://ui.adsabs.harvard.edu/abs/1972ApJ...177..619S}
}

@ARTICLE{Turner1991,
       author = {{Turner}, B.~E.},
        title = "{A Molecular Line Survey of Sagittarius B2 and Orion--KL from 70 to 115 GHz. II. Analysis of the Data}",
      journal = {\apjs},
         year = 1991,
        month = jun,
       volume = {76},
        pages = {617},
          doi = {10.1086/191577},
       adsurl = {https://ui.adsabs.harvard.edu/abs/1991ApJS...76..617T}
}

@article{Martín2008,
doi = {10.1086/533409},
url = {https://dx.doi.org/10.1086/533409},
year = {2008},
month = {may},
publisher = {},
volume = {678},
number = {1},
pages = {245},
author = {Martín, Sergio and Requena-Torres, M. A. and Martín-Pintado, J. and Mauersberger, R.},
title = {Tracing Shocks and Photodissociation in the Galactic Center Region*},
journal = {The Astrophysical Journal}
}

@ARTICLE{MacDonald1996,
       author = {{MacDonald}, G.~H. and {Gibb}, A.~G. and {Habing}, R.~J. and {Millar}, T.~J.},
        title = "{A 330-360 GHz spectral survey of G 34.3+0.15. I. Data and physical analysis.}",
      journal = {\aaps},
         year = 1996,
        month = oct,
       volume = {119},
        pages = {333-367},
       adsurl = {https://ui.adsabs.harvard.edu/abs/1996A&AS..119..333M}
}

@ARTICLE{Helmich1997,
       author = {{Helmich}, F.~P. and {van Dishoeck}, E.~F.},
        title = "{Physical and chemical variations within the W3 star-forming region. II. The 345 GHz spectral line survey.}",
      journal = {\aaps},
         year = 1997,
        month = aug,
       volume = {124},
        pages = {205-253},
          doi = {10.1051/aas:1997357},
       adsurl = {https://ui.adsabs.harvard.edu/abs/1997A&AS..124..205H}
}

@ARTICLE{Bisschop2008,
       author = {{Bisschop}, S.~E. and {J{\o}rgensen}, J.~K. and {Bourke}, T.~L. and {Bottinelli}, S. and {van Dishoeck}, E.~F.},
        title = "{An interferometric study of the low-mass protostar IRAS 16293-2422: small scale organic chemistry}",
      journal = {\aap},
         year = 2008,
        month = sep,
       volume = {488},
       number = {3},
        pages = {959-968},
          doi = {10.1051/0004-6361:200809673},
archivePrefix = {arXiv},
       eprint = {0807.1447},
 primaryClass = {astro-ph},
       adsurl = {https://ui.adsabs.harvard.edu/abs/2008A&A...488..959B}
}

@ARTICLE{Rodriguez2010,
       author = {{Rodr{\'\i}guez-Fern{\'a}ndez}, N.~J. and {Tafalla}, M. and {Gueth}, F. and {Bachiller}, R.},
        title = "{HNCO enhancement by shocks in the L1157 molecular outflow}",
      journal = {\aap},
         year = 2010,
        month = jun,
       volume = {516},
          eid = {A98},
        pages = {A98},
          doi = {10.1051/0004-6361/201013997},
archivePrefix = {arXiv},
       eprint = {1003.4219},
 primaryClass = {astro-ph.GA},
       adsurl = {https://ui.adsabs.harvard.edu/abs/2010A&A...516A..98R}
}

@ARTICLE{Lopez2015,
       author = {{L{\'o}pez-Sepulcre}, A. and {Jaber}, Ali A. and {Mendoza}, E. and {Lefloch}, B. and {Ceccarelli}, C. and {Vastel}, C. and {Bachiller}, R. and {Cernicharo}, J. and {Codella}, C. and {Kahane}, C. and {Kama}, M. and {Tafalla}, M.},
        title = "{Shedding light on the formation of the pre-biotic molecule formamide with ASAI}",
      journal = {\mnras},
         year = 2015,
        month = may,
       volume = {449},
       number = {3},
        pages = {2438-2458},
          doi = {10.1093/mnras/stv377},
archivePrefix = {arXiv},
       eprint = {1502.05762},
 primaryClass = {astro-ph.SR},
       adsurl = {https://ui.adsabs.harvard.edu/abs/2015MNRAS.449.2438L}
}

@ARTICLE{Biver2006,
       author = {{Biver}, N. and {Bockel{\'e}e-Morvan}, D. and {Crovisier}, J. and {Lis}, D.~C. and {Moreno}, R. and {Colom}, P. and {Henry}, F. and {Herpin}, F. and {Paubert}, G. and {Womack}, M.},
        title = "{Radio wavelength molecular observations of comets C/1999 T1 (McNaught-Hartley), C/2001 A2 (LINEAR), C/2000 WM$_{1}$ (LINEAR) and 153P/Ikeya-Zhang}",
      journal = {\aap},
         year = 2006,
        month = apr,
       volume = {449},
       number = {3},
        pages = {1255-1270},
          doi = {10.1051/0004-6361:20053849},
       adsurl = {https://ui.adsabs.harvard.edu/abs/2006A&A...449.1255B}
}

@ARTICLE{Oberg2021review,
       author = {{{\"O}berg}, Karin I. and {Bergin}, Edwin A.},
        title = "{Astrochemistry and compositions of planetary systems}",
      journal = {\physrep},
         year = 2021,
        month = jan,
       volume = {893},
        pages = {1-48},
          doi = {10.1016/j.physrep.2020.09.004},
archivePrefix = {arXiv},
       eprint = {2010.03529},
 primaryClass = {astro-ph.EP},
       adsurl = {https://ui.adsabs.harvard.edu/abs/2021PhR...893....1O}
}

@article{Gaia2018,
	author = {{Gaia Collaboration} and {Brown, A. G. A.} and {Vallenari, A.} and {Prusti, T.} and {de Bruijne, J. H. J.} and {Babusiaux, C.} and {Bailer-Jones, C. A. L.} and {Biermann, M.} and {Evans, D. W.} and {Eyer, L.} and {Jansen, F.} and {Jordi, C.} and {Klioner, S. A.} and {Lammers, U.} and {Lindegren, L.} and {Luri, X.} and {Mignard, F.} and {Panem, C.} and {Pourbaix, D.} and {Randich, S.} and {Sartoretti, P.} and {Siddiqui, H. I.} and {Soubiran, C.} and {van Leeuwen, F.} and {Walton, N. A.} and {Arenou, F.} and {Bastian, U.} and {Cropper, M.} and {Drimmel, R.} and {Katz, D.} and {Lattanzi, M. G.} and {Bakker, J.} and {Cacciari, C.} and {Castañeda, J.} and {Chaoul, L.} and {Cheek, N.} and {De Angeli, F.} and {Fabricius, C.} and {Guerra, R.} and {Holl, B.} and {Masana, E.} and {Messineo, R.} and {Mowlavi, N.} and {Nienartowicz, K.} and {Panuzzo, P.} and {Portell, J.} and {Riello, M.} and {Seabroke, G. M.} and {Tanga, P.} and {Thévenin, F.} and {Gracia-Abril, G.} and {Comoretto, G.} and {Garcia-Reinaldos, M.} and {Teyssier, D.} and {Altmann, M.} and {Andrae, R.} and {Audard, M.} and {Bellas-Velidis, I.} and {Benson, K.} and {Berthier, J.} and {Blomme, R.} and {Burgess, P.} and {Busso, G.} and {Carry, B.} and {Cellino, A.} and {Clementini, G.} and {Clotet, M.} and {Creevey, O.} and {Davidson, M.} and {De Ridder, J.} and {Delchambre, L.} and {Dell’Oro, A.} and {Ducourant, C.} and {Fernández-Hernández, J.} and {Fouesneau, M.} and {Frémat, Y.} and {Galluccio, L.} and {García-Torres, M.} and {González-Núñez, J.} and {González-Vidal, J. J.} and {Gosset, E.} and {Guy, L. P.} and {Halbwachs, J.-L.} and {Hambly, N. C.} and {Harrison, D. L.} and {Hernández, J.} and {Hestroffer, D.} and {Hodgkin, S. T.} and {Hutton, A.} and {Jasniewicz, G.} and {Jean-Antoine-Piccolo, A.} and {Jordan, S.} and {Korn, A. J.} and {Krone-Martins, A.} and {Lanzafame, A. C.} and {Lebzelter, T.} and {Löffler, W.} and {Manteiga, M.} and {Marrese, P. M.} and {Martín-Fleitas, J. M.} and {Moitinho, A.} and {Mora, A.} and {Muinonen, K.} and {Osinde, J.} and {Pancino, E.} and {Pauwels, T.} and {Petit, J.-M.} and {Recio-Blanco, A.} and {Richards, P. J.} and {Rimoldini, L.} and {Robin, A. C.} and {Sarro, L. M.} and {Siopis, C.} and {Smith, M.} and {Sozzetti, A.} and {Süveges, M.} and {Torra, J.} and {van Reeven, W.} and {Abbas, U.} and {Abreu Aramburu, A.} and {Accart, S.} and {Aerts, C.} and {Altavilla, G.} and {Álvarez, M. A.} and {Alvarez, R.} and {Alves, J.} and {Anderson, R. I.} and {Andrei, A. H.} and {Anglada Varela, E.} and {Antiche, E.} and {Antoja, T.} and {Arcay, B.} and {Astraatmadja, T. L.} and {Bach, N.} and {Baker, S. G.} and {Balaguer-Núñez, L.} and {Balm, P.} and {Barache, C.} and {Barata, C.} and {Barbato, D.} and {Barblan, F.} and {Barklem, P. S.} and {Barrado, D.} and {Barros, M.} and {Barstow, M. A.} and {Bartholomé Muñoz, S.} and {Bassilana, J.-L.} and {Becciani, U.} and {Bellazzini, M.} and {Berihuete, A.} and {Bertone, S.} and {Bianchi, L.} and {Bienaymé, O.} and {Blanco-Cuaresma, S.} and {Boch, T.} and {Boeche, C.} and {Bombrun, A.} and {Borrachero, R.} and {Bossini, D.} and {Bouquillon, S.} and {Bourda, G.} and {Bragaglia, A.} and {Bramante, L.} and {Breddels, M. A.} and {Bressan, A.} and {Brouillet, N.} and {Brüsemeister, T.} and {Brugaletta, E.} and {Bucciarelli, B.} and {Burlacu, A.} and {Busonero, D.} and {Butkevich, A. G.} and {Buzzi, R.} and {Caffau, E.} and {Cancelliere, R.} and {Cannizzaro, G.} and {Cantat-Gaudin, T.} and {Carballo, R.} and {Carlucci, T.} and {Carrasco, J. M.} and {Casamiquela, L.} and {Castellani, M.} and {Castro-Ginard, A.} and {Charlot, P.} and {Chemin, L.} and {Chiavassa, A.} and {Cocozza, G.} and {Costigan, G.} and {Cowell, S.} and {Crifo, F.} and {Crosta, M.} and {Crowley, C.} and {Cuypers†, J.} and {Dafonte, C.} and {Damerdji, Y.} and {Dapergolas, A.} and {David, P.} and {David, M.} and {de Laverny, P.} and {De Luise, F.} and {De March, R.} and {de Martino, D.} and {de Souza, R.} and {de Torres, A.} and {Debosscher, J.} and {del Pozo, E.} and {Delbo, M.} and {Delgado, A.} and {Delgado, H. E.} and {Di Matteo, P.} and {Diakite, S.} and {Diener, C.} and {Distefano, E.} and {Dolding, C.} and {Drazinos, P.} and {Durán, J.} and {Edvardsson, B.} and {Enke, H.} and {Eriksson, K.} and {Esquej, P.} and {Eynard Bontemps, G.} and {Fabre, C.} and {Fabrizio, M.} and {Faigler, S.} and {Falcão, A. J.} and {Farràs Casas, M.} and {Federici, L.} and {Fedorets, G.} and {Fernique, P.} and {Figueras, F.} and {Filippi, F.} and {Findeisen, K.} and {Fonti, A.} and {Fraile, E.} and {Fraser, M.} and {Frézouls, B.} and {Gai, M.} and {Galleti, S.} and {Garabato, D.} and {García-Sedano, F.} and {Garofalo, A.} and {Garralda, N.} and {Gavel, A.} and {Gavras, P.} and {Gerssen, J.} and {Geyer, R.} and {Giacobbe, P.} and {Gilmore, G.} and {Girona, S.} and {Giuffrida, G.} and {Glass, F.} and {Gomes, M.} and {Granvik, M.} and {Gueguen, A.} and {Guerrier, A.} and {Guiraud, J.} and {Gutiérrez-Sánchez, R.} and {Haigron, R.} and {Hatzidimitriou, D.} and {Hauser, M.} and {Haywood, M.} and {Heiter, U.} and {Helmi, A.} and {Heu, J.} and {Hilger, T.} and {Hobbs, D.} and {Hofmann, W.} and {Holland, G.} and {Huckle, H. E.} and {Hypki, A.} and {Icardi, V.} and {Janßen, K.} and {Jevardat de Fombelle, G.} and {Jonker, P. G.} and {Juhász, Á. L.} and {Julbe, F.} and {Karampelas, A.} and {Kewley, A.} and {Klar, J.} and {Kochoska, A.} and {Kohley, R.} and {Kolenberg, K.} and {Kontizas, M.} and {Kontizas, E.} and {Koposov, S. E.} and {Kordopatis, G.} and {Kostrzewa-Rutkowska, Z.} and {Koubsky, P.} and {Lambert, S.} and {Lanza, A. F.} and {Lasne, Y.} and {Lavigne, J.-B.} and {Le Fustec, Y.} and {Le Poncin-Lafitte, C.} and {Lebreton, Y.} and {Leccia, S.} and {Leclerc, N.} and {Lecoeur-Taibi, I.} and {Lenhardt, H.} and {Leroux, F.} and {Liao, S.} and {Licata, E.} and {Lindstrøm, H. E. P.} and {Lister, T. A.} and {Livanou, E.} and {Lobel, A.} and {López, M.} and {Managau, S.} and {Mann, R. G.} and {Mantelet, G.} and {Marchal, O.} and {Marchant, J. M.} and {Marconi, M.} and {Marinoni, S.} and {Marschalkó, G.} and {Marshall, D. J.} and {Martino, M.} and {Marton, G.} and {Mary, N.} and {Massari, D.} and {Matijevič, G.} and {Mazeh, T.} and {McMillan, P. J.} and {Messina, S.} and {Michalik, D.} and {Millar, N. R.} and {Molina, D.} and {Molinaro, R.} and {Molnár, L.} and {Montegriffo, P.} and {Mor, R.} and {Morbidelli, R.} and {Morel, T.} and {Morris, D.} and {Mulone, A. F.} and {Muraveva, T.} and {Musella, I.} and {Nelemans, G.} and {Nicastro, L.} and {Noval, L.} and {O’Mullane, W.} and {Ordénovic, C.} and {Ordóñez-Blanco, D.} and {Osborne, P.} and {Pagani, C.} and {Pagano, I.} and {Pailler, F.} and {Palacin, H.} and {Palaversa, L.} and {Panahi, A.} and {Pawlak, M.} and {Piersimoni, A. M.} and {Pineau, F.-X.} and {Plachy, E.} and {Plum, G.} and {Poggio, E.} and {Poujoulet, E.} and {Prša, A.} and {Pulone, L.} and {Racero, E.} and {Ragaini, S.} and {Rambaux, N.} and {Ramos-Lerate, M.} and {Regibo, S.} and {Reylé, C.} and {Riclet, F.} and {Ripepi, V.} and {Riva, A.} and {Rivard, A.} and {Rixon, G.} and {Roegiers, T.} and {Roelens, M.} and {Romero-Gómez, M.} and {Rowell, N.} and {Royer, F.} and {Ruiz-Dern, L.} and {Sadowski, G.} and {Sagristà Sellés, T.} and {Sahlmann, J.} and {Salgado, J.} and {Salguero, E.} and {Sanna, N.} and {Santana-Ros, T.} and {Sarasso, M.} and {Savietto, H.} and {Schultheis, M.} and {Sciacca, E.} and {Segol, M.} and {Segovia, J. C.} and {Ségransan, D.} and {Shih, I-C.} and {Siltala, L.} and {Silva, A. F.} and {Smart, R. L.} and {Smith, K. W.} and {Solano, E.} and {Solitro, F.} and {Sordo, R.} and {Soria Nieto, S.} and {Souchay, J.} and {Spagna, A.} and {Spoto, F.} and {Stampa, U.} and {Steele, I. A.} and {Steidelmüller, H.} and {Stephenson, C. A.} and {Stoev, H.} and {Suess, F. F.} and {Surdej, J.} and {Szabados, L.} and {Szegedi-Elek, E.} and {Tapiador, D.} and {Taris, F.} and {Tauran, G.} and {Taylor, M. B.} and {Teixeira, R.} and {Terrett, D.} and {Teyssandier, P.} and {Thuillot, W.} and {Titarenko, A.} and {Torra Clotet, F.} and {Turon, C.} and {Ulla, A.} and {Utrilla, E.} and {Uzzi, S.} and {Vaillant, M.} and {Valentini, G.} and {Valette, V.} and {van Elteren, A.} and {Van Hemelryck, E.} and {van Leeuwen, M.} and {Vaschetto, M.} and {Vecchiato, A.} and {Veljanoski, J.} and {Viala, Y.} and {Vicente, D.} and {Vogt, S.} and {von Essen, C.} and {Voss, H.} and {Votruba, V.} and {Voutsinas, S.} and {Walmsley, G.} and {Weiler, M.} and {Wertz, O.} and {Wevers, T.} and {Wyrzykowski, Ł.} and {Yoldas, A.} and {Žerjal, M.} and {Ziaeepour, H.} and {Zorec, J.} and {Zschocke, S.} and {Zucker, S.} and {Zurbach, C.} and {Zwitter, T.}},
	title = {Gaia Data Release 2 - Summary of the contents and survey properties},
	DOI= "10.1051/0004-6361/201833051",
	url= "https://doi.org/10.1051/0004-6361/201833051",
	journal = {A\&A},
	year = 2018,
	volume = 616,
	pages = "A1",
}

@ARTICLE{Isella2007,
       author = {{Isella}, A. and {Testi}, L. and {Natta}, A. and {Neri}, R. and {Wilner}, D. and {Qi}, C.},
        title = "{Millimeter imaging of HD 163296: probing the disk structure and kinematics}",
      journal = {\aap},
         year = 2007,
        month = jul,
       volume = {469},
       number = {1},
        pages = {213-222},
          doi = {10.1051/0004-6361:20077385},
archivePrefix = {arXiv},
       eprint = {0704.0616},
 primaryClass = {astro-ph},
       adsurl = {https://ui.adsabs.harvard.edu/abs/2007A&A...469..213I}
}

@ARTICLE{Tilling2012,
       author = {{Tilling}, I. and {Woitke}, P. and {Meeus}, G. and {Mora}, A. and {Montesinos}, B. and {Riviere-Marichalar}, P. and {Eiroa}, C. and {Thi}, W. -F. and {Isella}, A. and {Roberge}, A. and {Martin-Zaidi}, C. and {Kamp}, I. and {Pinte}, C. and {Sandell}, G. and {Vacca}, W.~D. and {M{\'e}nard}, F. and {Mendigut{\'\i}a}, I. and {Duch{\^e}ne}, G. and {Dent}, W.~R.~F. and {Aresu}, G. and {Meijerink}, R. and {Spaans}, M.},
        title = "{Gas modelling in the disc of HD 163296}",
      journal = {\aap},
         year = 2012,
        month = feb,
       volume = {538},
          eid = {A20},
        pages = {A20},
          doi = {10.1051/0004-6361/201116919},
archivePrefix = {arXiv},
       eprint = {1111.2549},
 primaryClass = {astro-ph.SR},
       adsurl = {https://ui.adsabs.harvard.edu/abs/2012A&A...538A..20T}
}

@ARTICLE{Powell2019,
       author = {{Powell}, Diana and {Murray-Clay}, Ruth and {P{\'e}rez}, Laura M. and {Schlichting}, Hilke E. and {Rosenthal}, Mickey},
        title = "{New Constraints From Dust Lines on the Surface Densities of Protoplanetary Disks}",
      journal = {\apj},
         year = 2019,
        month = jun,
       volume = {878},
       number = {2},
          eid = {116},
        pages = {116},
          doi = {10.3847/1538-4357/ab20ce},
archivePrefix = {arXiv},
       eprint = {1905.03252},
 primaryClass = {astro-ph.EP},
       adsurl = {https://ui.adsabs.harvard.edu/abs/2019ApJ...878..116P}
}

@ARTICLE{Kama2020,
       author = {{Kama}, M. and {Trapman}, L. and {Fedele}, D. and {Bruderer}, S. and {Hogerheijde}, M.~R. and {Miotello}, A. and {van Dishoeck}, E.~F. and {Clarke}, C. and {Bergin}, E.~A.},
        title = "{Mass constraints for 15 protoplanetary discs from HD 1-0}",
      journal = {\aap},
         year = 2020,
        month = feb,
       volume = {634},
          eid = {A88},
        pages = {A88},
          doi = {10.1051/0004-6361/201937124},
archivePrefix = {arXiv},
       eprint = {1912.11883},
 primaryClass = {astro-ph.EP},
       adsurl = {https://ui.adsabs.harvard.edu/abs/2020A&A...634A..88K}
}

@ARTICLE{Isella2016,
       author = {{Isella}, Andrea and {Guidi}, Greta and {Testi}, Leonardo and {Liu}, Shangfei and {Li}, Hui and {Li}, Shengtai and {Weaver}, Erik and {Boehler}, Yann and {Carperter}, John M. and {De Gregorio-Monsalvo}, Itziar and {Manara}, Carlo F. and {Natta}, Antonella and {P{\'e}rez}, Laura M. and {Ricci}, Luca and {Sargent}, Anneila and {Tazzari}, Marco and {Turner}, Neal},
        title = "{Ringed Structures of the HD 163296 Protoplanetary Disk Revealed by ALMA}",
      journal = {\prl},
         year = 2016,
        month = dec,
       volume = {117},
       number = {25},
          eid = {251101},
        pages = {251101},
          doi = {10.1103/PhysRevLett.117.251101},
       adsurl = {https://ui.adsabs.harvard.edu/abs/2016PhRvL.117y1101I}
}

@ARTICLE{Isella2018,
       author = {{Isella}, Andrea and {Huang}, Jane and {Andrews}, Sean M. and {Dullemond}, Cornelis P. and {Birnstiel}, Tilman and {Zhang}, Shangjia and {Zhu}, Zhaohuan and {Guzm{\'a}n}, Viviana V. and {P{\'e}rez}, Laura M. and {Bai}, Xue-Ning and {Benisty}, Myriam and {Carpenter}, John M. and {Ricci}, Luca and {Wilner}, David J.},
        title = "{The Disk Substructures at High Angular Resolution Project (DSHARP). IX. A High-definition Study of the HD 163296 Planet-forming Disk}",
      journal = {\apjl},
         year = 2018,
        month = dec,
       volume = {869},
       number = {2},
          eid = {L49},
        pages = {L49},
          doi = {10.3847/2041-8213/aaf747},
archivePrefix = {arXiv},
       eprint = {1812.04047},
 primaryClass = {astro-ph.SR},
       adsurl = {https://ui.adsabs.harvard.edu/abs/2018ApJ...869L..49I}
}

@ARTICLE{Pinte2018,
       author = {{Pinte}, C. and {Price}, D.~J. and {M{\'e}nard}, F. and {Duch{\^e}ne}, G. and {Dent}, W.~R.~F. and {Hill}, T. and {de Gregorio-Monsalvo}, I. and {Hales}, A. and {Mentiplay}, D.},
        title = "{Kinematic Evidence for an Embedded Protoplanet in a Circumstellar Disk}",
      journal = {\apjl},
         year = 2018,
        month = jun,
       volume = {860},
       number = {1},
          eid = {L13},
        pages = {L13},
          doi = {10.3847/2041-8213/aac6dc},
archivePrefix = {arXiv},
       eprint = {1805.10293},
 primaryClass = {astro-ph.SR},
       adsurl = {https://ui.adsabs.harvard.edu/abs/2018ApJ...860L..13P}
}

@ARTICLE{Pinte2020,
       author = {{Pinte}, C. and {Price}, D.~J. and {M{\'e}nard}, F. and {Duch{\^e}ne}, G. and {Christiaens}, V. and {Andrews}, S.~M. and {Huang}, J. and {Hill}, T. and {van der Plas}, G. and {Perez}, L.~M. and {Isella}, A. and {Boehler}, Y. and {Dent}, W.~R.~F. and {Mentiplay}, D. and {Loomis}, R.~A.},
        title = "{Nine Localized Deviations from Keplerian Rotation in the DSHARP Circumstellar Disks: Kinematic Evidence for Protoplanets Carving the Gaps}",
      journal = {\apjl},
         year = 2020,
        month = feb,
       volume = {890},
       number = {1},
          eid = {L9},
        pages = {L9},
          doi = {10.3847/2041-8213/ab6dda},
archivePrefix = {arXiv},
       eprint = {2001.07720},
 primaryClass = {astro-ph.SR},
       adsurl = {https://ui.adsabs.harvard.edu/abs/2020ApJ...890L...9P}
}

@ARTICLE{Teague2019,
       author = {{Teague}, Richard and {Bae}, Jaehan and {Bergin}, Edwin A.},
        title = "{Meridional flows in the disk around a young star}",
      journal = {\nat},
         year = 2019,
        month = oct,
       volume = {574},
       number = {7778},
        pages = {378-381},
          doi = {10.1038/s41586-019-1642-0},
archivePrefix = {arXiv},
       eprint = {1910.06980},
 primaryClass = {astro-ph.EP},
       adsurl = {https://ui.adsabs.harvard.edu/abs/2019Natur.574..378T}
}

@ARTICLE{Teague2018,
       author = {{Teague}, Richard and {Bae}, Jaehan and {Bergin}, Edwin A. and {Birnstiel}, Tilman and {Foreman-Mackey}, Daniel},
        title = "{A Kinematical Detection of Two Embedded Jupiter-mass Planets in HD 163296}",
      journal = {\apjl},
         year = 2018,
        month = jun,
       volume = {860},
       number = {1},
          eid = {L12},
        pages = {L12},
          doi = {10.3847/2041-8213/aac6d7},
archivePrefix = {arXiv},
       eprint = {1805.10290},
 primaryClass = {astro-ph.EP},
       adsurl = {https://ui.adsabs.harvard.edu/abs/2018ApJ...860L..12T}
}

@ARTICLE{Rodenkirch2021,
       author = {{Rodenkirch}, P.~J. and {Rometsch}, T. and {Dullemond}, C.~P. and {Weber}, P. and {Kley}, W.},
        title = "{Modeling the nonaxisymmetric structure in the HD 163296 disk with planet-disk interaction}",
      journal = {\aap},
         year = 2021,
        month = mar,
       volume = {647},
          eid = {A174},
        pages = {A174},
          doi = {10.1051/0004-6361/202038484},
archivePrefix = {arXiv},
       eprint = {2012.09217},
 primaryClass = {astro-ph.EP},
       adsurl = {https://ui.adsabs.harvard.edu/abs/2021A&A...647A.174R}
}

@ARTICLE{Andrews2018,
       author = {{Andrews}, Sean M. and {Huang}, Jane and {P{\'e}rez}, Laura M. and {Isella}, Andrea and {Dullemond}, Cornelis P. and {Kurtovic}, Nicol{\'a}s T. and {Guzm{\'a}n}, Viviana V. and {Carpenter}, John M. and {Wilner}, David J. and {Zhang}, Shangjia and {Zhu}, Zhaohuan and {Birnstiel}, Tilman and {Bai}, Xue-Ning and {Benisty}, Myriam and {Hughes}, A. Meredith and {{\"O}berg}, Karin I. and {Ricci}, Luca},
        title = "{The Disk Substructures at High Angular Resolution Project (DSHARP). I. Motivation, Sample, Calibration, and Overview}",
      journal = {\apjl},
         year = 2018,
        month = dec,
       volume = {869},
       number = {2},
          eid = {L41},
        pages = {L41},
          doi = {10.3847/2041-8213/aaf741},
archivePrefix = {arXiv},
       eprint = {1812.04040},
 primaryClass = {astro-ph.SR},
       adsurl = {https://ui.adsabs.harvard.edu/abs/2018ApJ...869L..41A}
}

@inproceedings{casa,
	adsurl = {https://ui.adsabs.harvard.edu/abs/2007ASPC..376..127M},
	Author = {{McMullin}, J.~P. and {Waters}, B. and {Schiebel}, D. and {Young}, W. and {Golap}, K.},
	Booktitle = {Astronomical Data Analysis Software and Systems XVI},
	Editor = {{Shaw}, R.~A. and {Hill}, F. and {Bell}, D.~J.},
	Month = oct,
	Pages = {127},
	Series = {Astronomical Society of the Pacific Conference Series},
	Title = {{CASA Architecture and Applications}},
	Volume = {376},
	Year = 2007}

@misc{kepmask,
doi = {10.5281/ZENODO.4321137},
url = {https://zenodo.org/record/4321137},
author = {Teague, Rich},
title = {richteague/keplerian\_mask: Initial Release},
publisher = {Zenodo},
year = {2020},
copyright = {Open Access}
}

@article{astropy:2013,
Adsurl = {http://adsabs.harvard.edu/abs/2013A%26A...558A..33A},
Archiveprefix = {arXiv},
Author = {{Astropy Collaboration} and {Robitaille}, T.~P. and {Tollerud}, E.~J. and {Greenfield}, P. and {Droettboom}, M. and {Bray}, E. and {Aldcroft}, T. and {Davis}, M. and {Ginsburg}, A. and {Price-Whelan}, A.~M. and {Kerzendorf}, W.~E. and {Conley}, A. and {Crighton}, N. and {Barbary}, K. and {Muna}, D. and {Ferguson}, H. and {Grollier}, F. and {Parikh}, M.~M. and {Nair}, P.~H. and {Unther}, H.~M. and {Deil}, C. and {Woillez}, J. and {Conseil}, S. and {Kramer}, R. and {Turner}, J.~E.~H. and {Singer}, L. and {Fox}, R. and {Weaver}, B.~A. and {Zabalza}, V. and {Edwards}, Z.~I. and {Azalee Bostroem}, K. and {Burke}, D.~J. and {Casey}, A.~R. and {Crawford}, S.~M. and {Dencheva}, N. and {Ely}, J. and {Jenness}, T. and {Labrie}, K. and {Lim}, P.~L. and {Pierfederici}, F. and {Pontzen}, A. and {Ptak}, A. and {Refsdal}, B. and {Servillat}, M. and {Streicher}, O.},
Doi = {10.1051/0004-6361/201322068},
Eid = {A33},
Eprint = {1307.6212},
Journal = {\aap},
Month = oct,
Pages = {A33},
Primaryclass = {astro-ph.IM},
Title = {{Astropy: A community Python package for astronomy}},
Volume = 558,
Year = 2013}

@ARTICLE{astropy:2018,
       author = {{Astropy Collaboration} and {Price-Whelan}, A.~M. and
         {Sip{\H{o}}cz}, B.~M. and {G{\"u}nther}, H.~M. and {Lim}, P.~L. and
         {Crawford}, S.~M. and {Conseil}, S. and {Shupe}, D.~L. and
         {Craig}, M.~W. and {Dencheva}, N. and {Ginsburg}, A. and {Vand
        erPlas}, J.~T. and {Bradley}, L.~D. and {P{\'e}rez-Su{\'a}rez}, D. and
         {de Val-Borro}, M. and {Aldcroft}, T.~L. and {Cruz}, K.~L. and
         {Robitaille}, T.~P. and {Tollerud}, E.~J. and {Ardelean}, C. and
         {Babej}, T. and {Bach}, Y.~P. and {Bachetti}, M. and {Bakanov}, A.~V. and
         {Bamford}, S.~P. and {Barentsen}, G. and {Barmby}, P. and
         {Baumbach}, A. and {Berry}, K.~L. and {Biscani}, F. and {Boquien}, M. and
         {Bostroem}, K.~A. and {Bouma}, L.~G. and {Brammer}, G.~B. and
         {Bray}, E.~M. and {Breytenbach}, H. and {Buddelmeijer}, H. and
         {Burke}, D.~J. and {Calderone}, G. and {Cano Rodr{\'\i}guez}, J.~L. and
         {Cara}, M. and {Cardoso}, J.~V.~M. and {Cheedella}, S. and {Copin}, Y. and
         {Corrales}, L. and {Crichton}, D. and {D'Avella}, D. and {Deil}, C. and
         {Depagne}, {\'E}. and {Dietrich}, J.~P. and {Donath}, A. and
         {Droettboom}, M. and {Earl}, N. and {Erben}, T. and {Fabbro}, S. and
         {Ferreira}, L.~A. and {Finethy}, T. and {Fox}, R.~T. and
         {Garrison}, L.~H. and {Gibbons}, S.~L.~J. and {Goldstein}, D.~A. and
         {Gommers}, R. and {Greco}, J.~P. and {Greenfield}, P. and
         {Groener}, A.~M. and {Grollier}, F. and {Hagen}, A. and {Hirst}, P. and
         {Homeier}, D. and {Horton}, A.~J. and {Hosseinzadeh}, G. and {Hu}, L. and
         {Hunkeler}, J.~S. and {Ivezi{\'c}}, {\v{Z}}. and {Jain}, A. and
         {Jenness}, T. and {Kanarek}, G. and {Kendrew}, S. and {Kern}, N.~S. and
         {Kerzendorf}, W.~E. and {Khvalko}, A. and {King}, J. and {Kirkby}, D. and
         {Kulkarni}, A.~M. and {Kumar}, A. and {Lee}, A. and {Lenz}, D. and
         {Littlefair}, S.~P. and {Ma}, Z. and {Macleod}, D.~M. and
         {Mastropietro}, M. and {McCully}, C. and {Montagnac}, S. and
         {Morris}, B.~M. and {Mueller}, M. and {Mumford}, S.~J. and {Muna}, D. and
         {Murphy}, N.~A. and {Nelson}, S. and {Nguyen}, G.~H. and
         {Ninan}, J.~P. and {N{\"o}the}, M. and {Ogaz}, S. and {Oh}, S. and
         {Parejko}, J.~K. and {Parley}, N. and {Pascual}, S. and {Patil}, R. and
         {Patil}, A.~A. and {Plunkett}, A.~L. and {Prochaska}, J.~X. and
         {Rastogi}, T. and {Reddy Janga}, V. and {Sabater}, J. and
         {Sakurikar}, P. and {Seifert}, M. and {Sherbert}, L.~E. and
         {Sherwood-Taylor}, H. and {Shih}, A.~Y. and {Sick}, J. and
         {Silbiger}, M.~T. and {Singanamalla}, S. and {Singer}, L.~P. and
         {Sladen}, P.~H. and {Sooley}, K.~A. and {Sornarajah}, S. and
         {Streicher}, O. and {Teuben}, P. and {Thomas}, S.~W. and
         {Tremblay}, G.~R. and {Turner}, J.~E.~H. and {Terr{\'o}n}, V. and
         {van Kerkwijk}, M.~H. and {de la Vega}, A. and {Watkins}, L.~L. and
         {Weaver}, B.~A. and {Whitmore}, J.~B. and {Woillez}, J. and
         {Zabalza}, V. and {Astropy Contributors}},
        title = "{The Astropy Project: Building an Open-science Project and Status of the v2.0 Core Package}",
      journal = {\aj},
         year = 2018,
        month = sep,
       volume = {156},
       number = {3},
          eid = {123},
        pages = {123},
          doi = {10.3847/1538-3881/aabc4f},
archivePrefix = {arXiv},
       eprint = {1801.02634},
 primaryClass = {astro-ph.IM},
       adsurl = {https://ui.adsabs.harvard.edu/abs/2018AJ....156..123A}
}

@ARTICLE{astropy:2022,
       author = {{Astropy Collaboration} and {Price-Whelan}, Adrian M. and {Lim}, Pey Lian and {Earl}, Nicholas and {Starkman}, Nathaniel and {Bradley}, Larry and {Shupe}, David L. and {Patil}, Aarya A. and {Corrales}, Lia and {Brasseur}, C.~E. and {N{"o}the}, Maximilian and {Donath}, Axel and {Tollerud}, Erik and {Morris}, Brett M. and {Ginsburg}, Adam and {Vaher}, Eero and {Weaver}, Benjamin A. and {Tocknell}, James and {Jamieson}, William and {van Kerkwijk}, Marten H. and {Robitaille}, Thomas P. and {Merry}, Bruce and {Bachetti}, Matteo and {G{"u}nther}, H. Moritz and {Aldcroft}, Thomas L. and {Alvarado-Montes}, Jaime A. and {Archibald}, Anne M. and {B{'o}di}, Attila and {Bapat}, Shreyas and {Barentsen}, Geert and {Baz{'a}n}, Juanjo and {Biswas}, Manish and {Boquien}, M{'e}d{'e}ric and {Burke}, D.~J. and {Cara}, Daria and {Cara}, Mihai and {Conroy}, Kyle E. and {Conseil}, Simon and {Craig}, Matthew W. and {Cross}, Robert M. and {Cruz}, Kelle L. and {D'Eugenio}, Francesco and {Dencheva}, Nadia and {Devillepoix}, Hadrien A.~R. and {Dietrich}, J{"o}rg P. and {Eigenbrot}, Arthur Davis and {Erben}, Thomas and {Ferreira}, Leonardo and {Foreman-Mackey}, Daniel and {Fox}, Ryan and {Freij}, Nabil and {Garg}, Suyog and {Geda}, Robel and {Glattly}, Lauren and {Gondhalekar}, Yash and {Gordon}, Karl D. and {Grant}, David and {Greenfield}, Perry and {Groener}, Austen M. and {Guest}, Steve and {Gurovich}, Sebastian and {Handberg}, Rasmus and {Hart}, Akeem and {Hatfield-Dodds}, Zac and {Homeier}, Derek and {Hosseinzadeh}, Griffin and {Jenness}, Tim and {Jones}, Craig K. and {Joseph}, Prajwel and {Kalmbach}, J. Bryce and {Karamehmetoglu}, Emir and {Ka{l}uszy{'n}ski}, Miko{l}aj and {Kelley}, Michael S.~P. and {Kern}, Nicholas and {Kerzendorf}, Wolfgang E. and {Koch}, Eric W. and {Kulumani}, Shankar and {Lee}, Antony and {Ly}, Chun and {Ma}, Zhiyuan and {MacBride}, Conor and {Maljaars}, Jakob M. and {Muna}, Demitri and {Murphy}, N.~A. and {Norman}, Henrik and {O'Steen}, Richard and {Oman}, Kyle A. and {Pacifici}, Camilla and {Pascual}, Sergio and {Pascual-Granado}, J. and {Patil}, Rohit R. and {Perren}, Gabriel I. and {Pickering}, Timothy E. and {Rastogi}, Tanuj and {Roulston}, Benjamin R. and {Ryan}, Daniel F. and {Rykoff}, Eli S. and {Sabater}, Jose and {Sakurikar}, Parikshit and {Salgado}, Jes{'u}s and {Sanghi}, Aniket and {Saunders}, Nicholas and {Savchenko}, Volodymyr and {Schwardt}, Ludwig and {Seifert-Eckert}, Michael and {Shih}, Albert Y. and {Jain}, Anany Shrey and {Shukla}, Gyanendra and {Sick}, Jonathan and {Simpson}, Chris and {Singanamalla}, Sudheesh and {Singer}, Leo P. and {Singhal}, Jaladh and {Sinha}, Manodeep and {Sip{H{o}}cz}, Brigitta M. and {Spitler}, Lee R. and {Stansby}, David and {Streicher}, Ole and {{{S}}umak}, Jani and {Swinbank}, John D. and {Taranu}, Dan S. and {Tewary}, Nikita and {Tremblay}, Grant R. and {Val-Borro}, Miguel de and {Van Kooten}, Samuel J. and {Vasovi{'c}}, Zlatan and {Verma}, Shresth and {de Miranda Cardoso}, Jos{'e} Vin{'i}cius and {Williams}, Peter K.~G. and {Wilson}, Tom J. and {Winkel}, Benjamin and {Wood-Vasey}, W.~M. and {Xue}, Rui and {Yoachim}, Peter and {Zhang}, Chen and {Zonca}, Andrea and {Astropy Project Contributors}},
        title = "{The Astropy Project: Sustaining and Growing a Community-oriented Open-source Project and the Latest Major Release (v5.0) of the Core Package}",
      journal = {\apj},
         year = 2022,
        month = aug,
       volume = {935},
       number = {2},
          eid = {167},
        pages = {167},
          doi = {10.3847/1538-4357/ac7c74},
archivePrefix = {arXiv},
       eprint = {2206.14220},
 primaryClass = {astro-ph.IM},
       adsurl = {https://ui.adsabs.harvard.edu/abs/2022ApJ...935..167A}
}

@Article{numpy,
 title         = {Array programming with {NumPy}},
 author        = {Charles R. Harris and K. Jarrod Millman and St{\'{e}}fan J.
                 van der Walt and Ralf Gommers and Pauli Virtanen and David
                 Cournapeau and Eric Wieser and Julian Taylor and Sebastian
                 Berg and Nathaniel J. Smith and Robert Kern and Matti Picus
                 and Stephan Hoyer and Marten H. van Kerkwijk and Matthew
                 Brett and Allan Haldane and Jaime Fern{\'{a}}ndez del
                 R{\'{i}}o and Mark Wiebe and Pearu Peterson and Pierre
                 G{\'{e}}rard-Marchant and Kevin Sheppard and Tyler Reddy and
                 Warren Weckesser and Hameer Abbasi and Christoph Gohlke and
                 Travis E. Oliphant},
 year          = {2020},
 month         = sep,
 journal       = {Nature},
 volume        = {585},
 number        = {7825},
 pages         = {357--362},
 doi           = {10.1038/s41586-020-2649-2},
 publisher     = {Springer Science and Business Media {LLC}},
 url           = {https://doi.org/10.1038/s41586-020-2649-2}
}

@article{matplotlib,
	Adsurl = {https://ui.adsabs.harvard.edu/abs/2007CSE.....9...90H},
	Author = {{Hunter}, John D.},
	Doi = {10.1109/MCSE.2007.55},
	Journal = {Computing in Science and Engineering},
	Month = may,
	Number = {3},
	Pages = {90-95},
	Title = {{Matplotlib: A 2D Graphics Environment}},
	Volume = {9},
	Year = 2007}

@ARTICLE{Marel2018,
       author = {{van der Marel}, Nienke and {Williams}, Jonathan P. and {Bruderer}, Simon},
        title = "{Rings and Gaps in Protoplanetary Disks: Planets or Snowlines?}",
      journal = {\apjl},
         year = 2018,
        month = nov,
       volume = {867},
       number = {1},
          eid = {L14},
        pages = {L14},
          doi = {10.3847/2041-8213/aae88e},
archivePrefix = {arXiv},
       eprint = {1810.05614},
 primaryClass = {astro-ph.EP},
       adsurl = {https://ui.adsabs.harvard.edu/abs/2018ApJ...867L..14V}
}

@ARTICLE{Kastner2015,
       author = {{Kastner}, Joel H. and {Qi}, Chunhua and {Gorti}, Uma and {Hily-Blant}, Pierre and {Oberg}, Karin and {Forveille}, Thierry and {Andrews}, Sean and {Wilner}, David},
        title = "{A Ring of C$_{2}$H in the Molecular Disk Orbiting TW Hya}",
      journal = {\apj},
         year = 2015,
        month = jun,
       volume = {806},
       number = {1},
          eid = {75},
        pages = {75},
          doi = {10.1088/0004-637X/806/1/75},
archivePrefix = {arXiv},
       eprint = {1504.05980},
 primaryClass = {astro-ph.SR},
       adsurl = {https://ui.adsabs.harvard.edu/abs/2015ApJ...806...75K}
}

@ARTICLE{Kama2016,
       author = {{Kama}, M. and {Bruderer}, S. and {van Dishoeck}, E.~F. and {Hogerheijde}, M. and {Folsom}, C.~P. and {Miotello}, A. and {Fedele}, D. and {Belloche}, A. and {G{\"u}sten}, R. and {Wyrowski}, F.},
        title = "{Volatile-carbon locking and release in protoplanetary disks. A study of TW Hya and HD 100546}",
      journal = {\aap},
         year = 2016,
        month = aug,
       volume = {592},
          eid = {A83},
        pages = {A83},
          doi = {10.1051/0004-6361/201526991},
archivePrefix = {arXiv},
       eprint = {1605.05093},
 primaryClass = {astro-ph.EP},
       adsurl = {https://ui.adsabs.harvard.edu/abs/2016A&A...592A..83K}
}

@ARTICLE{Bergin2016,
       author = {{Bergin}, Edwin A. and {Du}, Fujun and {Cleeves}, L. Ilsedore and {Blake}, G.~A. and {Schwarz}, K. and {Visser}, R. and {Zhang}, K.},
        title = "{Hydrocarbon Emission Rings in Protoplanetary Disks Induced by Dust Evolution}",
      journal = {\apj},
         year = 2016,
        month = nov,
       volume = {831},
       number = {1},
          eid = {101},
        pages = {101},
          doi = {10.3847/0004-637X/831/1/101},
archivePrefix = {arXiv},
       eprint = {1609.06337},
 primaryClass = {astro-ph.EP},
       adsurl = {https://ui.adsabs.harvard.edu/abs/2016ApJ...831..101B}
}

@ARTICLE{Miotello2019,
       author = {{Miotello}, A. and {Facchini}, S. and {van Dishoeck}, E.~F. and {Cazzoletti}, P. and {Testi}, L. and {Williams}, J.~P. and {Ansdell}, M. and {van Terwisga}, S. and {van der Marel}, N.},
        title = "{Bright C$_{2}$H emission in protoplanetary discs in Lupus: high volatile C/O > 1 ratios}",
      journal = {\aap},
         year = 2019,
        month = nov,
       volume = {631},
          eid = {A69},
        pages = {A69},
          doi = {10.1051/0004-6361/201935441},
archivePrefix = {arXiv},
       eprint = {1909.04477},
 primaryClass = {astro-ph.SR},
       adsurl = {https://ui.adsabs.harvard.edu/abs/2019A&A...631A..69M}
}

@ARTICLE{Fadul2025,
       author = {{Fadul}, Abubakar M.~A. and {Schwarz}, Kamber R. and {van'T Hoff}, Merel L.~R. and {Huang}, Jane and {Bergner}, Jennifer B. and {Suhasaria}, Tushar and {Calahan}, Jenny K.},
        title = "{A Deep Search for Complex Organic Molecules toward the Protoplanetary Disk of V883 Ori}",
      journal = {\aj},
         year = 2025,
        month = jun,
       volume = {169},
       number = {6},
          eid = {307},
        pages = {307},
          doi = {10.3847/1538-3881/adc998},
archivePrefix = {arXiv},
       eprint = {2504.06005},
 primaryClass = {astro-ph.SR},
       adsurl = {https://ui.adsabs.harvard.edu/abs/2025AJ....169..307F}
}

@ARTICLE{Law2025,
       author = {{Law}, Charles J. and {Le Gal}, Romane and {Yamato}, Yoshihide and {Zhang}, Ke and {Guzm{\'a}n}, Viviana V. and {Hern{\'a}ndez-Vera}, Claudio and {Cleeves}, L. Ilsedore and {Guidi}, Greta and {Booth}, Alice S.},
        title = "{A Multiline Analysis of the Distribution and Excitation of CS and H$_{2}$CS in the HD 163296 Disk}",
      journal = {\apj},
         year = 2025,
        month = may,
       volume = {985},
       number = {1},
          eid = {84},
        pages = {84},
          doi = {10.3847/1538-4357/adc304},
archivePrefix = {arXiv},
       eprint = {2503.16605},
 primaryClass = {astro-ph.EP},
       adsurl = {https://ui.adsabs.harvard.edu/abs/2025ApJ...985...84L}
}

@ARTICLE{Birnstiel2018,
       author = {{Birnstiel}, Tilman and {Dullemond}, Cornelis P. and {Zhu}, Zhaohuan and {Andrews}, Sean M. and {Bai}, Xue-Ning and {Wilner}, David J. and {Carpenter}, John M. and {Huang}, Jane and {Isella}, Andrea and {Benisty}, Myriam and {P{\'e}rez}, Laura M. and {Zhang}, Shangjia},
        title = "{The Disk Substructures at High Angular Resolution Project (DSHARP). V. Interpreting ALMA Maps of Protoplanetary Disks in Terms of a Dust Model}",
      journal = {\apjl},
         year = 2018,
        month = dec,
       volume = {869},
       number = {2},
          eid = {L45},
        pages = {L45},
          doi = {10.3847/2041-8213/aaf743},
archivePrefix = {arXiv},
       eprint = {1812.04043},
 primaryClass = {astro-ph.SR},
       adsurl = {https://ui.adsabs.harvard.edu/abs/2018ApJ...869L..45B}
}

@ARTICLE{Wild2011,
       author = {{Wild}, Vivienne and {Charlot}, St{\'e}phane and {Brinchmann}, Jarle and {Heckman}, Timothy and {Vince}, Oliver and {Pacifici}, Camilla and {Chevallard}, Jacopo},
        title = "{Empirical determination of the shape of dust attenuation curves in star-forming galaxies}",
      journal = {\mnras},
         year = 2011,
        month = nov,
       volume = {417},
       number = {3},
        pages = {1760-1786},
          doi = {10.1111/j.1365-2966.2011.19367.x},
archivePrefix = {arXiv},
       eprint = {1106.1646},
 primaryClass = {astro-ph.CO},
       adsurl = {https://ui.adsabs.harvard.edu/abs/2011MNRAS.417.1760W}
}

@ARTICLE{Molliere2022,
       author = {{Molli{\`e}re}, Paul and {Molyarova}, Tamara and {Bitsch}, Bertram and {Henning}, Thomas and {Schneider}, Aaron and {Kreidberg}, Laura and {Eistrup}, Christian and {Burn}, Remo and {Nasedkin}, Evert and {Semenov}, Dmitry and {Mordasini}, Christoph and {Schlecker}, Martin and {Schwarz}, Kamber R. and {Lacour}, Sylvestre and {Nowak}, Mathias and {Schulik}, Matth{\"a}us},
        title = "{Interpreting the Atmospheric Composition of Exoplanets: Sensitivity to Planet Formation Assumptions}",
      journal = {\apj},
         year = 2022,
        month = jul,
       volume = {934},
       number = {1},
          eid = {74},
        pages = {74},
          doi = {10.3847/1538-4357/ac6a56},
archivePrefix = {arXiv},
       eprint = {2204.13714},
 primaryClass = {astro-ph.EP},
       adsurl = {https://ui.adsabs.harvard.edu/abs/2022ApJ...934...74M}
}

@ARTICLE{Madhusudhan2014,
       author = {{Madhusudhan}, Nikku and {Amin}, Mustafa A. and {Kennedy}, Grant M.},
        title = "{Toward Chemical Constraints on Hot Jupiter Migration}",
      journal = {\apjl},
         year = 2014,
        month = oct,
       volume = {794},
       number = {1},
          eid = {L12},
        pages = {L12},
          doi = {10.1088/2041-8205/794/1/L12},
archivePrefix = {arXiv},
       eprint = {1408.3668},
 primaryClass = {astro-ph.EP},
       adsurl = {https://ui.adsabs.harvard.edu/abs/2014ApJ...794L..12M}
}

@ARTICLE{Schneider2021,
       author = {{Schneider}, Aaron David and {Bitsch}, Bertram},
        title = "{How drifting and evaporating pebbles shape giant planets. I. Heavy element content and atmospheric C/O}",
      journal = {\aap},
         year = 2021,
        month = oct,
       volume = {654},
          eid = {A71},
        pages = {A71},
          doi = {10.1051/0004-6361/202039640},
archivePrefix = {arXiv},
       eprint = {2105.13267},
 primaryClass = {astro-ph.EP},
       adsurl = {https://ui.adsabs.harvard.edu/abs/2021A&A...654A..71S}
}

@ARTICLE{Mordasini2016,
       author = {{Mordasini}, C. and {van Boekel}, R. and {Molli{\`e}re}, P. and {Henning}, Th. and {Benneke}, Bj{\"o}rn},
        title = "{The Imprint of Exoplanet Formation History on Observable Present-day Spectra of Hot Jupiters}",
      journal = {\apj},
         year = 2016,
        month = nov,
       volume = {832},
       number = {1},
          eid = {41},
        pages = {41},
          doi = {10.3847/0004-637X/832/1/41},
archivePrefix = {arXiv},
       eprint = {1609.03019},
 primaryClass = {astro-ph.EP},
       adsurl = {https://ui.adsabs.harvard.edu/abs/2016ApJ...832...41M}
}

@ARTICLE{Thiabaud2015,
       author = {{Thiabaud}, A. and {Marboeuf}, U. and {Alibert}, Y. and {Leya}, I. and {Mezger}, K.},
        title = "{Gas composition of the main volatile elements in protoplanetary discs and its implication for planet formation}",
      journal = {\aap},
         year = 2015,
        month = feb,
       volume = {574},
          eid = {A138},
        pages = {A138},
          doi = {10.1051/0004-6361/201424868},
       adsurl = {https://ui.adsabs.harvard.edu/abs/2015A&A...574A.138T}
}

@ARTICLE{Bitsch2022,
       author = {{Bitsch}, Bertram and {Schneider}, Aaron David and {Kreidberg}, Laura},
        title = "{How drifting and evaporating pebbles shape giant planets. III. The formation of WASP-77A b and {\ensuremath{\tau}} Bo{\"o}tis b}",
      journal = {\aap},
         year = 2022,
        month = sep,
       volume = {665},
          eid = {A138},
        pages = {A138},
          doi = {10.1051/0004-6361/202243345},
archivePrefix = {arXiv},
       eprint = {2207.06077},
 primaryClass = {astro-ph.EP},
       adsurl = {https://ui.adsabs.harvard.edu/abs/2022A&A...665A.138B}
}

@ARTICLE{Lodders2003,
       author = {{Lodders}, Katharina},
        title = "{Solar System Abundances and Condensation Temperatures of the Elements}",
      journal = {\apj},
         year = 2003,
        month = jul,
       volume = {591},
       number = {2},
        pages = {1220-1247},
          doi = {10.1086/375492},
       adsurl = {https://ui.adsabs.harvard.edu/abs/2003ApJ...591.1220L}
}

@ARTICLE{Calahan2023,
       author = {{Calahan}, Jenny K. and {Bergin}, Edwin A. and {Bosman}, Arthur D. and {Rich}, Evan A. and {Andrews}, Sean M. and {Bergner}, Jennifer B. and {Cleeves}, L. Ilsedore and {Guzm{\'a}n}, Viviana V. and {Huang}, Jane and {Ilee}, John D. and {Law}, Charles J. and {Le Gal}, Romane and {{\"O}berg}, Karin I. and {Teague}, Richard and {Walsh}, Catherine and {Wilner}, David J. and {Zhang}, Ke},
        title = "{UV-driven chemistry as a signpost of late-stage planet formation}",
      journal = {Nature Astronomy},
         year = 2023,
        month = jan,
       volume = {7},
        pages = {49-56},
          doi = {10.1038/s41550-022-01831-8},
archivePrefix = {arXiv},
       eprint = {2212.05539},
 primaryClass = {astro-ph.EP},
       adsurl = {https://ui.adsabs.harvard.edu/abs/2023NatAs...7...49C}
}

@ARTICLE{Oberg2011_2,
       author = {{{\"O}berg}, Karin I. and {Murray-Clay}, Ruth and {Bergin}, Edwin A.},
        title = "{The Effects of Snowlines on C/O in Planetary Atmospheres}",
      journal = {\apjl},
         year = 2011,
        month = dec,
       volume = {743},
       number = {1},
          eid = {L16},
        pages = {L16},
          doi = {10.1088/2041-8205/743/1/L16},
archivePrefix = {arXiv},
       eprint = {1110.5567},
 primaryClass = {astro-ph.GA},
       adsurl = {https://ui.adsabs.harvard.edu/abs/2011ApJ...743L..16O}
}

@ARTICLE{Qi2013,
       author = {{Qi}, Chunhua and {{\"O}berg}, Karin I. and {Wilner}, David J. and {Rosenfeld}, Katherine A.},
        title = "{First Detection of c-C$_{3}$H$_{2}$ in a Circumstellar Disk}",
      journal = {\apjl},
         year = 2013,
        month = mar,
       volume = {765},
       number = {1},
          eid = {L14},
        pages = {L14},
          doi = {10.1088/2041-8205/765/1/L14},
archivePrefix = {arXiv},
       eprint = {1302.0251},
 primaryClass = {astro-ph.GA},
       adsurl = {https://ui.adsabs.harvard.edu/abs/2013ApJ...765L..14Q}
}

@ARTICLE{Oberg2015Natur,
       author = {{{\"O}berg}, Karin I. and {Guzm{\'a}n}, Viviana V. and {Furuya}, Kenji and {Qi}, Chunhua and {Aikawa}, Yuri and {Andrews}, Sean M. and {Loomis}, Ryan and {Wilner}, David J.},
        title = "{The comet-like composition of a protoplanetary disk as revealed by complex cyanides}",
      journal = {\nat},
         year = 2015,
        month = apr,
       volume = {520},
       number = {7546},
        pages = {198-201},
          doi = {10.1038/nature14276},
archivePrefix = {arXiv},
       eprint = {1505.06347},
 primaryClass = {astro-ph.GA},
       adsurl = {https://ui.adsabs.harvard.edu/abs/2015Natur.520..198O}
}

@ARTICLE{Favre2018,
       author = {{Favre}, C{\'e}cile and {Fedele}, Davide and {Semenov}, Dmitry and {Parfenov}, Sergey and {Codella}, Claudio and {Ceccarelli}, Cecilia and {Bergin}, Edwin A. and {Chapillon}, Edwige and {Testi}, Leonardo and {Hersant}, Franck and {Lefloch}, Bertrand and {Fontani}, Francesco and {Blake}, Geoffrey A. and {Cleeves}, L. Ilsedore and {Qi}, Chunhua and {Schwarz}, Kamber R. and {Taquet}, Vianney},
        title = "{First Detection of the Simplest Organic Acid in a Protoplanetary Disk}",
      journal = {\apjl},
         year = 2018,
        month = jul,
       volume = {862},
       number = {1},
          eid = {L2},
        pages = {L2},
          doi = {10.3847/2041-8213/aad046},
archivePrefix = {arXiv},
       eprint = {1807.05768},
 primaryClass = {astro-ph.SR},
       adsurl = {https://ui.adsabs.harvard.edu/abs/2018ApJ...862L...2F}
}

@ARTICLE{Loomis2020,
       author = {{Loomis}, Ryan A. and {{\"O}berg}, Karin I. and {Andrews}, Sean M. and {Bergin}, Edwin and {Bergner}, Jennifer and {Blake}, Geoffrey A. and {Cleeves}, L. Ilsedore and {Czekala}, Ian and {Huang}, Jane and {Le Gal}, Romane and {M{\'e}nard}, Francois and {Pegues}, Jamila and {Qi}, Chunhua and {Walsh}, Catherine and {Williams}, Jonathan P. and {Wilner}, David J.},
        title = "{An Unbiased ALMA Spectral Survey of the LkCa 15 and MWC 480 Protoplanetary Disks}",
      journal = {\apj},
         year = 2020,
        month = apr,
       volume = {893},
       number = {2},
          eid = {101},
        pages = {101},
          doi = {10.3847/1538-4357/ab7cc8},
archivePrefix = {arXiv},
       eprint = {2006.16187},
 primaryClass = {astro-ph.SR},
       adsurl = {https://ui.adsabs.harvard.edu/abs/2020ApJ...893..101L}
}

@ARTICLE{Booth2024a,
       author = {{Booth}, Alice S. and {Leemker}, Margot and {van Dishoeck}, Ewine F. and {Evans}, Lucy and {Ilee}, John D. and {Kama}, Mihkel and {Keyte}, Luke and {Law}, Charles J. and {van der Marel}, Nienke and {Nomura}, Hideko and {Notsu}, Shota and {{\"O}berg}, Karin and {Temmink}, Milou and {Walsh}, Catherine},
        title = "{An ALMA Molecular Inventory of Warm Herbig Ae Disks. I. Molecular Rings, Asymmetries, and Complexity in the HD 100546 Disk}",
      journal = {\aj},
         year = 2024,
        month = apr,
       volume = {167},
       number = {4},
          eid = {164},
        pages = {164},
          doi = {10.3847/1538-3881/ad2700},
archivePrefix = {arXiv},
       eprint = {2402.04001},
 primaryClass = {astro-ph.EP},
       adsurl = {https://ui.adsabs.harvard.edu/abs/2024AJ....167..164B}
}

@ARTICLE{Booth2024b,
       author = {{Booth}, Alice S. and {Temmink}, Milou and {van Dishoeck}, Ewine F. and {Evans}, Lucy and {Ilee}, John D. and {Kama}, Mihkel and {Keyte}, Luke and {Law}, Charles J. and {Leemker}, Margot and {van der Marel}, Nienke and {Nomura}, Hideko and {Notsu}, Shota and {{\"O}berg}, Karin and {Walsh}, Catherine},
        title = "{An ALMA Molecular Inventory of Warm Herbig Ae Disks. II. Abundant Complex Organics and Volatile Sulphur in the IRS 48 Disk}",
      journal = {\aj},
         year = 2024,
        month = apr,
       volume = {167},
       number = {4},
          eid = {165},
        pages = {165},
          doi = {10.3847/1538-3881/ad26ff},
archivePrefix = {arXiv},
       eprint = {2402.04002},
 primaryClass = {astro-ph.EP},
       adsurl = {https://ui.adsabs.harvard.edu/abs/2024AJ....167..165B}
}

@ARTICLE{Brunken2022,
       author = {{Brunken}, Nashanty G.~C. and {Booth}, Alice S. and {Leemker}, Margot and {Nazari}, Pooneh and {van der Marel}, Nienke and {van Dishoeck}, Ewine F.},
        title = "{A major asymmetric ice trap in a planet-forming disk. III. First detection of dimethyl ether}",
      journal = {\aap},
         year = 2022,
        month = mar,
       volume = {659},
          eid = {A29},
        pages = {A29},
          doi = {10.1051/0004-6361/202142981},
archivePrefix = {arXiv},
       eprint = {2203.02936},
 primaryClass = {astro-ph.EP},
       adsurl = {https://ui.adsabs.harvard.edu/abs/2022A&A...659A..29B}
}

@ARTICLE{Law2021_MAPSIII,
       author = {{Law}, Charles J. and {Loomis}, Ryan A. and {Teague}, Richard and {{\"O}berg}, Karin I. and {Czekala}, Ian and {Andrews}, Sean M. and {Huang}, Jane and {Aikawa}, Yuri and {Alarc{\'o}n}, Felipe and {Bae}, Jaehan and {Bergin}, Edwin A. and {Bergner}, Jennifer B. and {Boehler}, Yann and {Booth}, Alice S. and {Bosman}, Arthur D. and {Calahan}, Jenny K. and {Cataldi}, Gianni and {Cleeves}, L. Ilsedore and {Furuya}, Kenji and {Guzm{\'a}n}, Viviana V. and {Ilee}, John D. and {Le Gal}, Romane and {Liu}, Yao and {Long}, Feng and {M{\'e}nard}, Fran{\c{c}}ois and {Nomura}, Hideko and {Qi}, Chunhua and {Schwarz}, Kamber R. and {Sierra}, Anibal and {Tsukagoshi}, Takashi and {Yamato}, Yoshihide and {van't Hoff}, Merel L.~R. and {Walsh}, Catherine and {Wilner}, David J. and {Zhang}, Ke},
        title = "{Molecules with ALMA at Planet-forming Scales (MAPS). III. Characteristics of Radial Chemical Substructures}",
      journal = {\apjs},
         year = 2021,
        month = nov,
       volume = {257},
       number = {1},
          eid = {3},
        pages = {3},
          doi = {10.3847/1538-4365/ac1434},
archivePrefix = {arXiv},
       eprint = {2109.06210},
 primaryClass = {astro-ph.EP},
       adsurl = {https://ui.adsabs.harvard.edu/abs/2021ApJS..257....3L}
}

@ARTICLE{Calahan2021,
       author = {{Calahan}, Jenny K. and {Bergin}, Edwin A. and {Zhang}, Ke and {Schwarz}, Kamber R. and {{\"O}berg}, Karin I. and {Guzm{\'a}n}, Viviana V. and {Walsh}, Catherine and {Aikawa}, Yuri and {Alarc{\'o}n}, Felipe and {Andrews}, Sean M. and {Bae}, Jaehan and {Bergner}, Jennifer B. and {Booth}, Alice S. and {Bosman}, Arthur D. and {Cataldi}, Gianni and {Czekala}, Ian and {Huang}, Jane and {Ilee}, John D. and {Law}, Charles J. and {Le Gal}, Romane and {Long}, Feng and {Loomis}, Ryan A. and {M{\'e}nard}, Fran{\c{c}}ois and {Nomura}, Hideko and {Qi}, Chunhua and {Teague}, Richard and {van't Hoff}, Merel L.~R. and {Wilner}, David J. and {Yamato}, Yoshihide},
        title = "{Molecules with ALMA at Planet-forming Scales (MAPS). XVII. Determining the 2D Thermal Structure of the HD 163296 Disk}",
      journal = {\apjs},
         year = 2021,
        month = nov,
       volume = {257},
       number = {1},
          eid = {17},
        pages = {17},
          doi = {10.3847/1538-4365/ac143f},
archivePrefix = {arXiv},
       eprint = {2109.06202},
 primaryClass = {astro-ph.EP},
       adsurl = {https://ui.adsabs.harvard.edu/abs/2021ApJS..257...17C}
}

@ARTICLE{Pezzotta2025,
       author = {{Pezzotta}, V. and {Facchini}, S. and {Longarini}, C. and {Lodato}, G. and {Martire}, P.},
        title = "{The two-dimensional pressure structure of the HD 163296 protoplanetary disk as probed by multi-molecule kinematics}",
      journal = {\aap},
         year = 2025,
        month = feb,
       volume = {694},
          eid = {A108},
        pages = {A108},
          doi = {10.1051/0004-6361/202451307},
archivePrefix = {arXiv},
       eprint = {2501.05517},
 primaryClass = {astro-ph.EP},
       adsurl = {https://ui.adsabs.harvard.edu/abs/2025A&A...694A.108P}
}

@ARTICLE{Thi2004,
       author = {{Thi}, W. -F. and {van Zadelhoff}, G. -J. and {van Dishoeck}, E.~F.},
        title = "{Organic molecules in protoplanetary disks around T Tauri and Herbig Ae stars}",
      journal = {\aap},
         year = 2004,
        month = oct,
       volume = {425},
        pages = {955-972},
          doi = {10.1051/0004-6361:200400026},
archivePrefix = {arXiv},
       eprint = {astro-ph/0406577},
 primaryClass = {astro-ph},
       adsurl = {https://ui.adsabs.harvard.edu/abs/2004A&A...425..955T}
}

@ARTICLE{Qi2011,
       author = {{Qi}, Chunhua and {D'Alessio}, Paola and {{\"O}berg}, Karin I. and {Wilner}, David J. and {Hughes}, A. Meredith and {Andrews}, Sean M. and {Ayala}, Sandra},
        title = "{Resolving the CO Snow Line in the Disk around HD 163296}",
      journal = {\apj},
         year = 2011,
        month = oct,
       volume = {740},
       number = {2},
          eid = {84},
        pages = {84},
          doi = {10.1088/0004-637X/740/2/84},
archivePrefix = {arXiv},
       eprint = {1107.5061},
 primaryClass = {astro-ph.SR},
       adsurl = {https://ui.adsabs.harvard.edu/abs/2011ApJ...740...84Q}
}

@ARTICLE{Salinas2017,
       author = {{Salinas}, V.~N. and {Hogerheijde}, M.~R. and {Mathews}, G.~S. and {{\"O}berg}, K.~I. and {Qi}, C. and {Williams}, J.~P. and {Wilner}, D.~J.},
        title = "{DCO$^{+}$, DCN, and N$_{2}$D$^{+}$ reveal three different deuteration regimes in the disk around the Herbig Ae star HD 163296}",
      journal = {\aap},
         year = 2017,
        month = oct,
       volume = {606},
          eid = {A125},
        pages = {A125},
          doi = {10.1051/0004-6361/201731223},
archivePrefix = {arXiv},
       eprint = {1707.06475},
 primaryClass = {astro-ph.SR},
       adsurl = {https://ui.adsabs.harvard.edu/abs/2017A&A...606A.125S}
}

@ARTICLE{Bergner2019,
       author = {{Bergner}, Jennifer B. and {{\"O}berg}, Karin I. and {Bergin}, Edwin A. and {Loomis}, Ryan A. and {Pegues}, Jamila and {Qi}, Chunhua},
        title = "{A Survey of C$_{2}$H, HCN, and C$^{18}$O in Protoplanetary Disks}",
      journal = {\apj},
         year = 2019,
        month = may,
       volume = {876},
       number = {1},
          eid = {25},
        pages = {25},
          doi = {10.3847/1538-4357/ab141e},
archivePrefix = {arXiv},
       eprint = {1904.09315},
 primaryClass = {astro-ph.EP},
       adsurl = {https://ui.adsabs.harvard.edu/abs/2019ApJ...876...25B}
}

@ARTICLE{Izquierdo2022,
       author = {{Izquierdo}, Andr{\'e}s F. and {Facchini}, Stefano and {Rosotti}, Giovanni P. and {van Dishoeck}, Ewine F. and {Testi}, Leonardo},
        title = "{A New Planet Candidate Detected in a Dust Gap of the Disk around HD 163296 through Localized Kinematic Signatures: An Observational Validation of the DISCMINER}",
      journal = {\apj},
         year = 2022,
        month = mar,
       volume = {928},
       number = {1},
          eid = {2},
        pages = {2},
          doi = {10.3847/1538-4357/ac474d},
archivePrefix = {arXiv},
       eprint = {2111.06367},
 primaryClass = {astro-ph.EP},
       adsurl = {https://ui.adsabs.harvard.edu/abs/2022ApJ...928....2I}
}

@article{Alarcón2022,
doi = {10.3847/2041-8213/aca6e6},
url = {https://dx.doi.org/10.3847/2041-8213/aca6e6},
year = {2022},
month = {dec},
publisher = {The American Astronomical Society},
volume = {941},
number = {2},
pages = {L24},
author = {Alarcón, Felipe and Bergin, Edwin A. and Teague, Richard},
title = {A Localized Kinematic Structure Detected in Atomic Carbon Emission Spatially Coincident with a Proposed Protoplanet in the HD 163296 Disk},
journal = {The Astrophysical Journal Letters}
}

@ARTICLE{Calcino2022,
       author = {{Calcino}, Josh and {Hilder}, Thomas and {Price}, Daniel J. and {Pinte}, Christophe and {Bollati}, Francesco and {Lodato}, Giuseppe and {Norfolk}, Brodie J.},
        title = "{Mapping the Planetary Wake in HD 163296 with Kinematics}",
      journal = {\apjl},
         year = 2022,
        month = apr,
       volume = {929},
       number = {2},
          eid = {L25},
        pages = {L25},
          doi = {10.3847/2041-8213/ac64a7},
archivePrefix = {arXiv},
       eprint = {2111.07416},
 primaryClass = {astro-ph.EP},
       adsurl = {https://ui.adsabs.harvard.edu/abs/2022ApJ...929L..25C}
}

@ARTICLE{Salinas2018,
       author = {{Salinas}, V.~N. and {Hogerheijde}, M.~R. and {Murillo}, N.~M. and {Mathews}, G.~S. and {Qi}, C. and {Williams}, J.~P. and {Wilner}, D.~J.},
        title = "{Exploring DCO$^{+}$ as a tracer of thermal inversion in the disk around the Herbig Ae star HD 163296}",
      journal = {\aap},
         year = 2018,
        month = aug,
       volume = {616},
          eid = {A45},
        pages = {A45},
          doi = {10.1051/0004-6361/201731745},
archivePrefix = {arXiv},
       eprint = {1806.02950},
 primaryClass = {astro-ph.SR},
       adsurl = {https://ui.adsabs.harvard.edu/abs/2018A&A...616A..45S}
}

@ARTICLE{Taniguchi2023,
       author = {{Taniguchi}, Kotomi and {Sanhueza}, Patricio and {Olguin}, Fernando A. and {Gorai}, Prasanta and {Das}, Ankan and {Nakamura}, Fumitaka and {Saito}, Masao and {Zhang}, Qizhou and {Lu}, Xing and {Li}, Shanghuo and {Chen}, Huei-Ru Vivien},
        title = "{Digging into the Interior of Hot Cores with the ALMA (DIHCA). III. The Chemical Link between NH$_{2}$CHO, HNCO, and H$_{2}$CO}",
      journal = {\apj},
         year = 2023,
        month = jun,
       volume = {950},
       number = {1},
          eid = {57},
        pages = {57},
          doi = {10.3847/1538-4357/acca1d},
archivePrefix = {arXiv},
       eprint = {2304.00267},
 primaryClass = {astro-ph.GA},
       adsurl = {https://ui.adsabs.harvard.edu/abs/2023ApJ...950...57T}
}

@ARTICLE{Bergner2017,
       author = {{Bergner}, Jennifer B. and {{\"O}berg}, Karin I. and {Garrod}, Robin T. and {Graninger}, Dawn M.},
        title = "{Complex Organic Molecules toward Embedded Low-mass Protostars}",
      journal = {\apj},
         year = 2017,
        month = jun,
       volume = {841},
       number = {2},
          eid = {120},
        pages = {120},
          doi = {10.3847/1538-4357/aa72f6},
archivePrefix = {arXiv},
       eprint = {1705.05338},
 primaryClass = {astro-ph.SR},
       adsurl = {https://ui.adsabs.harvard.edu/abs/2017ApJ...841..120B}
}

@ARTICLE{CASATeam2022,
       author = {{CASA Team} and {Bean}, Ben and {Bhatnagar}, Sanjay and {Castro}, Sandra and {Donovan Meyer}, Jennifer and {Emonts}, Bjorn and {Garcia}, Enrique and {Garwood}, Robert and {Golap}, Kumar and {Gonzalez Villalba}, Justo and {Harris}, Pamela and {Hayashi}, Yohei and {Hoskins}, Josh and {Hsieh}, Mingyu and {Jagannathan}, Preshanth and {Kawasaki}, Wataru and {Keimpema}, Aard and {Kettenis}, Mark and {Lopez}, Jorge and {Marvil}, Joshua and {Masters}, Joseph and {McNichols}, Andrew and {Mehringer}, David and {Miel}, Renaud and {Moellenbrock}, George and {Montesino}, Federico and {Nakazato}, Takeshi and {Ott}, Juergen and {Petry}, Dirk and {Pokorny}, Martin and {Raba}, Ryan and {Rau}, Urvashi and {Schiebel}, Darrell and {Schweighart}, Neal and {Sekhar}, Srikrishna and {Shimada}, Kazuhiko and {Small}, Des and {Steeb}, Jan-Willem and {Sugimoto}, Kanako and {Suoranta}, Ville and {Tsutsumi}, Takahiro and {van Bemmel}, Ilse M. and {Verkouter}, Marjolein and {Wells}, Akeem and {Xiong}, Wei and {Szomoru}, Arpad and {Griffith}, Morgan and {Glendenning}, Brian and {Kern}, Jeff},
        title = "{CASA, the Common Astronomy Software Applications for Radio Astronomy}",
      journal = {\pasp},
         year = 2022,
        month = nov,
       volume = {134},
       number = {1041},
          eid = {114501},
        pages = {114501},
          doi = {10.1088/1538-3873/ac9642},
archivePrefix = {arXiv},
       eprint = {2210.02276},
 primaryClass = {astro-ph.IM},
       adsurl = {https://ui.adsabs.harvard.edu/abs/2022PASP..134k4501C}
}

@ARTICLE{Huang2018b,
       author = {{Huang}, Jane and {Andrews}, Sean M. and {Dullemond}, Cornelis P. and {Isella}, Andrea and {P{\'e}rez}, Laura M. and {Guzm{\'a}n}, Viviana V. and {{\"O}berg}, Karin I. and {Zhu}, Zhaohuan and {Zhang}, Shangjia and {Bai}, Xue-Ning and {Benisty}, Myriam and {Birnstiel}, Tilman and {Carpenter}, John M. and {Hughes}, A. Meredith and {Ricci}, Luca and {Weaver}, Erik and {Wilner}, David J.},
        title = "{The Disk Substructures at High Angular Resolution Project (DSHARP). II. Characteristics of Annular Substructures}",
      journal = {\apjl},
         year = 2018,
        month = dec,
       volume = {869},
       number = {2},
          eid = {L42},
        pages = {L42},
          doi = {10.3847/2041-8213/aaf740},
archivePrefix = {arXiv},
       eprint = {1812.04041},
 primaryClass = {astro-ph.EP},
       adsurl = {https://ui.adsabs.harvard.edu/abs/2018ApJ...869L..42H}
}

@ARTICLE{Teague2021,
       author = {{Teague}, Richard and {Bae}, Jaehan and {Aikawa}, Yuri and {Andrews}, Sean M. and {Bergin}, Edwin A. and {Bergner}, Jennifer B. and {Boehler}, Yann and {Booth}, Alice S. and {Bosman}, Arthur D. and {Cataldi}, Gianni and {Czekala}, Ian and {Guzm{\'a}n}, Viviana V. and {Huang}, Jane and {Ilee}, John D. and {Law}, Charles J. and {Le Gal}, Romane and {Long}, Feng and {Loomis}, Ryan A. and {M{\'e}nard}, Fran{\c{c}}ois and {{\"O}berg}, Karin I. and {P{\'e}rez}, Laura M. and {Schwarz}, Kamber R. and {Sierra}, Anibal and {Walsh}, Catherine and {Wilner}, David J. and {Yamato}, Yoshihide and {Zhang}, Ke},
        title = "{Molecules with ALMA at Planet-forming Scales (MAPS). XVIII. Kinematic Substructures in the Disks of HD 163296 and MWC 480}",
      journal = {\apjs},
         year = 2021,
        month = nov,
       volume = {257},
       number = {1},
          eid = {18},
        pages = {18},
          doi = {10.3847/1538-4365/ac1438},
archivePrefix = {arXiv},
       eprint = {2109.06218},
 primaryClass = {astro-ph.EP},
       adsurl = {https://ui.adsabs.harvard.edu/abs/2021ApJS..257...18T}
}

@ARTICLE{Fairlamb2015,
       author = {{Fairlamb}, J.~R. and {Oudmaijer}, R.~D. and {Mendigut{\'\i}a}, I. and {Ilee}, J.~D. and {van den Ancker}, M.~E.},
        title = "{A spectroscopic survey of Herbig Ae/Be stars with X-shooter - I. Stellar parameters and accretion rates}",
      journal = {\mnras},
         year = 2015,
        month = oct,
       volume = {453},
       number = {1},
        pages = {976-1001},
          doi = {10.1093/mnras/stv1576},
archivePrefix = {arXiv},
       eprint = {1507.05967},
 primaryClass = {astro-ph.SR},
       adsurl = {https://ui.adsabs.harvard.edu/abs/2015MNRAS.453..976F}
}

@ARTICLE{Sierra2021,
       author = {{Sierra}, Anibal and {P{\'e}rez}, Laura M. and {Zhang}, Ke and {Law}, Charles J. and {Guzm{\'a}n}, Viviana V. and {Qi}, Chunhua and {Bosman}, Arthur D. and {{\"O}berg}, Karin I. and {Andrews}, Sean M. and {Long}, Feng and {Teague}, Richard and {Booth}, Alice S. and {Walsh}, Catherine and {Wilner}, David J. and {M{\'e}nard}, Fran{\c{c}}ois and {Cataldi}, Gianni and {Czekala}, Ian and {Bae}, Jaehan and {Huang}, Jane and {Bergner}, Jennifer B. and {Ilee}, John D. and {Benisty}, Myriam and {Le Gal}, Romane and {Loomis}, Ryan A. and {Tsukagoshi}, Takashi and {Liu}, Yao and {Yamato}, Yoshihide and {Aikawa}, Yuri},
        title = "{Molecules with ALMA at Planet-forming Scales (MAPS). XIV. Revealing Disk Substructures in Multiwavelength Continuum Emission}",
      journal = {\apjs},
         year = 2021,
        month = nov,
       volume = {257},
       number = {1},
          eid = {14},
        pages = {14},
          doi = {10.3847/1538-4365/ac1431},
archivePrefix = {arXiv},
       eprint = {2109.06433},
 primaryClass = {astro-ph.EP},
       adsurl = {https://ui.adsabs.harvard.edu/abs/2021ApJS..257...14S}
}

@ARTICLE{Codella2017,
       author = {{Codella}, C. and {Ceccarelli}, C. and {Caselli}, P. and {Balucani}, N. and {Barone}, V. and {Fontani}, F. and {Lefloch}, B. and {Podio}, L. and {Viti}, S. and {Feng}, S. and {Bachiller}, R. and {Bianchi}, E. and {Dulieu}, F. and {Jim{\'e}nez-Serra}, I. and {Holdship}, J. and {Neri}, R. and {Pineda}, J.~E. and {Pon}, A. and {Sims}, I. and {Spezzano}, S. and {Vasyunin}, A.~I. and {Alves}, F. and {Bizzocchi}, L. and {Bottinelli}, S. and {Caux}, E. and {Chac{\'o}n-Tanarro}, A. and {Choudhury}, R. and {Coutens}, A. and {Favre}, C. and {Hily-Blant}, P. and {Kahane}, C. and {Jaber Al-Edhari}, A. and {Laas}, J. and {L{\'o}pez-Sepulcre}, A. and {Ospina}, J. and {Oya}, Y. and {Punanova}, A. and {Puzzarini}, C. and {Quenard}, D. and {Rimola}, A. and {Sakai}, N. and {Skouteris}, D. and {Taquet}, V. and {Testi}, L. and {Theul{\'e}}, P. and {Ugliengo}, P. and {Vastel}, C. and {Vazart}, F. and {Wiesenfeld}, L. and {Yamamoto}, S.},
        title = "{Seeds of Life in Space (SOLIS). II. Formamide in protostellar shocks: Evidence for gas-phase formation}",
      journal = {\aap},
         year = 2017,
        month = sep,
       volume = {605},
          eid = {L3},
        pages = {L3},
          doi = {10.1051/0004-6361/201731249},
archivePrefix = {arXiv},
       eprint = {1708.04663},
 primaryClass = {astro-ph.EP},
       adsurl = {https://ui.adsabs.harvard.edu/abs/2017A&A...605L...3C}
}

@ARTICLE{Duan2025,
       author = {{Duan}, Chunguo and {Gou}, Qian and {Liu}, Tie and {Xu}, Fengwei and {Xu}, Xuefang and {Lan}, Junlin and {Wang}, Ke and {Pagani}, Laurent and {Quan}, Donghui and {Wang}, Junzhi and {Liu}, Xunchuan and {He}, Mingwei},
        title = "{An ALMA Study of Molecular Complexity in the Hot Core G336.99-00.03 MM1}",
      journal = {\apj},
         year = 2025,
        month = jul,
       volume = {988},
       number = {1},
          eid = {95},
        pages = {95},
          doi = {10.3847/1538-4357/addbd6},
archivePrefix = {arXiv},
       eprint = {2505.17403},
 primaryClass = {astro-ph.GA},
       adsurl = {https://ui.adsabs.harvard.edu/abs/2025ApJ...988...95D}
}

@ARTICLE{Booth2023_hd169142,
       author = {{Booth}, Alice S. and {Law}, Charles J. and {Temmink}, Milou and {Leemker}, Margot and {Mac{\'\i}as}, Enrique},
        title = "{Tracing snowlines and C/O ratio in a planet-hosting disk. ALMA molecular line observations towards the HD 169142 disk}",
      journal = {\aap},
         year = 2023,
        month = oct,
       volume = {678},
          eid = {A146},
        pages = {A146},
          doi = {10.1051/0004-6361/202346974},
archivePrefix = {arXiv},
       eprint = {2308.07910},
 primaryClass = {astro-ph.EP},
       adsurl = {https://ui.adsabs.harvard.edu/abs/2023A&A...678A.146B}
}

@ARTICLE{Booth2025_hd100453,
       author = {{Booth}, Alice S. and {W{\"o}lfer}, Lisa and {Temmink}, Milou and {Calahan}, Jenny and {Evans}, Lucy and {Law}, Charles J. and {Leemker}, Margot and {Notsu}, Shota and {{\"O}berg}, Karin and {Walsh}, Catherine},
        title = "{Ice Sublimation in the Dynamic HD 100453 Disk Reveals a Rich Reservoir of Inherited Complex Organics}",
      journal = {\apjl},
         year = 2025,
        month = jun,
       volume = {986},
       number = {1},
          eid = {L9},
        pages = {L9},
          doi = {10.3847/2041-8213/adc7b2},
archivePrefix = {arXiv},
       eprint = {2504.14023},
 primaryClass = {astro-ph.EP},
       adsurl = {https://ui.adsabs.harvard.edu/abs/2025ApJ...986L...9B}
}

@ARTICLE{Vandermarel2021_IRS48,
       author = {{van der Marel}, Nienke and {Booth}, Alice S. and {Leemker}, Margot and {van Dishoeck}, Ewine F. and {Ohashi}, Satoshi},
        title = "{A major asymmetric ice trap in a planet-forming disk. I. Formaldehyde and methanol}",
      journal = {\aap},
         year = 2021,
        month = jul,
       volume = {651},
          eid = {L5},
        pages = {L5},
          doi = {10.1051/0004-6361/202141051},
archivePrefix = {arXiv},
       eprint = {2104.08906},
 primaryClass = {astro-ph.EP},
       adsurl = {https://ui.adsabs.harvard.edu/abs/2021A&A...651L...5V}
}

@ARTICLE{Booth2024_100546,
       author = {{Booth}, Alice S. and {Leemker}, Margot and {van Dishoeck}, Ewine F. and {Evans}, Lucy and {Ilee}, John D. and {Kama}, Mihkel and {Keyte}, Luke and {Law}, Charles J. and {van der Marel}, Nienke and {Nomura}, Hideko and {Notsu}, Shota and {{\"O}berg}, Karin and {Temmink}, Milou and {Walsh}, Catherine},
        title = "{An ALMA Molecular Inventory of Warm Herbig Ae Disks. I. Molecular Rings, Asymmetries, and Complexity in the HD 100546 Disk}",
      journal = {\aj},
         year = 2024,
        month = apr,
       volume = {167},
       number = {4},
          eid = {164},
        pages = {164},
          doi = {10.3847/1538-3881/ad2700},
archivePrefix = {arXiv},
       eprint = {2402.04001},
 primaryClass = {astro-ph.EP},
       adsurl = {https://ui.adsabs.harvard.edu/abs/2024AJ....167..164B}
}

@ARTICLE{Romero-Mirza2023,
       author = {{Romero-Mirza}, Carlos E. and {{\"O}berg}, Karin I. and {Law}, Charles J. and {Teague}, Richard and {Aikawa}, Yuri and {Bergner}, Jennifer B. and {Wilner}, David J. and {Huang}, Jane and {Guzm{\'a}n}, Viviana V. and {Cleeves}, L. Ilsedore},
        title = "{Cold Deuterium Fractionation in the Nearest Planet-forming Disk}",
      journal = {\apj},
         year = 2023,
        month = jan,
       volume = {943},
       number = {1},
          eid = {35},
        pages = {35},
          doi = {10.3847/1538-4357/aca765},
archivePrefix = {arXiv},
       eprint = {2212.06912},
 primaryClass = {astro-ph.EP},
       adsurl = {https://ui.adsabs.harvard.edu/abs/2023ApJ...943...35R}
}

@ARTICLE{Paneque-Carreno2023,
       author = {{Paneque-Carre{\~n}o}, T. and {Miotello}, A. and {van Dishoeck}, E.~F. and {Tabone}, B. and {Izquierdo}, A.~F. and {Facchini}, S.},
        title = "{Directly tracing the vertical stratification of molecules in protoplanetary disks}",
      journal = {\aap},
         year = 2023,
        month = jan,
       volume = {669},
          eid = {A126},
        pages = {A126},
          doi = {10.1051/0004-6361/202244428},
archivePrefix = {arXiv},
       eprint = {2210.01130},
 primaryClass = {astro-ph.EP},
       adsurl = {https://ui.adsabs.harvard.edu/abs/2023A&A...669A.126P}
}

@ARTICLE{Keyte2024,
       author = {{Keyte}, Luke and {Kama}, Mihkel and {Chuang}, Ko-Ju and {Cleeves}, L. Ilsedore and {Drozdovskaya}, Maria N. and {Furuya}, Kenji and {Rawlings}, Jonathan and {Shorttle}, Oliver},
        title = "{Spatially resolving the volatile sulfur abundance in the HD 100546 protoplanetary disc}",
      journal = {\mnras},
         year = 2024,
        month = feb,
       volume = {528},
       number = {1},
        pages = {388-407},
          doi = {10.1093/mnras/stae019},
archivePrefix = {arXiv},
       eprint = {2312.13997},
 primaryClass = {astro-ph.EP},
       adsurl = {https://ui.adsabs.harvard.edu/abs/2024MNRAS.528..388K}
}

@ARTICLE{Keyte2024_b,
       author = {{Keyte}, Luke and {Kama}, Mihkel and {Booth}, Alice S. and {Law}, Charles J. and {Leemker}, Margot},
        title = "{Volatile composition of the HD 169142 disc and its embedded planet}",
      journal = {\mnras},
         year = 2024,
        month = nov,
       volume = {534},
       number = {4},
        pages = {3576-3594},
          doi = {10.1093/mnras/stae2314},
archivePrefix = {arXiv},
       eprint = {2410.01418},
 primaryClass = {astro-ph.EP},
       adsurl = {https://ui.adsabs.harvard.edu/abs/2024MNRAS.534.3576K}
}
\bibliographystyle{aasjournal}

\end{document}